\documentclass[11pt]{article}
\usepackage{amsmath,amsfonts,amssymb,graphicx,epsfig,pdflscape,multirow,theorem}
\usepackage{arydshln,cite}
\numberwithin{equation}{section}
\graphicspath{{./eps/}}
\usepackage{cite}
\usepackage[usenames]{color}
\usepackage{pstricks}
\usepackage{pst-plot}
\usepackage[backref=false]{hyperref}
\hypersetup{
colorlinks=true,
citecolor=red,
linkcolor=darkblue,
urlcolor=darkblue
}

\long\def\ignore#1{}
\definecolor{darkblue}{rgb}{0,0,.8}
\definecolor{red}{rgb}{1,0,0}
\definecolor{purple}{rgb}{1,0,1}
\definecolor{coloroflink}{rgb}{0.7,0,1}
\definecolor{coloroflink}{rgb}{0.180392, 0.545098, 0.341176}
\definecolor{darkpurple}{rgb}{1,.2,1}
\definecolor{pink}{rgb}{1,.7,.7}
\definecolor{lightblue}{rgb}{.61,.61,1}
\definecolor{midblue}{rgb}{.7,.7,1}
\definecolor{lightlightblue}{rgb}{.9,.9,1}
\definecolor{lightestblue}{rgb}{.96,.96,1}
\definecolor{lightpurple}{rgb}{1,.65,1}
\definecolor{darkgreen}{rgb}{0.180392, 0.545098, 0.341176}

\theorembodyfont{\itshape} 
\theoremheaderfont{\scshape}
\theoremstyle{plain}  
\newtheorem{Lemme}{Lemma}[section]
\newtheorem{Theoreme}[Lemme]{Theorem}
\newtheorem{Proposition}[Lemme]{Proposition}
\newtheorem{Lemma}[Lemme]{Lemma}
\newtheorem{Corollaire}[Lemme]{Corollary}

\numberwithin{equation}{section}

\newcommand{\nc}{\newcommand}
\nc{\bib}{\bibitem}
\nc{\be}{\begin{equation}}
\nc{\ee}{\end{equation}}
\nc{\nn}{\nonumber\\ }
\nc{\chit}{\raisebox{0.25ex}{$\chi$}}
\nc{\chih}{\raisebox{0.25ex}{$\hat\chi$}}
\nc{\ir}{\mathrm{i}}

\nc{\Db}{\mbox{\boldmath $D$}}
\nc{\funkyDb}{\boldsymbol{\mathcal D}}
\nc{\Tb}{\mbox{\boldmath $T$}}
\nc{\Ib}{\mbox{\boldmath $I$}}

\def\wobbly{
\rput{90}(0,0){
\psplot[linecolor=coloroflink,linewidth=1.5pt]
    {0}{0.11}{x 0.0005 div sin 25 div x mul 0.1 div}
\psplot[linecolor=coloroflink,linewidth=1.5pt]
    {0.1}{0.5}{x 0.0005 div sin 25 div }}       
}
\def\wobblytwo{
\rput{90}(0,0){
\psplot[linecolor=coloroflink,linewidth=1.25pt]
    {0}{0.11}{x 0.0005 div sin 25 div x mul 0.1 div}
\psplot[linecolor=coloroflink,linewidth=1.25pt]
    {0.1}{0.5}{x 0.0005 div sin 25 div }}       
}

\def\dimvcoord{
\pspolygon[fillstyle=solid,linewidth=0pt,linecolor=white]
(-0.2,0)(-0.2,1.)(-0.199605,1.01256)(-0.198423,1.02507)(-0.196457,1.03748)(-0.193717,1.04974)(-0.190211,1.0618)(-0.185955,1.07362)(-0.180965,1.08516)(-0.175261,1.09635)(-0.168866,1.10717)(-0.161803,1.11756)(-0.154103,1.12748)(-0.145794,1.13691)(-0.136909,1.14579)(-0.127485,1.1541)(-0.117557,1.1618)(-0.107165,1.16887)(-0.0963507,1.17526)(-0.0851559,1.18097)(-0.0736249,1.18596)(-0.0618034,1.19021)(-0.049738,1.19372)(-0.0374763,1.19646)(-0.0250666,1.19842)(-0.0125581,1.19961)(0.,1.2)(0.0125581,1.19961)(0.0250666,1.19842)(0.0374763,1.19646)(0.049738,1.19372)(0.0618034,1.19021)(0.0736249,1.18596)(0.0851559,1.18097)(0.0963507,1.17526)(0.107165,1.16887)(0.117557,1.1618)(0.127485,1.1541)(0.136909,1.14579)(0.145794,1.13691)(0.154103,1.12748)(0.161803,1.11756)(0.168866,1.10717)(0.175261,1.09635)(0.180965,1.08516)(0.185955,1.07362)(0.190211,1.0618)(0.193717,1.04974)(0.196457,1.03748)(0.198423,1.02507)(0.199605,1.01256)(0.2,1.)(0.2,0.)(0.199605,-0.0125581)(0.198423,-0.0250666)(0.196457,-0.0374763)(0.193717,-0.049738)(0.190211,-0.0618034)(0.185955,-0.0736249)(0.180965,-0.0851559)(0.175261,-0.0963507)(0.168866,-0.107165)(0.161803,-0.117557)(0.154103,-0.127485)(0.145794,-0.136909)(0.136909,-0.145794)(0.127485,-0.154103)(0.117557,-0.161803)(0.107165,-0.168866)(0.0963507,-0.175261)(0.0851559,-0.180965)(0.0736249,-0.185955)(0.0618034,-0.190211)(0.049738,-0.193717)(0.0374763,-0.196457)(0.0250666,-0.198423)(0.0125581,-0.199605)(0.,-0.2)(-0.0125581,-0.199605)(-0.0250666,-0.198423)(-0.0374763,-0.196457)(-0.049738,-0.193717)(-0.0618034,-0.190211)(-0.0736249,-0.185955)(-0.0851559,-0.180965)(-0.0963507,-0.175261)(-0.107165,-0.168866)(-0.117557,-0.161803)(-0.127485,-0.154103)(-0.136909,-0.145794)(-0.145794,-0.136909)(-0.154103,-0.127485)(-0.161803,-0.117557)(-0.168866,-0.107165)(-0.175261,-0.0963507)(-0.180965,-0.0851559)(-0.185955,-0.0736249)(-0.190211,-0.0618034)(-0.193717,-0.049738)(-0.196457,-0.0374763)(-0.198423,-0.0250666)(-0.199605,-0.0125581)(-0.2,0.)
}

\def\dimhcoord{
\pspolygon[fillstyle=solid,linewidth=0pt,linecolor=white]
(0,-0.2)(1.,-0.2)(1.01256,-0.199605)(1.02507,-0.198423)(1.03748,-0.196457)(1.04974,-0.193717)(1.0618,-0.190211)(1.07362,-0.185955)(1.08516,-0.180965)(1.09635,-0.175261)(1.10717,-0.168866)(1.11756,-0.161803)(1.12748,-0.154103)(1.13691,-0.145794)(1.14579,-0.136909)(1.1541,-0.127485)(1.1618,-0.117557)(1.16887,-0.107165)(1.17526,-0.0963507)(1.18097,-0.0851559)(1.18596,-0.0736249)(1.19021,-0.0618034)(1.19372,-0.049738)(1.19646,-0.0374763)(1.19842,-0.0250666)(1.19961,-0.0125581)(1.2,0.)(1.19961,0.0125581)(1.19842,0.0250666)(1.19646,0.0374763)(1.19372,0.049738)(1.19021,0.0618034)(1.18596,0.0736249)(1.18097,0.0851559)(1.17526,0.0963507)(1.16887,0.107165)(1.1618,0.117557)(1.1541,0.127485)(1.14579,0.136909)(1.13691,0.145794)(1.12748,0.154103)(1.11756,0.161803)(1.10717,0.168866)(1.09635,0.175261)(1.08516,0.180965)(1.07362,0.185955)(1.0618,0.190211)(1.04974,0.193717)(1.03748,0.196457)(1.02507,0.198423)(1.01256,0.199605)(1.,0.2)(0.,0.2)(-0.0125581,0.199605)(-0.0250666,0.198423)(-0.0374763,0.196457)(-0.049738,0.193717)(-0.0618034,0.190211)(-0.0736249,0.185955)(-0.0851559,0.180965)(-0.0963507,0.175261)(-0.107165,0.168866)(-0.117557,0.161803)(-0.127485,0.154103)(-0.136909,0.145794)(-0.145794,0.136909)(-0.154103,0.127485)(-0.161803,0.117557)(-0.168866,0.107165)(-0.175261,0.0963507)(-0.180965,0.0851559)(-0.185955,0.0736249)(-0.190211,0.0618034)(-0.193717,0.049738)(-0.196457,0.0374763)(-0.198423,0.0250666)(-0.199605,0.0125581)(-0.2,0.)(-0.199605,-0.0125581)(-0.198423,-0.0250666)(-0.196457,-0.0374763)(-0.193717,-0.049738)(-0.190211,-0.0618034)(-0.185955,-0.0736249)(-0.180965,-0.0851559)(-0.175261,-0.0963507)(-0.168866,-0.107165)(-0.161803,-0.117557)(-0.154103,-0.127485)(-0.145794,-0.136909)(-0.136909,-0.145794)(-0.127485,-0.154103)(-0.117557,-0.161803)(-0.107165,-0.168866)(-0.0963507,-0.175261)(-0.0851559,-0.180965)(-0.0736249,-0.185955)(-0.0618034,-0.190211)(-0.049738,-0.193717)(-0.0374763,-0.196457)(-0.0250666,-0.198423)(-0.0125581,-0.199605)(0.,-0.2)
}

\def\halfdimcoordup{
\pspolygon[fillstyle=solid,linewidth=0pt,linecolor=white](-0.2,0)(-0.2,0.5)(0.2,0.5)(0.2,0.)(0.199605,-0.0125581)(0.198423,-0.0250666)(0.196457,-0.0374763)(0.193717,-0.049738)(0.190211,-0.0618034)(0.185955,-0.0736249)(0.180965,-0.0851559)(0.175261,-0.0963507)(0.168866,-0.107165)(0.161803,-0.117557)(0.154103,-0.127485)(0.145794,-0.136909)(0.136909,-0.145794)(0.127485,-0.154103)(0.117557,-0.161803)(0.107165,-0.168866)(0.0963507,-0.175261)(0.0851559,-0.180965)(0.0736249,-0.185955)(0.0618034,-0.190211)(0.049738,-0.193717)(0.0374763,-0.196457)(0.0250666,-0.198423)(0.0125581,-0.199605)(0.,-0.2)(-0.0125581,-0.199605)(-0.0250666,-0.198423)(-0.0374763,-0.196457)(-0.049738,-0.193717)(-0.0618034,-0.190211)(-0.0736249,-0.185955)(-0.0851559,-0.180965)(-0.0963507,-0.175261)(-0.107165,-0.168866)(-0.117557,-0.161803)(-0.127485,-0.154103)(-0.136909,-0.145794)(-0.145794,-0.136909)(-0.154103,-0.127485)(-0.161803,-0.117557)(-0.168866,-0.107165)(-0.175261,-0.0963507)(-0.180965,-0.0851559)(-0.185955,-0.0736249)(-0.190211,-0.0618034)(-0.193717,-0.049738)(-0.196457,-0.0374763)(-0.198423,-0.0250666)(-0.199605,-0.0125581)(-0.2,0.)
}

\def\halfdimcoorddown{
\rput(0,-1){\pspolygon[fillstyle=solid,linewidth=0pt,linecolor=white](-0.2,0.5)
(-0.2,1.)(-0.199605,1.01256)(-0.198423,1.02507)(-0.196457,1.03748)(-0.193717,1.04974)(-0.190211,1.0618)(-0.185955,1.07362)(-0.180965,1.08516)(-0.175261,1.09635)(-0.168866,1.10717)(-0.161803,1.11756)(-0.154103,1.12748)(-0.145794,1.13691)(-0.136909,1.14579)(-0.127485,1.1541)(-0.117557,1.1618)(-0.107165,1.16887)(-0.0963507,1.17526)(-0.0851559,1.18097)(-0.0736249,1.18596)(-0.0618034,1.19021)(-0.049738,1.19372)(-0.0374763,1.19646)(-0.0250666,1.19842)(-0.0125581,1.19961)(0.,1.2)(0.0125581,1.19961)(0.0250666,1.19842)(0.0374763,1.19646)(0.049738,1.19372)(0.0618034,1.19021)(0.0736249,1.18596)(0.0851559,1.18097)(0.0963507,1.17526)(0.107165,1.16887)(0.117557,1.1618)(0.127485,1.1541)(0.136909,1.14579)(0.145794,1.13691)(0.154103,1.12748)(0.161803,1.11756)(0.168866,1.10717)(0.175261,1.09635)(0.180965,1.08516)(0.185955,1.07362)(0.190211,1.0618)(0.193717,1.04974)(0.196457,1.03748)(0.198423,1.02507)(0.199605,1.01256)(0.2,1.)(0.2,0.5)}
}

\def\arxiv#1#2{\href{http://arxiv.org/abs/#1}{arXiv:#1 #2}}

\def\a{\alpha}
\def\b{\beta}
\newcommand\bea{\begin{eqnarray}}
\newcommand\eea{\end{eqnarray}}

\def\ra{\rangle}

\def\dimerv{
\psset{fillcolor=lightblue}\dimvcoord
}

\def\dimerh{
\psset{fillcolor=lightpurple}\dimhcoord
}

\def\halfdimervup{
\psset{fillcolor=lightblue}\halfdimcoordup
}

\def\halfdimervdown{
\psset{fillcolor=lightblue}\halfdimcoorddown
}

\def\uarrow{
\psline[linecolor=blue,linewidth=1.2pt,arrowscale=0.8,arrowinset=0.1]{->}(0,0.15)(0,0.7)
}

\def\darrow{
\psline[linecolor=purple,linewidth=1.2pt,arrowscale=0.8,arrowinset=0.1]{->}(0,0.7)(0,0.15)
}

\makeatletter
\def\Ddots{\mathinner{\mkern1mu\raise\p@
\vbox{\kern7\p@\hbox{.}}\mkern2mu
\raise4\p@\hbox{.}\mkern2mu\raise7\p@\hbox{.}\mkern1mu}}
\makeatother

\makeatletter
\def\Ddots{\mathinner{\mkern1mu\raise\p@
\vbox{\kern7\p@\hbox{.}}\mkern2mu
\raise4\p@\hbox{.}\mkern2mu\raise7\p@\hbox{.}\mkern1mu}}
\makeatother

\begin{document}

\topmargin -15mm
\oddsidemargin 05mm

%
%

\title{\mbox{}\vspace{-.2in}
\bf 
Dimer representations of the Temperley-Lieb algebra
}

\date{}
\maketitle

\begin{center}
{\vspace{-5mm}\Large Alexi Morin-Duchesne$^\ast$,\, J{\o}rgen Rasmussen$^\natural$,\, Philippe Ruelle$^\ast$}
\\[.5cm]
{\em {}$^\ast$Institut de Recherche en Math\'ematique et Physique\\ Universit\'e Catholique de Louvain, Louvain-la-Neuve, B-1348, Belgium}
\\[.2cm]
{\em {}$^\natural$School of Mathematics and Physics, University of Queensland}\\
{\em St Lucia, Brisbane, Queensland 4072, Australia}
\\[.4cm] 
{\tt alexi.morin-duchesne\,@\,uclouvain.be}
\qquad
{\tt j.rasmussen\,@\,uq.edu.au}
\quad
{\tt philippe.ruelle\,@\,uclouvain.be}
\end{center}

%
%

\vspace{0.5cm}

\begin{abstract}
A new spin-chain representation of the Temperley-Lieb algebra ${\rm TL}_n(\beta=0)$ is introduced and related to the dimer model. 
Unlike the usual XXZ spin-chain representations of dimension $2^n$, this dimer representation is of dimension $2^{n-1}$. A detailed 
analysis of its structure is presented and found to yield indecomposable zigzag modules. 
\vspace{0.2cm}

\noindent Keywords: dimer model, Temperley-Lieb algebra, quantum spin-chains, representation theory.
\end{abstract}

%
%


\tableofcontents
\clearpage

\section{Introduction}
\label{sec:Introduction}

The classical dimer model describes perfect domino tilings or coverings of a lattice by $1\times2$ and $2\times1$ rectangles. 
It can be traced back to a paper by Fowler and Rushbrooke~\cite{FR37} from 1937, with many fundamental results~\cite{Kas61,TF61} 
obtained in the 1960's, see also the review~\cite{Kenyon}.
The transfer matrix approach by Lieb~\cite{Lieb67}, in particular, uses tools of statistical mechanics to describe the combinatorial 
problem on the square lattice and was recently revisited~\cite{RR12} in a study of the conformal properties arising in the continuum 
scaling limit of the model. Lieb's approach is based on a map from dimer configurations to spin configurations and thus opens
the door to study the dimer model using the machinery of spin-chains.

The open Heisenberg and XXZ spin-chains, in particular, are known~\cite{PS90} to yield representations of the 
Temperley-Lieb algebra TL$_n(\beta)$~\cite{TL,Jones} where $n$ is the number of sites and $\beta$ the loop fugacity.
These spin-chain representations are constructed in terms of Pauli matrices acting on $(\mathbb{C}^2)^{\otimes n}$ and are thus
of dimension $2^n$. 
However, these representations have not found applications in the dimer model.

Here we offer a new spin-chain representation of ${\rm TL}_n(\beta=0)$. 
It is also constructed in terms of Pauli matrices, but unlike the familiar spin-chain representations, it acts on one fewer spin-$\frac{1}{2}$ 
site and is therefore only of dimension $2^{n-1}$.
It is additionally distinguished by the property that each of the Temperley-Lieb generators $e_j$,  $2 \le j \le n-2$, acts on three 
consecutive spin sites instead of the usual two as in the open XXZ spin-chain. The generators $e_1$ and $e_{n-1}$ act only on two spin 
sites each.

This new spin-chain representation is furthermore shown to be linked to the dimer transfer matrix of Lieb. 
Following the separation of the dimer configuration space into
sectors~\cite{RR12}, the corresponding separation of the 
spin-chain representation yields a family of ${\rm TL}_n(0)$ modules 
$E_{n-1}^v$ labeled by the total magnetisation $v$. A detailed analysis of these dimer representations 
is presented and found to yield indecomposable (zigzag) modules. 
These modules are known~\cite{Westbury} in the representation theory of the 
Temperley-Lieb algebra~\cite{MartinBook,GL98,RSA1204} and can be constructed as quotients of 
direct sums of projective modules in the XXZ spin-chain models~\cite{GV12}.
However, the new spin-chain representations seem to be the first examples, coming directly from a physical model, in which the zigzag modules appear as direct summands.

The layout of this paper is as follows.
In Section~\ref{sec:Dimers}, we review some basics of the dimer model, Lieb's transfer matrix approach 
and the variation index operator separating the configuration space into sectors.
In Section~\ref{sec:TL}, we review the basics of the Temperley-Lieb algebra ${\rm TL}_n(\beta)$, parts of its representation theory and the
specialisation to $\beta=0$.
In Section~\ref{sec:Dimersrep}, we introduce the new spin-chain representation of ${\rm TL}_n(0)$ and relate it to the dimer model.
We also present the main result on the structure of the ${\rm TL}_n(0)$ modules $E_{n-1}^v$, but defer the proof to 
Section~\ref{sec:Structure}.
In preparation for that, in Section~\ref{sec:Intertwiners}, we introduce three families of homomorphisms intertwining the various 
representations. Section~\ref{sec:Conclusion} contains some concluding remarks.

\section{Dimer model}
\label{sec:Dimers}

Here we briefly review some basics of the dimer model. The presentation follows the one in~\cite{RR12}.

\subsection{Statistical model}

\begin{figure}[t]
\psset{unit=.7cm}
\begin{center}
\begin{pspicture}(-0.2,-0.2)(8.2,5.2)
\multiput(0,0)(0,1){6}{\psline[linecolor=gray]{-}(0,0)(8,0)}\multiput(0,0)(1,0){9}{\psline[linecolor=gray]{-}(0,-0.5)(0,5.5)}
\rput(0,0){\dimerh}
\rput(0,1){\dimerh}
\rput(1,2){\dimerh}
\rput(1,3){\dimerh}
\rput(2,4){\dimerh}
\rput(3,5){\dimerh}
\rput(2,1){\dimerh}
\rput(3,0){\dimerh}
\rput(7,5){\dimerh}
\rput(6,3){\dimerh}
\rput(6,4){\dimerh}
\rput(7,2){\dimerh}
\rput(0,2){\dimerv}
\rput(0,4){\dimerv}
\rput(1,4){\dimerv}
\rput(3,2){\dimerv}
\rput(4,1){\dimerv}
\rput(4,3){\dimerv}
\rput(5,3){\dimerv}
\rput(5,1){\dimerv}
\rput(8,3){\dimerv}
\rput(6,1){\dimerv}
\rput(7,0){\dimerv}
\rput(8,0){\dimerv}
\rput(2,0){\halfdimervdown}
\rput(2,5){\halfdimervup}
\rput(5,0){\halfdimervdown}
\rput(5,5){\halfdimervup}
\rput(6,0){\halfdimervdown}
\rput(6,5){\halfdimervup}
\multiput(0,0)(0,1){6}{\multiput(0,0)(1,0){9}{\pscircle[linewidth=0.025,fillstyle=solid,fillcolor=gray](0,0){0.08}}}
\end{pspicture} 
\qquad \qquad
\begin{pspicture}(-0.2,-0.2)(8.2,5.2)
\rput(0,5){\darrow}\rput(1,5){\darrow}\rput(2,5){\uarrow}\rput(3,5){\darrow}\rput(4,5){\darrow}\rput(5,5){\uarrow}\rput(6,5){\uarrow}\rput(7,5){\darrow}\rput(8,5){\darrow}
\rput(0,4){\uarrow}\rput(1,4){\uarrow}\rput(2,4){\darrow}\rput(3,4){\darrow}\rput(4,4){\darrow}\rput(5,4){\darrow}\rput(6,4){\darrow}\rput(7,4){\darrow}\rput(8,4){\darrow}
\rput(0,3){\darrow}\rput(1,3){\darrow}\rput(2,3){\darrow}\rput(3,3){\darrow}\rput(4,3){\uarrow}\rput(5,3){\uarrow}\rput(6,3){\darrow}\rput(7,3){\darrow}\rput(8,3){\uarrow}
\rput(0,2){\uarrow}\rput(1,2){\darrow}\rput(2,2){\darrow}\rput(3,2){\uarrow}\rput(4,2){\darrow}\rput(5,2){\darrow}\rput(6,2){\darrow}\rput(7,2){\darrow}\rput(8,2){\darrow}
\rput(0,1){\darrow}\rput(1,1){\darrow}\rput(2,1){\darrow}\rput(3,1){\darrow}\rput(4,1){\uarrow}\rput(5,1){\uarrow}\rput(6,1){\uarrow}\rput(7,1){\darrow}\rput(8,1){\darrow}
\rput(0,0){\darrow}\rput(1,0){\darrow}\rput(2,0){\darrow}\rput(3,0){\darrow}\rput(4,0){\darrow}\rput(5,0){\darrow}\rput(6,0){\darrow}\rput(7,0){\uarrow}\rput(8,0){\uarrow}
\multiput(0,0)(0,1){6}{\multiput(0,0)(1,0){9}{\pscircle[linewidth=0.025,fillstyle=solid,fillcolor=gray](0,0){0.08}}}
\end{pspicture}
\caption{A dimer covering of the $6\times9$ cylinder and its corresponding spin configuration. 
Each spin is attached to the site located just below it.}
\label{fig:dimerconf}
\end{center}
\end{figure}
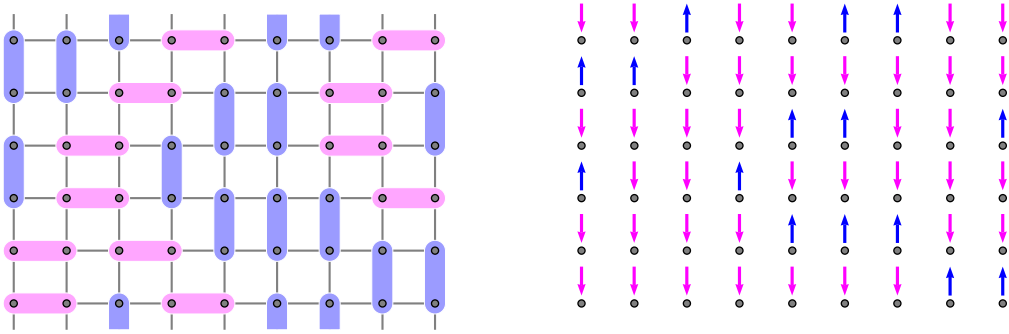

The dimer model discussed here is defined on an $M \times N$ rectangular grid with $M$ and $N$ respectively counting the 
number of rows and columns. Vertices 
on this lattice are referred to as sites and are labeled by their position $(i,j)$ with $i = 1, \dots, N$ and $j = 1, \dots, M$. A dimer is a 
$2\times1$ or a $1\times 2$ small bridge that covers two adjacent sites. In a dimer covering, every site is occupied by a 
single dimer, horizontal or vertical, and because each dimer covers exactly two sites, 
the set of dimer coverings is non-empty only if the product $MN$ is even. The 
boundary condition we are concerned with is that of a horizontal cylinder, meaning that for $i=1,\dots,N$, the sites $(i,1)$ and $(i,M)$ are 
considered as neighbours and can be covered by the same dimer. Sites on this cylindrical lattice have four neighbours to which they can 
be connected by a dimer, except for the sites $(1,j)$ and $(N,j)$ which have only three. The left panel of Figure~\ref{fig:dimerconf} shows 
an example of a dimer covering of the $6\times 9$ cylinder.

It is customary to assign to each covering a weight $\alpha^h$ where $h$ is the number of horizontal dimers while $\alpha\in\mathbb C$ 
is a free parameter measuring the relative preference for horizontal dimers over vertical ones. The partition function of the dimer model is 
the sum of the weights over all possible coverings,
\be
 Z(\a) = \sum_{\textrm{coverings}}\hspace{-0.15cm} \alpha^h,
\ee
so the total number of dimer coverings of the $M \times N$ cylinder is given by $Z(\alpha=1)$.

\subsection{Transfer matrix approach}

The use of a transfer matrix to calculate partition functions for the dimer model dates back to a 1967 paper by Lieb~\cite{Lieb67}. 
The first step is to build a map from dimer coverings to spin configurations. To every site in a given covering, one thus assigns an 
up-arrow if a dimer connects it with the site immediately above it, and a down-arrow otherwise. 
This map is illustrated in Figure~\ref{fig:dimerconf}.

The same map is also well defined locally. It sends a row of the dimer covering (of length $N$) to an element of the canonical basis of 
$(\mathbb C^2)^{\otimes N}$, the vector space spanned by $N$ $\tfrac12$-spins. For instance, 
\begin{equation}
\psset{unit=.7cm}
\begin{pspicture}(-0.2,-0.2)(3.2,0.2)
\rput(0,0){\halfdimervup}
\rput(1,0){\dimerh}
\rput(3,0){\halfdimervdown}
\multiput(0,0)(1,0){4}{\pscircle[linewidth=0.025,fillstyle=solid,fillcolor=gray](0,0){0.08}}
\end{pspicture}
\quad \rightarrow \quad 
\begin{pspicture}(-0.2,-0.2)(3.2,0.7)
\rput(0,0){\uarrow}
\rput(1,0){\darrow}
\rput(2,0){\darrow}
\rput(3,0){\darrow}
\multiput(0,0)(1,0){4}{\pscircle[linewidth=0.025,fillstyle=solid,fillcolor=gray](0,0){0.08}}
\end{pspicture}
\label{eq:example4}
\end{equation}

In general, the canonical basis
of $(\mathbb C^2)^{\otimes N}$
consists of elements of the form $| s \rangle = | s_1 s_2 \dots s_N \rangle$ with 
$s_i \in \{\uparrow, \downarrow\}$. Writing 
$\uparrow\;=\left(\begin{smallmatrix} 1\\ 0 \end{smallmatrix}\right)$ and 
$\downarrow\;=\left(\begin{smallmatrix} 0\\ 1\end{smallmatrix}\right)$, 
the dynamics at the level of the spin-chain is described using the Pauli matrices
\be
\sigma^x = \begin{pmatrix} 0 & 1 \\ 1 & 0 \end{pmatrix}, \quad 
\sigma^y = \begin{pmatrix} 0 & - \ir \\ \ir & 0 \end{pmatrix}, \quad
\sigma^z = \begin{pmatrix} 1 & 0 \\ 0 & -1 \end{pmatrix}, \quad
\sigma^+ = \begin{pmatrix} 0 & 1 \\ 0 & 0 \end{pmatrix}, \quad
\sigma^- = \begin{pmatrix} 0 & 0 \\ 1 & 0 \end{pmatrix}
\ee
and the corresponding operators on $(\mathbb C^2)^{\otimes N}$, 
\be
 \sigma^a_j = \underbrace{\mathbb I_2 \otimes \dots \otimes \mathbb I_2}_{j-1} \otimes \; \sigma^a \otimes 
  \underbrace{\mathbb I_2 \otimes \dots \otimes \mathbb I_2}_{N-j}, \qquad \big(a \in \{x,y,z,+,-\}\big),\qquad 
  \mathbb I_2=\begin{pmatrix} 1&0\\ 0&1 \end{pmatrix},
\ee 
that only modify $s_j$, the $j$-th component in the tensored space.

The transfer matrix acts on any state $|s \rangle$ by constructing all possible states in the row of spins just above it in a way consistent 
with the dimer coverings. This is done in two steps. The first step reverses all the spins by applying the operator 
\be 
 V_1 = \prod_{j=1}^N \sigma^x_j
\label{eq:V1}
\ee
on the spin state $|s\rangle$. At the level of the dimers, for sites where a dimer was connecting upwards, this transformation produces the 
top part of the dimer. For the other sites, those that were occupied by the top half of a vertical dimer or by half a horizontal one, $V_1$ 
produces a new dimer directed upwards. For example,
\medskip
\be
\psset{unit=.7cm}
\begin{pspicture}(-0.2,-0.2)(3.2,0.2)
\rput(0,0){\halfdimervup}
\rput(1,0){\dimerh}
\rput(3,0){\halfdimervdown}
\multiput(0,0)(1,0){4}{\pscircle[linewidth=0.025,fillstyle=solid,fillcolor=gray](0,0){0.08}}
\end{pspicture}
\qquad \overset{V_1}{\longrightarrow} \qquad 
\begin{pspicture}(-0.2,-0.2)(3.2,0.2)
\rput(0,0){\halfdimervdown}
\rput(1,0){\halfdimervup}
\rput(2,0){\halfdimervup}
\rput(3,0){\halfdimervup}
\multiput(0,0)(1,0){4}{\pscircle[linewidth=0.025,fillstyle=solid,fillcolor=gray](0,0){0.08}}
\end{pspicture}
\label{eq:transferexample}
\ee

\medskip
The second step is to replace the configuration just produced by a linear combination of states that takes into account the fact that sites
where upward-pointing dimers were just produced, if adjacent, can instead be occupied by a horizontal dimer. At the level of the 
spin-chain, this is implemented by the operator
\be 
 V_3 = \prod_{j=1}^{N-1} (\mathbb I+\alpha\; \sigma^-_j\sigma^-_{j+1}) = \exp\Big(\sum_{j=1}^{N-1}\alpha\; \sigma^-_j\sigma^-_{j+1}\Big)
\ee
which also incorporates the weight $\alpha$ of each added horizontal dimer. In the previous example,
\medskip
\be
\psset{unit=.7cm}
\begin{pspicture}(-0.3,-0.2)(3.3,0.2)
\rput(0,0){\halfdimervdown}
\rput(1,0){\halfdimervup}
\rput(2,0){\halfdimervup}
\rput(3,0){\halfdimervup}
\multiput(0,0)(1,0){4}{\pscircle[linewidth=0.025,fillstyle=solid,fillcolor=gray](0,0){0.08}}
\end{pspicture}
\quad \overset{V_3}{\longrightarrow} \quad 
\begin{pspicture}(-0.3,-0.2)(3.3,0.2)
\rput(0,0){\halfdimervdown}
\rput(1,0){\halfdimervup}
\rput(2,0){\halfdimervup}
\rput(3,0){\halfdimervup}
\multiput(0,0)(1,0){4}{\pscircle[linewidth=0.025,fillstyle=solid,fillcolor=gray](0,0){0.08}}
\end{pspicture} \ +  \alpha \
\begin{pspicture}(-0.3,-0.2)(3.3,0.2)
\rput(0,0){\halfdimervdown}
\rput(1,0){\dimerh}
\rput(3,0){\halfdimervup}
\multiput(0,0)(1,0){4}{\pscircle[linewidth=0.025,fillstyle=solid,fillcolor=gray](0,0){0.08}}
\end{pspicture}
 \ +  \alpha \
\begin{pspicture}(-0.3,-0.2)(3.3,0.2)
\rput(0,0){\halfdimervdown}
\rput(1,0){\halfdimervup}
\rput(2,0){\dimerh}
\multiput(0,0)(1,0){4}{\pscircle[linewidth=0.025,fillstyle=solid,fillcolor=gray](0,0){0.08}}
\end{pspicture}
\ee

\medskip\noindent
and the result reproduces all possible coverings of the row above the first configuration in \eqref{eq:transferexample}. The transfer matrix 
is the product $V_3V_1$,
\be
 T(\alpha) = \exp\Big(\sum_{j=1}^{N-1}\alpha\; \sigma^-_j\sigma^-_{j+1}\Big) \prod_{j=1}^N \sigma^x_j .
\ee
It is not hard to see that, for $\alpha \in \mathbb R$, 
$T(\alpha)$ is real and symmetric, thus rendering it diagonalisable with real eigenvalues. The 
partition function on the horizontal $M \times N$ cylinder is obtained by taking the trace of $T(\alpha)$ to the power $M$,
\be
Z(\a) = \textrm{Tr}\ T^M(\alpha).
\label{eq:Ztr}
\ee
The computation of $Z(\a)$ is thus reduced to the calculation of the eigenvalues of $T(\alpha)$, a matrix of dimension 
$2^N$. The full diagonalisation of (the square of) $T(\a)$ has been worked out in~\cite{RR12} using the techniques of~\cite{Lieb67}.

The relation \eqref{eq:Ztr} is subtle as the local map from rows of dimers to 
$(\mathbb C^2)^{\otimes N}$ is not one-to-one. 
For spin configurations with two or more adjacent down spins, the pre-image in terms of 
dimer row configurations is not unique, since both horizontal dimers and pairs of adjacent top-halves of vertical dimers are sent to down-arrows. This implies that the map is not locally injective. 

It was remarked without proof by Lieb \cite{Lieb67} that taking the trace assigns the correct weights and multiplicities to each dimer configuration. As this is nontrivial, let us outline why it is true.
The map from dimer coverings to spin states can be seen locally as a map from occupation states of vertical edges to spin states in 
$(\mathbb C^2)^{\otimes N}$. For \eqref{eq:example4}, for instance, this alternative interpretation corresponds to
\be
\psset{unit=.7cm}
\begin{pspicture}[shift=-0.7](-0.2,-0.3)(3.2,1.5)
\pspolygon[fillstyle=solid,fillcolor=lightgray,linecolor=white](-0.35,0.22)(3.35,0.22)(3.35,0.78)(-0.35,0.78)
\rput(0,0){\dimerv}
\rput(1,1){\halfdimervup}
\rput(1,0){\dimerh}
\rput(2,1){\dimerh}
\rput(3,0){\halfdimervdown}
\multiput(0,0)(1,0){4}{\pscircle[linewidth=0.025,fillstyle=solid,fillcolor=gray](0,0){0.08}}
\multiput(0,1)(1,0){4}{\pscircle[linewidth=0.025,fillstyle=solid,fillcolor=gray](0,0){0.08}}
\end{pspicture}
\quad \rightarrow \quad 
\begin{pspicture}[shift=-0.6](-0.2,-0.2)(3.2,1.2)
\pspolygon[fillstyle=solid,fillcolor=lightgray,linecolor=white](-0.35,0.20)(3.35,0.20)(3.35,0.80)(-0.35,0.80)
\rput(0,0.075){\uarrow}
\rput(1,0.075){\darrow}
\rput(2,0.075){\darrow}
\rput(3,0.075){\darrow}
\multiput(0,0)(1,0){4}{\pscircle[linewidth=0.025,fillstyle=solid,fillcolor=gray](0,0){0.08}}
\multiput(0,1)(1,0){4}{\pscircle[linewidth=0.025,fillstyle=solid,fillcolor=gray](0,0){0.08}}
\end{pspicture}
\vspace{.1cm}
\ee
At the level of edge occupations, it is 
readily seen that this map is injective and thus one-to-one locally, 
while the transfer matrix $T(\alpha)$
maps between consecutive rows of occupation states in a way consistent with the possible dimer coverings of the adjacent 
nodes. Taking the trace imposes that the same edge state appears as the in- and out-state and 
correctly yields the dimer partition function on the cylinder.

For reasons that will become clear later, we will henceforth work with the squared transfer matrix $T^2(\alpha)$ which can be 
conveniently written in a form where the reflection operator $V_1$ no longer appears,
\be 
 T^2(\alpha) =  \exp\Big(\sum_{j=1}^{N-1}\alpha\; \sigma^-_j\sigma^-_{j+1}\Big)  
  \exp\Big(\sum_{j=1}^{N-1}\alpha\; \sigma^+_j\sigma^+_{j+1}\Big).
\label{eq:T2+-}
\ee

\subsection{Variation index} 

From \eqref{eq:T2+-}, one can expand $T^2(\alpha)$ into sums and products of the operators 
\be
 \mathcal O^-_j = \sigma^-_j \sigma^-_{j+1},\qquad \mathcal O^+_j = \sigma^+_j \sigma^+_{j+1}, \qquad j = 1, \dots, N-1,
\ee 
which can therefore be viewed as building blocks in the transfer matrix construction. Because each of these operators consists of a pair
of $\sigma^+$ or a pair of $\sigma^-$ matrices acting on neighbouring sites, the operator
\be
\mathcal V = \tfrac12 \sum_{j=1}^N (-1)^j \sigma^z_j 
\ee
commutes with $\mathcal O^\pm_j$ and anticommutes with $V_1$~\cite{RR12}. 
As a consequence, $\mathcal V$ anticommutes with $T(\a)$ and therefore commutes with $T^2(\alpha)$,
\be 
\{\mathcal V,T(\a) \} = 0,\qquad [\mathcal V,T^2(\a)] = 0.
\ee 
Eigenspaces of $\mathcal V$ are thus stable under the action of $\mathcal O^\pm_j$. Under the action of $T^2(\alpha)$, the space
$(\mathbb{C}^2)^{\otimes N}$ then splits into {\em sectors} labeled by the eigenvalues $v$ of $\mathcal V$. These eigenvalues are of the 
form 
\be
v \in \{-\tfrac N 2, -\tfrac N 2 + 1, \dots, \tfrac N 2 - 1, \tfrac N 2 \}\label{eq:vvalues}
\ee
and take integer or half-integer values for $N$ even or odd, respectively.

Each element in the canonical basis of $(\mathbb C^2)^{\otimes N}$ is an eigenstate of $\mathcal V$.
For example, the states $| \!\uparrow \downarrow \uparrow \downarrow \cdots \ \rangle$ and 
$| \! \downarrow \uparrow \downarrow \uparrow \cdots \ \rangle$ each form one-dimensional subspaces corresponding to 
$v = -\tfrac N 2$ and $v = \tfrac N 2$, respectively. Their pre-images in terms of dimers are
\medskip
\be
\psset{unit=.7cm}
\begin{pspicture}(-0.2,-0.2)(4.2,0.2)
\rput(0,0){\halfdimervup}
\rput(1,0){\halfdimervdown}
\rput(2,0){\halfdimervup}
\rput(3,0){\halfdimervdown}
\rput(4,0){$\dots$}
\multiput(0,0)(1,0){4}{\pscircle[linewidth=0.025,fillstyle=solid,fillcolor=gray](0,0){0.08}}
\end{pspicture}
\qquad {\rm and}\qquad
\begin{pspicture}(-0.2,-0.2)(4.2,0.2)
\rput(0,0){\halfdimervdown}
\rput(1,0){\halfdimervup}
\rput(2,0){\halfdimervdown}
\rput(3,0){\halfdimervup}
\rput(4,0){$\dots$}
\multiput(0,0)(1,0){4}{\pscircle[linewidth=0.025,fillstyle=solid,fillcolor=gray](0,0){0.08}}
\end{pspicture}
\ee

\medskip\noindent
and it is obvious why $T^2(\alpha)$ leaves each of the two subspaces invariant. 

In general, upon acting on a spin state, the operator $\mathcal V$ is a measure of the number of times two adjacent spins have different 
orientations and it was baptised the {\em variation index} in~\cite{RR12}. The eigenspaces of $\mathcal V$ with eigenvalue $v$ are 
here denoted by $\bar E_N^v$ and have dimension
\be
 \dim \bar E_N^v = \begin{pmatrix} N \\ \tfrac N 2 - v\end{pmatrix}.
\label{dimENv}
\ee
It was moreover shown in~\cite{RR12} that each $\bar E_N^v$ forms an orbit under the action of the operators $\mathcal O_j^\pm$.

We note that the cylinder partition function \eqref{eq:Ztr} is equivalently obtained by first computing the trace of $T^M(\alpha)$ restricted to $\bar E_N^v$ and then taking the sum over all values of $v$, see \eqref{eq:vvalues}.
Other boundary conditions may be considered, for instance those pertaining to a rectangular domain, in which case 
the partition function is typically written as a matrix element of a power of the transfer matrix~\cite{BPPR}. 
Compared to the cylinder case, the main difference is that the degrees of freedom propagating from 
the in- to the out-state now belong to the subspace $v=0$ of the full configuration space.

\section{Temperley-Lieb algebra}
\label{sec:TL}

A brief review is presented of the Temperley-Lieb (TL) algebra~\cite{TL,Jones} on $n$ sites, $\mathrm{TL}_n(\beta)$, and some of its representation theory~\cite{MartinBook,Westbury,GL98,RSA1204}. Particular emphasis is put on aspects relevant for the discussion in Section~\ref{sec:Dimersrep} where the loop fugacity vanishes, $\beta=0$. It is noted that the corresponding TL algebra TL$_n(0)$ also underlies the critical dense polymer model
in~\cite{PR07}.

\subsection{Connectivities and algebraic relations}

The elementary objects spanning the TL$_n(\beta)$ algebra are {\em connectivities}. Let us draw a rectangle with $n$ nodes on the top 
edge and another $n$ on the bottom one. A connectivity is then a pairwise connection of these nodes by non-intersecting loop segments. 
For example, 
\be 
 c_1 =\!
\begin{pspicture}[shift=-0.45](-0.2,-0.5)(3.0,0.5)
\pspolygon[fillstyle=solid,fillcolor=lightlightblue](0,-0.5)(2.8,-0.5)(2.8,0.5)(0,0.5)
\psarc[linecolor=blue,linewidth=1.5pt]{-}(0.8,-0.5){0.2}{0}{180}
\psarc[linecolor=blue,linewidth=1.5pt]{-}(1.2,0.5){0.2}{180}{0}
\psarc[linecolor=blue,linewidth=1.5pt]{-}(2.4,0.5){0.2}{180}{0}
\psbezier[linecolor=blue,linewidth=1.5pt]{-}(0.2,-0.5)(0.2,0)(1.4,0)(1.4,-0.5)
\psbezier[linecolor=blue,linewidth=1.5pt]{-}(0.2,0.5)(0.2,-0.0)(1.8,0.0)(1.8,-0.5)
\psbezier[linecolor=blue,linewidth=1.5pt]{-}(0.6,0.5)(0.6,-0.0)(2.2,0.0)(2.2,-0.5)
\psbezier[linecolor=blue,linewidth=1.5pt]{-}(1.8,0.5)(1.8,0)(2.6,0)(2.6,-0.5)
\end{pspicture} 
\label{eq:c1}
\ee
is a connectivity in TL$_7(\beta)$. 
Two connectivities are considered equal if there exists a continuous deformation mapping one into the other 
while preserving the positions of the nodes.
In general, the number of connectivities is given by the Catalan number
\be
\dim\mathrm{TL}_n(\beta)=\frac{1}{n+1} \begin{pmatrix} 2n \\ n \end{pmatrix}.
\ee
The addition of connectivities is commutative, and linear combinations of these objects are called {\em tangles}. 
The TL$_n(\beta)$ algebra is then the vector space spanned by the connectivities, endowed with the following rule for the multiplication 
of connectivities. Let $c_1$ and $c_2$ be two connectivities in TL$_n(\beta)$. The product $c_1c_2$ is obtained by drawing $c_1$ 
under $c_2$, identifying the top edge of $c_1$ with the bottom edge of $c_2$ in such a way that the $n$ nodes of each of the identified 
edges coincide. In the resulting diagram, the intermediate edge, along which $c_1$ and $c_2$ were glued together, is removed. This 
produces a new rectangle where loop segments connect the nodes on the top and bottom edges. If closed loops are formed in the 
process, they are removed and replaced by the multiplicative factor of $\beta^\ell$ where $\ell$ is the number of removed loops. 
This multiplication prescription is illustrated by
\be c_1 c_2=\! 
\begin{pspicture}[shift=-0.9](-0.2,-0.5)(3.0,1.5)
\pspolygon[fillstyle=solid,fillcolor=lightlightblue](0,-0.5)(2.8,-0.5)(2.8,0.5)(0,0.5)
\psarc[linecolor=blue,linewidth=1.5pt]{-}(0.8,-0.5){0.2}{0}{180}
\psarc[linecolor=blue,linewidth=1.5pt]{-}(1.2,0.5){0.2}{180}{0}
\psarc[linecolor=blue,linewidth=1.5pt]{-}(2.4,0.5){0.2}{180}{0}
\psbezier[linecolor=blue,linewidth=1.5pt]{-}(0.2,-0.5)(0.2,0)(1.4,0)(1.4,-0.5)
\psbezier[linecolor=blue,linewidth=1.5pt]{-}(0.2,0.5)(0.2,-0.0)(1.8,0.0)(1.8,-0.5)
\psbezier[linecolor=blue,linewidth=1.5pt]{-}(0.6,0.5)(0.6,-0.0)(2.2,0.0)(2.2,-0.5)
\psbezier[linecolor=blue,linewidth=1.5pt]{-}(1.8,0.5)(1.8,0)(2.6,0)(2.6,-0.5)
\pspolygon[fillstyle=solid,fillcolor=lightlightblue](0,1.5)(2.8,1.5)(2.8,0.5)(0,0.5)
\psarc[linecolor=blue,linewidth=1.5pt]{-}(1.2,0.5){0.2}{0}{180}
\psarc[linecolor=blue,linewidth=1.5pt]{-}(2.4,0.5){0.2}{0}{180}
\psarc[linecolor=blue,linewidth=1.5pt]{-}(0.4,1.5){0.2}{180}{0}
\psarc[linecolor=blue,linewidth=1.5pt]{-}(1.6,1.5){0.2}{180}{0}
\psarc[linecolor=blue,linewidth=1.5pt]{-}(2.4,1.5){0.2}{180}{0}
\psbezier[linecolor=blue,linewidth=1.5pt]{-}(0.6,0.5)(0.6,1)(1.8,1)(1.8,0.5)
\psbezier[linecolor=blue,linewidth=1.5pt]{-}(0.2,0.5)(0.2,1.0)(1.0,1.0)(1.0,1.5)
\end{pspicture}
=\, \beta^2
\begin{pspicture}[shift=-0.45](-0.2,-0.5)(3.0,0.5)
\pspolygon[fillstyle=solid,fillcolor=lightlightblue](0,-0.5)(2.8,-0.5)(2.8,0.5)(0,0.5)
\psarc[linecolor=blue,linewidth=1.5pt]{-}(0.8,-0.5){0.2}{0}{180}
\psarc[linecolor=blue,linewidth=1.5pt]{-}(2.4,-0.5){0.2}{0}{180}
\psarc[linecolor=blue,linewidth=1.5pt]{-}(0.4,0.5){0.2}{180}{0}
\psarc[linecolor=blue,linewidth=1.5pt]{-}(1.6,0.5){0.2}{180}{0}
\psarc[linecolor=blue,linewidth=1.5pt]{-}(2.4,0.5){0.2}{180}{0}
\psbezier[linecolor=blue,linewidth=1.5pt]{-}(0.2,-0.5)(0.2,0)(1.4,0)(1.4,-0.5)
\psbezier[linecolor=blue,linewidth=1.5pt]{-}(1.8,-0.5)(1.8,0.0)(1.0,0.0)(1.0,0.5)
\end{pspicture} 
\ee

The algebra TL$_n(\beta)$ is defined alternatively in terms of a restricted set of generators, 
\be
\mathrm{TL}_n(\beta)=\big\langle I,\,e_j ;\,j=1,\ldots,n-1\big\rangle,\qquad
I =\!
\begin{pspicture}[shift=-0.55](-0.2,-0.65)(2.2,0.45)
\pspolygon[fillstyle=solid,fillcolor=lightlightblue](0,-0.35)(2,-0.35)(2,0.35)(0,0.35)
\rput(1.4,0.0){\small$...$}
\psline[linecolor=blue,linewidth=1.5pt]{-}(0.2,0.35)(0.2,-0.35)\rput(0.2,-0.55){$_1$}
\psline[linecolor=blue,linewidth=1.5pt]{-}(0.6,0.35)(0.6,-0.35)\rput(0.6,-0.55){$_2$}
\psline[linecolor=blue,linewidth=1.5pt]{-}(1.0,0.35)(1.0,-0.35)\rput(1.0,-0.55){$_3$}
\psline[linecolor=blue,linewidth=1.5pt]{-}(1.8,0.35)(1.8,-0.35)\rput(1.8,-0.55){$_n$}
\end{pspicture} 
\qquad
 e_j=\!
 \begin{pspicture}[shift=-0.55](-0.2,-0.65)(3.4,0.45)
 \pspolygon[fillstyle=solid,fillcolor=lightlightblue](0,-0.35)(3.2,-0.35)(3.2,0.35)(0,0.35)
\rput(0.6,0.0){\small$...$}
\rput(2.6,0.0){\small$...$}
\psline[linecolor=blue,linewidth=1.5pt]{-}(0.2,0.35)(0.2,-0.35)\rput(0.2,-0.55){$_1$}
\psline[linecolor=blue,linewidth=1.5pt]{-}(1.0,0.35)(1.0,-0.35)
\psline[linecolor=blue,linewidth=1.5pt]{-}(2.2,0.35)(2.2,-0.35)
\psline[linecolor=blue,linewidth=1.5pt]{-}(3.0,0.35)(3.0,-0.35)\rput(3.0,-0.55){$_{n}$}
\psarc[linecolor=blue,linewidth=1.5pt]{-}(1.6,0.35){0.2}{180}{0}\rput(1.35,-0.55){$_j$}
\psarc[linecolor=blue,linewidth=1.5pt]{-}(1.6,-0.35){0.2}{0}{180}\rput(1.85,-0.55){$_{j+1}$}
\end{pspicture} 
\ee
where the multiplication rules yield the defining relations
\be
 IA=AI=A,\qquad e_j^2=\beta e_j, \qquad e_j e_{j\pm1} e_j = e_j, \qquad e_i e_j = e_j e_i, \qquad (|i-j|>1)
\label{eq:TLedef}
\ee
with $A\in\{I,e_j ;\,j=1,\ldots,n-1\}$. 
For instance, the connectivity in \eqref{eq:c1} can be written as 
\be 
 c_1 = e_2e_1e_3e_2e_4e_3e_5e_6.
\ee

\subsection{Link state representations}
\label{sec:lsr}

\paragraph{Link states}
Computing physical quantities in a statistical model based on the TL algebra 
typically requires working with representations rather than with the algebra itself.
The standard representations and composite representations discussed below 
are founded on the notion of {\em link states}. To introduce these states, let there be $n$ nodes on a horizontal line. 
A link state is then a diagram where some (possibly all or none) of these nodes are connected pairwise by non-intersecting
loop segments 
(half-arcs) that live above the horizontal line, while the remaining nodes are occupied by vertical line 
segments, called {\em defects}, that no loop segment can overarch. Because the 
half-arcs connect pairs of sites, the defect number is constrained to have the same parity as $n$. 
The set of link states on $n$ nodes with $d$ defects is denoted by $B_n^d$, and an example of a link state in $B_8^2$ is
\be
\begin{pspicture}[shift=-0.15](-0.0,0)(3.2,0.5)
\psline{-}(0,0)(3.2,0)
\psline[linecolor=coloroflink,linewidth=1.5pt]{-}(0.2,0)(0.2,0.5)
\psline[linecolor=coloroflink,linewidth=1.5pt]{-}(0.6,0)(0.6,0.5)
\psarc[linecolor=coloroflink,linewidth=1.5pt]{-}(1.6,0){0.2}{0}{180}
\psarc[linecolor=coloroflink,linewidth=1.5pt]{-}(2.4,0){0.2}{0}{180}
\psbezier[linecolor=coloroflink,linewidth=1.5pt]{-}(1,0)(1,0.7)(3,0.7)(3,0)
\end{pspicture}
\label{cB82}
\ee
In general, the cardinality of $B_n^d$ is
\be
 \big|B_n^d\big|= {n \choose \frac{n-d}{2}} - {n \choose \frac{n-d}{2}-1}.
\ee

\paragraph{Standard modules}
The {\em standard action} of a connectivity $c \in\mathrm{TL}_n(\beta)$
on a link state $w \in B_n^d$ closely resembles the rule given for the multiplication of two connectivities. To compute 
$cw$, one draws $w$ above $c$, erases the top horizontal edge of $c$, reads the new link state from the bottom $n$ nodes and 
replaces by a factor of $\beta$ each contractible loop closed in the process. An extra rule applies: if the number of defects has decreased 
(that is, if $cw\in B_n^{d'}$ where $d'<d$), the result is set to zero. 
On the element of $B_8^2$ depicted in (\ref{cB82}), this action is illustrated by 
\be
\begin{pspicture}[shift=-0.50](-0,-1)(3.2,0.5)
\pspolygon[fillstyle=solid,fillcolor=lightlightblue](0,-1)(3.2,-1)(3.2,0)(0,0)
\psline[linecolor=coloroflink,linewidth=1.5pt]{-}(0.2,0)(0.2,0.5)
\psline[linecolor=coloroflink,linewidth=1.5pt]{-}(0.6,0)(0.6,0.5)
\psarc[linecolor=coloroflink,linewidth=1.5pt]{-}(1.6,0){0.2}{0}{180}
\psarc[linecolor=coloroflink,linewidth=1.5pt]{-}(2.4,0){0.2}{0}{180}
\psbezier[linecolor=coloroflink,linewidth=1.5pt]{-}(1,0)(1,0.7)(3,0.7)(3,0)
\psline[linecolor=blue,linewidth=1.5pt]{-}(0.2,0)(0.2,-1)
\psline[linecolor=blue,linewidth=1.5pt]{-}(3,0)(3,-1)
\psarc[linecolor=blue,linewidth=1.5pt]{-}(0.8,-1){0.2}{0}{180}
\psarc[linecolor=blue,linewidth=1.5pt]{-}(2.4,-1){0.2}{0}{180}
\psarc[linecolor=blue,linewidth=1.5pt]{-}(2.0,0){0.2}{180}{0}
\psbezier[linecolor=blue,linewidth=1.5pt]{-}(1.4,0)(1.4,-0.5)(2.6,-0.5)(2.6,0)
\psbezier[linecolor=blue,linewidth=1.5pt]{-}(1,0)(1,-0.5)(1.8,-0.5)(1.8,-1)
\psbezier[linecolor=blue,linewidth=1.5pt]{-}(0.6,0)(0.6,-0.5)(1.4,-0.5)(1.4,-1)
\end{pspicture} \ = \beta \
\begin{pspicture}[shift=-0.10](-0.0,0)(3.2,0.5)
\psline{-}(0,0)(3.2,0)
\psline[linecolor=coloroflink,linewidth=1.5pt]{-}(0.2,0)(0.2,0.5)
\psline[linecolor=coloroflink,linewidth=1.5pt]{-}(1.4,0)(1.4,0.5)
\psarc[linecolor=coloroflink,linewidth=1.5pt]{-}(0.8,0){0.2}{0}{180}
\psarc[linecolor=coloroflink,linewidth=1.5pt]{-}(2.4,0){0.2}{0}{180}
\psbezier[linecolor=coloroflink,linewidth=1.5pt]{-}(1.8,0)(1.8,0.5)(3.0,0.5)(3.0,0)
\end{pspicture} \qquad \qquad
\begin{pspicture}[shift=-0.50](-0,-1)(3.2,0.5)
\pspolygon[fillstyle=solid,fillcolor=lightlightblue](0,-1)(3.2,-1)(3.2,0)(0,0)
\psline[linecolor=coloroflink,linewidth=1.5pt]{-}(0.2,0)(0.2,0.5)
\psline[linecolor=coloroflink,linewidth=1.5pt]{-}(0.6,0)(0.6,0.5)
\psarc[linecolor=coloroflink,linewidth=1.5pt]{-}(1.6,0){0.2}{0}{180}
\psarc[linecolor=coloroflink,linewidth=1.5pt]{-}(2.4,0){0.2}{0}{180}
\psbezier[linecolor=coloroflink,linewidth=1.5pt]{-}(1,0)(1,0.7)(3,0.7)(3,0)
\psarc[linecolor=blue,linewidth=1.5pt]{-}(1.2,-1){0.2}{0}{180}
\psarc[linecolor=blue,linewidth=1.5pt]{-}(2.4,-1){0.2}{0}{180}
\psarc[linecolor=blue,linewidth=1.5pt]{-}(0.4,0){0.2}{180}{0}
\psarc[linecolor=blue,linewidth=1.5pt]{-}(2.0,0){0.2}{180}{0}
\psbezier[linecolor=blue,linewidth=1.5pt]{-}(0.6,-1)(0.6,-0.5)(1.8,-0.5)(1.8,-1)
\psbezier[linecolor=blue,linewidth=1.5pt]{-}(0.2,-1)(0.2,-0.5)(1.0,-0.5)(1.0,0)
\psbezier[linecolor=blue,linewidth=1.5pt]{-}(1.4,0)(1.4,-0.5)(2.6,-0.5)(2.6,0)
\psline[linecolor=blue,linewidth=1.5pt]{-}(3,0)(3,-1)
\end{pspicture} \ = 0.
\ee
Finally, the action is linearly extended from $B_n^d$ to $\mathrm{span}\,B_n^d$. 
For each $0 \le d \le n$ with $n-d = 0 \; \textrm{mod} \; 2$, this defines a representation 
\be
 \rho_d \;:\; \mathrm{TL}_n(\b)\, \to\, \textrm{End}(\mathrm{span}\,B_n^d)
\ee
known as a {\em standard representation} of $\mathrm{TL}_n(\beta)$. The corresponding {\em standard module} over 
$\mathrm{TL}_n(\beta)$ is denoted by $V_n^d$ and its dimension is given by the cardinality of 
$B_n^d$,
\be
 \dim V_n^d = {n \choose \frac{n-d}{2}} - {n \choose \frac{n-d}{2}-1}.
\ee

The standard modules are known to be indecomposable for all $\beta$. For generic $\beta$, that is if $\beta$ cannot be written as 
$q + q^{-1}$ with $q^{2 \ell} = 1$ and 
$\ell \in \mathbb Z_{\ge 2}$, the algebra TL$_n(\beta)$ is semi-simple and 
all representations are fully reducible. In that case, the standard modules $V_n^d$ form a complete set of non-isomorphic irreducible 
modules~\cite{MartinBook, Westbury, GL98, RSA1204},
\be 
 V_n^d\,\simeq\,\mathcal I_n^d, \qquad 0\le d \le n, \qquad n-d = 0 \;\textrm{mod}\; 2.
\label{eq:standirr}
\ee 
Here and in the following, the irreducible modules over TL$_n(\beta)$ are denoted by $\mathcal I_n^d$ and thus labeled by the 
integer $d$. For non-generic $\beta$, i.e.\! $\beta=q+q^{-1}$ with $q$ as above, the representation theory is much more involved 
and includes reducible yet indecomposable representations. As discussed below in the case $\beta=0$, 
some of the standard representations are of this kind.

\paragraph{Composite modules}
The standard modules are not the only TL modules playing a role in our investigation of the dimer model.
Indeed, for every $0 \le d \le n$ with 
$n-d = 0 \; \textrm{mod} \; 2$, 
we consider 
\be
 \pi_d \;:\; \mathrm{TL}_n(\b)\,\to\, \mathrm{End}\big(\mathrm{span} \, (B_n^d \cup B_n^{d+2})\big).
\label{pid}
\ee
These {\em composite representations} are defined as follows.
On a link state in the subset $B_n^d$ appearing in (\ref{pid}), the action of $\mathrm{TL}_n(\beta)$ 
connectivities is identical to the one applied 
in the definition of the standard representation $\rho_d$. On a link state $w$ in the subset $B_n^{d+2}$, 
however, a different rule is prescribed. Again, one starts by drawing $w$ above the connectivity $c$. 
If the number of defects has not decreased in the resulting diagram, the new link state is obtained by reading from the 
bottom $n$ nodes and is multiplied by a factor of $\beta$ for each closed loop. If the number of defects has decreased by more than two, 
the result is set to zero. If the number of defects has decreased by exactly two, the result is set to zero unless the rightmost defect is one of 
the two annihilated defects. In this last case, the result is obtained by identifying the resulting link state in $B_n^{d}$ and 
adding the appropriate factors of $\beta$. Finally, this action is linearly extended to $\mathrm{span}(B_n^d \cup B_n^{d+2})$ 
and the ensuing TL$_n(\beta)$ module is denoted by $W_n^d$. 
It readily follows from the definition of the action that the standard module $V_n^d$ is a submodule of the composite 
module $W_n^d$ with the standard module $V_n^{d+2}$ appearing as the corresponding quotient module
\be
 W_n^d/V_n^d\,\simeq\,V_n^{d+2}.
\ee

To emphasise the special role of the rightmost defect in the subset $B_n^{d+2}$, we indicate it by a wavy line segment, 
as illustrated here:
\be
\begin{pspicture}[shift=-0.50](-0,-1)(3.2,0.5)
\psline{-}(0,0)(3.2,0)
\psline[linecolor=coloroflink,linewidth=1.5pt]{-}(1.0,0)(1.0,0.5)
\psline[linecolor=coloroflink,linewidth=1.5pt]{-}(1.4,0)(1.4,0.5)
\psarc[linecolor=coloroflink,linewidth=1.5pt]{-}(0.4,0){0.2}{0}{180}
\psarc[linecolor=coloroflink,linewidth=1.5pt]{-}(2.4,0){0.2}{0}{180}
\psline[linecolor=coloroflink,linewidth=1.5pt]{-}(1.8,0)(1.8,0.5)
\rput(3,0){\wobbly}
\pspolygon[fillstyle=solid,fillcolor=lightlightblue](0,-1)(3.2,-1)(3.2,0)(0,0)
\psline[linecolor=blue,linewidth=1.5pt]{-}(0.2,0)(0.2,-1)
\psarc[linecolor=blue,linewidth=1.5pt]{-}(0.8,0){-0.2}{0}{180}
\psarc[linecolor=blue,linewidth=1.5pt]{-}(1.6,-1){0.2}{0}{180}
\psarc[linecolor=blue,linewidth=1.5pt]{-}(2.8,-1){0.2}{0}{180}
\psarc[linecolor=blue,linewidth=1.5pt]{-}(2.4,0){-0.2}{0}{180}
\psbezier[linecolor=blue,linewidth=1.5pt]{-}(1.8,0)(1.8,-0.5)(3.0,-0.5)(3.0,0)
\psbezier[linecolor=blue,linewidth=1.5pt]{-}(1.0,-1)(1.0,-0.5)(2.2,-0.5)(2.2,-1)
\psbezier[linecolor=blue,linewidth=1.5pt]{-}(0.6,-1)(0.6,-0.5)(1.4,-0.5)(1.4,0)
\end{pspicture} \ = \beta \
\begin{pspicture}[shift=-0.10](-0.0,0)(3.2,0.5)
\psline{-}(0,0)(3.2,0)
\psline[linecolor=coloroflink,linewidth=1.5pt]{-}(0.2,0)(0.2,0.5)
\psline[linecolor=coloroflink,linewidth=1.5pt]{-}(0.6,0)(0.6,0.5)
\psarc[linecolor=coloroflink,linewidth=1.5pt]{-}(1.6,0){0.2}{0}{180}
\psarc[linecolor=coloroflink,linewidth=1.5pt]{-}(2.8,0){0.2}{0}{180}
\psbezier[linecolor=coloroflink,linewidth=1.5pt]{-}(1.0,0)(1.0,0.5)(2.2,0.5)(2.2,0)
\end{pspicture}
\qquad \qquad
\begin{pspicture}[shift=-0.50](-0,-1)(3.2,0.5)
\psline{-}(0,0)(3.2,0)
\psline[linecolor=coloroflink,linewidth=1.5pt]{-}(1.0,0)(1.0,0.5)
\psline[linecolor=coloroflink,linewidth=1.5pt]{-}(1.4,0)(1.4,0.5)
\psarc[linecolor=coloroflink,linewidth=1.5pt]{-}(0.4,0){0.2}{0}{180}
\psarc[linecolor=coloroflink,linewidth=1.5pt]{-}(2.8,0){0.2}{0}{180}
\psline[linecolor=coloroflink,linewidth=1.5pt]{-}(1.8,0)(1.8,0.5)
\rput(2.2,0){\wobbly}
\pspolygon[fillstyle=solid,fillcolor=lightlightblue](0,-1)(3.2,-1)(3.2,0)(0,0)
\psarc[linecolor=blue,linewidth=1.5pt]{-}(0.4,-1){0.2}{0}{180}
\psarc[linecolor=blue,linewidth=1.5pt]{-}(1.6,-1){0.2}{0}{180}
\psarc[linecolor=blue,linewidth=1.5pt]{-}(2.4,-1){0.2}{0}{180}
\psarc[linecolor=blue,linewidth=1.5pt]{-}(0.8,0){0.2}{180}{0}
\psarc[linecolor=blue,linewidth=1.5pt]{-}(2.8,0){0.2}{180}{0}
\psbezier[linecolor=blue,linewidth=1.5pt]{-}(2.2,0)(2.2,-0.5)(3.0,-0.5)(3.0,-1)
\psbezier[linecolor=blue,linewidth=1.5pt]{-}(1.8,0)(1.8,-0.5)(1.0,-0.5)(1.0,-1)
\psbezier[linecolor=blue,linewidth=1.5pt]{-}(0.2,0)(0.2,-0.5)(1.4,-0.5)(1.4,0)
\end{pspicture} \ = 0.
\ee

That $\pi_d$ is a representation is well known. For generic $\beta$, the associated module $W_n^d$
decomposes as a direct sum of the standard modules $V_n^d$ 
and $V_n^{d+2}$, while for non-generic $\beta$ the decomposition can be more intricate. The representations $\pi_d$ have appeared in 
the literature before. In~\cite{PRZ06,PR07,RP0707}, they take the form $(1,d+2) \otimes (1,2)$ and are used to probe fusion of 
boundary conditions of loop models and of the corresponding representations of the Virasoro algebra. 
The composite modules $W_n^d$ are also equivalent to the modules 
$\mathsf S_{\frac{d+1}2}[n_1] \times _f \mathsf S_{\frac12}[n_2]$ (with $n_1 + n_2 = n$) appearing in 
the fusion construction of~\cite{GV12}. Finally, $W_n^d$ can be alternatively constructed~\cite{RSA1204} as the module induced from 
$V_{n-1}^{d+1}$ or as the module restricted from $V_{n+1}^{d+1}$.

\subsection{Representation theory of TL$_n(0)$}
\label{sec:reptheor0}

The TL algebra $\mathrm{TL}_n(0)$ is non-generic as $\beta=q+q^{-1} = 0$ corresponds to $q= i$ and $\ell = 2$
with $q^{2\ell}=1$. 
As discussed in the recent review paper~\cite{RSA1204}, the corresponding representation theory depends critically on the parity of $n$.

An important class of modules not discussed above are the {\em principal indecomposable modules}. These are 
the modules appearing as the indecomposable direct summands in the decomposition of the regular representation of the TL algebra. 
They are also precisely the indecomposable projective modules. They do therefore not appear as proper quotients of larger 
indecomposable modules. For $\beta=0$, there are $\lfloor\frac{n+1}{2}\rfloor$ principal indecomposable modules and they are denoted 
here by $\mathcal P_n^d$ where $1\leq d\leq n$ and $n-d = 0 \; \textrm{mod} \; 2$.

\paragraph{$\boldsymbol n$ odd}
For $n$ odd, $\mathrm{TL}_n(0)$ is semi-simple, implying that all representations are fully reducible, i.e.~every module decomposes 
as a direct sum of irreducible modules. The complete set of non-isomorphic irreducible modules
is given by $\{\mathcal I_n^d,\; d = 1, 3, \dots, n\}$, with dimensions 
\be
\dim \mathcal I_n^d = \dim V_{n}^{d}= {n \choose \frac{n-d}{2}} - {n \choose \frac{n-d}{2}-1}.
\ee
In fact, the standard modules and the principal indecomposable modules are all irreducible,
\be
 V_n^d\,\simeq\, \mathcal P_n^d\,\simeq\,\mathcal I_n^d, \qquad d = 1, 3, \dots, n,
\label{PVI}
\ee
while the composite modules $W_n^d$ decompose as
\be 
 W_n^d\,\simeq\,\mathcal I_n^d \oplus \mathcal I_n^{d+2}, \qquad d = 1, 3, \dots, n.
\ee

\paragraph{$\boldsymbol n$ even}
For $n$ even, $\mathrm{TL}_n(0)$ is not semi-simple. The complete set of non-isomorphic irreducible modules is 
$\{\mathcal I_n^d,\; d = 2, 4, \dots, n\}$, with the dimensions given by
\be
 \dim \mathcal I_n^d = \dim V_{n-1}^{d-1}= {n-1 \choose \frac{n-d}{2}} - {n-1 \choose \frac{n-d}{2}-1}.
\ee
The standard modules are indecomposable and their structure patterns depend on $d$,
\be
 V_n^0\,\simeq\,\mathcal I_n^2, \qquad V_n^d\,\simeq\,\Big(\mathcal I_n^d 
\begin{pspicture}(0,0)(1,0)
\psline[linewidth=.8pt,arrowsize=3pt 2]{->}(0.2,0.1)(0.8,0.1)
\end{pspicture} 
\mathcal I_n^{d+2}\Big)\quad (d = 2,4, \dots, n-2),
\qquad V_n^n\,\simeq\,\mathcal I_n^n.
\label{eq:vnd}
\ee
For $d=0$ and $d=n$, the standard modules are thus irreducible, while for the intermediate values of $d$ (all even), 
they contain two composition factors, one of which ($\mathcal I_n^{d+2}$) is a proper submodule. 

More generally, the structure patterns of a module is described in terms of its {\it Loewy diagram} in which the composition factors of the 
module are 
vertices connected by arrows.
If an arrow points from the factor $\mathcal A$ to the factor 
$\mathcal B$, as in (\ref{eq:vnd}), vectors in $\mathcal B$ can be reached from vectors in $\mathcal A$ by 
the action of the TL algebra, whereas no vector in $\mathcal A$ can be reached from $\mathcal B$. 
Loewy diagrams are typically drawn with all arrows pointing downwards. It is nevertheless convenient occasionally to use 
horizontal arrows. To avoid confusion with regular maps, we use a different style of arrow (\begin{pspicture}(0.1,0)(0.9,0)
\psline[linewidth=.8pt,arrowsize=3pt 2]{->}(0.2,0.1)(0.8,0.1)
\end{pspicture} instead of $\rightarrow$) and include large parentheses around Loewy diagrams with horizontal arrows, 
as in \eqref{eq:vnd}.

For $n$ even, the structure patterns of the principal indecomposable modules are given by
\medskip
\be
\mathcal P_n^2\, \simeq
\begin{pspicture}[shift=-0.9](1,-1)(3,1)
\rput(1.5,1){$\mathcal I_n^{2}$}
\rput(1.5,-1){$\mathcal I_n^{2}$}
\rput(2.5,0){$\mathcal I_n^{4}$}
\psline[linewidth=.8pt,arrowsize=3pt 2]{->}(1.75,0.75)(2.25,0.25)
\psline[linewidth=.8pt,arrowsize=3pt 2]{->}(2.25,-0.25)(1.75,-0.75)
\end{pspicture}
\qquad 
\mathcal P_n^d\, \simeq
\begin{pspicture}[shift=-0.9](0,-1)(3,1)
\rput(0.5,0){$\mathcal I_n^{d-2}$}
\rput(1.5,1){$\mathcal I_n^{d}$}
\rput(1.5,-1){$\mathcal I_n^{d}$}
\rput(2.5,0){$\mathcal I_n^{d+2}$}
\psline[linewidth=.8pt,arrowsize=3pt 2]{->}(1.25,0.75)(0.75,0.25)
\psline[linewidth=.8pt,arrowsize=3pt 2]{->}(1.75,0.75)(2.25,0.25)
\psline[linewidth=.8pt,arrowsize=3pt 2]{->}(0.75,-0.25)(1.25,-0.75)
\psline[linewidth=.8pt,arrowsize=3pt 2]{->}(2.25,-0.25)(1.75,-0.75)
\end{pspicture}\quad (d = 4, \dots, n-2)
\qquad \quad
\mathcal P_n^n\, \simeq
\begin{pspicture}[shift=-0.9](0,-1)(2,1)
\rput(0.5,0){$\mathcal I_n^{n-2}$}
\rput(1.5,1){$\mathcal I_n^{n}$}
\rput(1.5,-1){$\mathcal I_n^{n}$}
\psline[linewidth=.8pt,arrowsize=3pt 2]{->}(1.25,0.75)(0.75,0.25)
\psline[linewidth=.8pt,arrowsize=3pt 2]{->}(0.75,-0.25)(1.25,-0.75)
\end{pspicture}
\label{eq:Ps}
\vspace{.15cm}
\ee
The module $\mathcal P_n^d\/$ has a submodule isomorphic to $V_n^{d-2}$ with the corresponding quotient module given by 
\be
 \mathcal P_n^d / V_n^{d-2}\,\simeq\,V_n^{d}.
\ee 
Crucially, for $\beta = 0$ and $n$ even, the composite modules realise the principal indecomposable modules,
\be
 W_n^{d-2}\,\simeq\,\mathcal P_n^d,\qquad d=2,4,\ldots,n.
\label{WP}
\ee
This can be deduced from~\cite{RSA1204}
where it is shown, as Proposition 6.3 and Proposition 8.2, respectively, that (for $\beta=0$, $n$ even and $d=2,\ldots,n$)
\be
 V_{n+1}^{d-1}\!\!\downarrow\;\simeq V_{n-1}^{d-1}\!\!\uparrow\qquad\mbox{and}\qquad
 V_{n-1}^{d-1}\!\!\uparrow\;\simeq \mathcal P_n^{d}.
\ee
Here $V_{n+1}^{d-1}\!\!\downarrow$ is defined as the restriction of $V_{n+1}^{d-1}$ to the action of ${\rm TL}_n \subset {\rm TL}_{n+1}$
(generated by the identity and the $e_j$ with $j<n$), while $V_{n-1}^{d-1}\!\!\uparrow$ is obtained by the induction of $V_{n-1}^{d-1}$ to a 
${\rm TL}_n$-module, following a recipe also used to compute fusion of TL representations~\cite{GV12}.
Now, $W_n^{d-2}$ is easily seen to be isomorphic to $V_{n+1}^{d-1}\!\!\downarrow$, with the bijective map given as follows.
If $w \in B_{n+1}^{d-1}$ has a defect at $n+1$, it cannot be displaced by the action of TL$_n$. This node and its defect are erased 
and the new state is a basis element in the subspace $V_n^{d-2}\subset W_n^{d-2}$. If the node $n+1$ is occupied by the right 
end of a half-arc, the node it connects to becomes occupied by a wavy defect, the node $n+1$ is again erased, and the new 
state is a basis element of the quotient $V_n^d \simeq W_n^{d-2}/ V_n^{d-2}$. The inverse map is constructed similarly,
thus establishing the isomorphism $W_n^{d-2}\simeq V_{n+1}^{d-1}\!\!\downarrow$ and hence (\ref{WP}).

\section[Dimer representations of TL$\boldsymbol{_n(0)}$]{Dimer representations of TL$\boldsymbol{_n(0)}$}
\label{sec:Dimersrep}

We show in Section~\ref{sec:screp} that the spin configuration space of the dimer model carries a representation of the 
TL algebra for $\b=0$ by constructing a map
\be 
 \tau\;:\; \mathrm{TL}_n(0) \,\to\,  \mathrm{End}\big((\mathbb C^{2})^{\otimes(n-1)}\big)
\label{tau}
\ee 
and relating it to the expression for the squared transfer matrix $T^2(\alpha)$ of the dimer model.
We refer to $\tau$ as the {\it dimer representation} of the TL$_n(0)$ algebra.
Its structure is exhibited in Section~\ref{sec:com}.

\subsection{Spin-chain representations}
\label{sec:screp}

\begin{Proposition}
Let $\tau$ in (\ref{tau}) be a linear map defined on the $n$ basic TL generators by
\be
 \tau(I) = \mathbb I, \qquad \tau(e_j) = \sigma^-_{j-1} \sigma^+_j + \sigma^+_j \sigma^-_{j+1}, \qquad j = 1, \dots, n-1,
\label{eq:simplertau}
\ee
where $\sigma^\pm_0\equiv\sigma^\pm_{n}\equiv0$, and on nontrivial words in $\mathrm{TL}_n(0)$ by the multiplication rule
\be 
\tau(cc')=\tau(c)\tau(c'),\qquad c,c'\in\mathrm{TL}_n(0).
\label{eq:rules}
\ee
The map $\tau$ is then a representation of $\mathrm{TL}_n(0)$.
\label{prop:tau}
\end{Proposition}

\medskip\noindent {\sc Proof } The proposition follows from the TL relations
\be
 [\tau(e_j)]^2 = 0, \qquad  \tau(e_j)\tau(e_{j\pm 1})\tau(e_j) = \tau(e_j), \qquad [\tau(e_i),\tau(e_j)] = 0,\qquad (|i-j|>1)
\label{eq:taurep}
\ee
which are verified straightforwardly. 
\hfill $\square$
\medskip

Noting that the configuration space upon which $T^2(\alpha)$ acts is $(\mathbb C^2)^{\otimes N}$ while the representation space of
$\tau$ is $(\mathbb C^2)^{\otimes(n-1)}$, we henceforth set $N=n-1$. 
The relationship between $\tau$ and the dimer model is then seen by rewriting $T^2(\alpha)$ in \eqref{eq:T2+-} as
\bea
 T^2(\alpha) \!\!\!&=&\!\!\! \prod_{j=1}^{N-1}\big(\mathbb I + \alpha\, \sigma^-_{j}\,\sigma^-_{j+1}\big) 
  \times \prod_{j=1}^{N-1}\big(\mathbb I + \alpha\, \sigma^+_{j}\,\sigma^+_{j+1}\big)\nn
 \!\!\!&=&\!\!\! \prod_{j=1}^{\lfloor\! \frac{n}2\! \rfloor} 
  \Big(\mathbb I + \alpha\, (\sigma^-_{2j-2}\,\sigma^-_{2j-1}+\sigma^-_{2j-1}\,\sigma^-_{2j})\Big) 
 \times \prod_{j=1}^{\lfloor\! \frac{n-1}2 \!\rfloor}
  \Big(\mathbb I+\alpha\, (\sigma^+_{2j-1}\,\sigma^+_{2j}+\sigma^+_{2j}\,\sigma^+_{2j+1})\Big)\nn
 \!\!\!&=&\!\!\! \prod_{j=1 \atop {\rm odd}}^{n-1}\big(\mathbb I+\a\,\bar\tau(e_j)\big)
  \times\prod_{j=1 \atop {\rm even}}^{n-1}\big(\mathbb I+\a\,\bar\tau(e_j)\big).
\label{eq:T2TL}
\eea
Here we have introduced
\be
 \bar\tau = U^{-1}\, \tau \, U, \qquad U = \prod_{j=1 \atop {\rm odd}}^{n-1} \sigma^x_{j},
\ee
that is,
\be 
\bar\tau(I) = \mathbb I, \qquad
\bar\tau(e_j) =  \left\{\begin{array}{ll} 
\!\!\sigma_{j-1}^- \sigma_{j}^-+ \sigma_{j}^- \sigma_{j+1}^-,& \quad j \ \textrm{odd},
\\[0.2cm]
\!\!\sigma_{j-1}^+ \sigma_{j}^++ \sigma_{j}^+ \sigma_{j+1}^+,& \quad j \ \textrm{even.}
\end{array}\right.
\label{eq:tau1}
\ee
This yields an equivalent and likewise $2^{n-1}$-dimensional representation of TL$_n(0)$, 
but one in which the odd spins have been reversed compared to $\tau$.

Although the representation $\bar\tau$ is the one directly linked to the dimer model, we find it convenient to work with $\tau$ instead 
as the definition of $\tau(e_j)$ in (\ref{eq:simplertau}) is independent of the parity of $j$. In the $\tau$ representation, the usual TL 
Hamiltonian 
\be
 H = \tau(\mathcal H) =-\sum_{j=1}^{n-1} \tau(e_j) = - \sum_{j=1}^{n-2} (\sigma^+_j \sigma^-_{j+1} +\sigma^-_{j} \sigma^+_{j+1} )
\ee
is a Heisenberg spin-chain Hamiltonian with no boundary magnetic field and is seen to arise as the linear term in the decomposition of 
$UT^2(\alpha)U^{-1}$ in powers of $\alpha$.

Under the  transformation $U$, the variation index operator becomes the total magnetisation,
\be
 U\,\mathcal V\,U^{-1}= \tfrac12\sum_{j=1}^{n-1} \sigma^z_j = S^z.
\ee 
From this, it is readily seen that the TL representation $\tau$ commutes with $S^z$,
\be
 [\tau(c),S^z] = 0, \qquad c \in {\rm TL}_n(0).
\label{tauSz}
\ee
Indeed, this general commutativity property follows from the relation (\ref{tauSz}) for the TL generators $c = e_j$, $j = 1, \dots,n-1$, 
and the multiplication rule \eqref{eq:rules}.

To the best of our knowledge, 
the representation of TL$_n(0)$ defined in Proposition~\ref{prop:tau} has not appeared in the literature before. 
Of course, there exist other spin-chain representations of TL$_n(\beta)$ and not just for $\beta=0$. 
A well-known example is the
representation
\be \chit\;:\; {\rm TL}_n(\beta = q+q^{-1}) \ \rightarrow\  (\mathbb C^2)^{\otimes n}\ee
related to the $6$-vertex model and the XXZ spin chain~\cite{ABBBQ87,PS90}, defined as
\be
\chit(e_j) = -\frac12\Big(\sigma_i^x\sigma_{i+1}^x+\sigma_i^y\sigma_{i+1}^y+\frac12(q+q^{-1})(\sigma_i^z\sigma_{i+1}^z-\mathbb I)+\frac12(q-q^{-1})(\sigma_i^z-\sigma_{i+1}^z)\Big).
\ee
However, even though they have common features, it is emphasised that the representation $\chit$ at $q=\ir$ 
(and $\beta = \ir + \ir^{-1} = 0$) and the one in \eqref{eq:simplertau} are {\em not isomorphic}.
Indeed, a key result of our analysis below demonstrates that the modules related to the dimer representations 
are structurally different from those appearing in the XXZ spin-chains. 
This follows immediately by comparing the structure of the dimer modules given in Theorem~\ref{thm:main} with the decomposition
of the XXZ modules~\cite{Martin92,RS90}
\be 
(\mathbb C^2)^{\otimes n} \simeq  \bigoplus_{d=1, 3, \dots}^{n-1} \mathcal (d+1)\, \mathcal I_n^d \quad (n \; \textrm{odd}),\qquad
(\mathbb C^2)^{\otimes n} \simeq \Big( \!
\bigoplus_{d=2, 4, \dots}^{n} \hspace{-0.1cm} \tfrac d2 \, \mathcal P_n^d\;\Big) \oplus \, \tfrac {n+2}2 \, \mathcal I^n_n \quad 
 (n \; \textrm{even}),
\ee
where the integers $d+1$, $\frac d2$ and $\frac {n+2}2$ indicate the multiplicities with which the corresponding modules appear. 
It is also recalled that the $\tau$ and XXZ representations are of different dimensions ($2^{n-1}$ and $2^n$, respectively)
and furthermore noted that $\tau(e_j)$ generally acts on three sites ($j-1$, $j$ and $j+1$), whereas $\chit(e_j)$ 
only acts on a pair of adjacent sites.

\subsection{Characterisation of modules}
\label{sec:com}

As a consequence of the commutativity (\ref{tauSz}), the TL representation $\tau$ on the full space $(\mathbb C^2)^{\otimes(n-1)}$ 
decomposes into a direct sum of representations labeled by the eigenvalues of $S^z$. The corresponding eigenspaces $E_{n-1}^v$, 
where $v = -\frac{n-1}2, -\frac{n-3}2,\ldots ,\frac{n-1}2$, are generated by spin states with fixed total magnetisation $v$,
\be
 (\mathbb C^2)^{\otimes(n-1)} = \bigoplus_v E_{n-1}^v\,, \qquad \tau \simeq \bigoplus_v \, \tau_v,
\ee
where the restriction of $\tau$ to $E_{n-1}^v$ is denoted by $\tau_v$. Evidently, from (\ref{dimENv}), the dimension of $E_{n-1}^v$ is 
\be
 \dim E_{n-1}^v={n-1 \choose \frac{n-1}2-v},
\ee
and it is readily verified that
\be
 \sum_{v=-\frac{n-1}{2}}^{\frac{n-1}{2}} \dim E_{n-1}^v=2^{n-1}.
\ee
From here onwards, we will consider $E_{n-1}^v$ as the TL$_n(0)$ module corresponding to $\tau_v$, and not simply as the 
eigenspace of $S^z$ associated to the eigenvalue $v$. 

Because states in $E_{n-1}^v$ and $E_{n-1}^{-v}$ are in bijective correspondence under the 
action of the spin reversal operator $V_1=V_1^{-1}=V_1^{\rm T}$, see \eqref{eq:V1}, it is not hard to see that the matrices $\tau_{-v}(e_j)$ 
and $\tau_v(e_j)^{\rm T}$ are similar,
\be
\tau_{-v}(e_j) = \big(V_1^{-1}\, \tau_{v}(e_j) \,V_1\big)^{\rm T} = V_1^{-1}\, \tau_v(e_j)^{\rm T} \, V_1.
\label{eq:transfv}
\ee
Here and in the following, the superscript T indicates the matrix transpose.
It is straightforward to verify that the matrices $\tau_v(e_j)^{\rm T}$ satisfy the defining relations \eqref{eq:TLedef}
of the TL algebra. In fact, the {\em contragredient representation} to $\tau(c)$, here denoted by 
\be
\tau^\star(c)=\tau(c^\dagger)^{\rm T},
\ee 
provides a representation of the full TL algebra, 
\be
 \tau^\star(c_1c_2)=\tau\big((c_1c_2)^\dagger\big)^{\rm T}=\tau(c_2^\dagger c_1^\dagger)^{\rm T}
  =\big(\tau(c_2^\dagger)\tau(c_1^\dagger)\big)^{\rm T}=\tau(c_1^\dagger)^{\rm T} \tau(c_2^\dagger)^{\rm T}
  =\tau^\star(c_1)\tau^\star(c_2).
\ee
Here the reflected (also called adjoint~\cite{RSA1204}) connectivity $c^\dagger$ is obtained from $c$ by interchanging the 
bottom and top edges, implying that the order of composition of reflected connectivities is reversed. 
Since $e_j^\dagger=e^{}_j$, it follows that the similarity relation (\ref{eq:transfv}) extends to a similarity relation involving
the contragredient representation restricted to fixed values of $v$,
\be
 \tau_{-v} = V_1^{-1}\, \tau^\star_v\, V_1.
\ee

In general, taking the contragredient of an indecomposable module not only replaces the irreducible composition factors by their 
contragredient counterparts, it also reverses the arrows (if any) between them. 
However, as all irreducible modules over TL$_n(\beta)$ are self-contragredient\footnote{In most cases, the irreducible modules 
$\mathcal I_n^d$ all have distinct dimensions. Because they exhaust the set of irreducible TL$_n$ modules, the module 
$(\mathcal I_n^d)^\star$, which is also irreducible, must be isomorphic to $\mathcal I_n^d$, implying that $\mathcal I_n^d$ is 
self-contragredient. The argument can be extended to the degenerate cases with a bit of work, but instead of providing a proof, 
we refer to the upcoming paper by Bellet\^ete, Ridout and Saint-Aubin~\cite{BRSA}.}, 
reversing the arrows alone yields the contragredient module. 
For this reason, the investigation of $\tau_v$ for $v \ge 0$ is sufficient to obtain the structure of $\tau_v$ for every $v$. 

Our main objective is to determine the module structure of $E_{n-1}^v$. As it is natural to compare these modules with the standard 
modules $V_n^d$, we tabulate the dimensions of $E_{n-1}^v$ and $V_n^d$ in Table~\ref{fig:Tab}.
\begin{table}
$\begin{array}{c|cccccccccc}
\multicolumn{11}{c}{\dim E_{n-1}^v} 
\\[0.2cm]
n {\setminus}v & 0 & \tfrac12 & 1 & \tfrac32 & 2 & \tfrac52 & 3 & \tfrac72 & 4  \\ \hline
1  & 1 &  &  &  \\
2  &  & 1 &  &  \\
3  & 2 &  & 1 &  \\
4  &  & 3 &  & 1 \\
5  & 6 &  & 4 &  & 1 \\
6  &  & 10 &  & 5 &  & 1 \\
7  & 20  &  & 15 &  & 6 & & 1 \\
8 &  & 35 &  & 21 & & 7 & & 1 \\
9 & 70 & &  56 &  & 28 & & 8 & & 1
\end{array}
\qquad \quad
\begin{array}{c|cccccccccc}
\multicolumn{11}{c}{\dim V_n^d}
\\[0.2cm]
n {\setminus}d & 0 & 1 & 2 & 3 & 4 & 5 & 6 & 7 & 8 & 9 \\ \hline
1 &  & 1 &  &  &  \\
2 & 1 &  & 1 &  &  \\
3 &  & 2 &  & 1 &  \\
4 & 2 &  & 3 &  & 1 \\
5 &  & 5 &  & 4 &  & 1 \\
6 & 5 &  & 9 &  & 5 &  & 1 \\
7 &  & 14  &  & 14 &  & 6 & & 1 \\
8 & 14 &  & 28 &  & 20 & & 7 & & 1 \\
9 & & 42 & & 48 & & 27 & & 8 & & 1 
\end{array}$
\caption{The dimensions of $E_{n-1}^v$ and $V_n^d$. 
The dimensions $\dim E_{n-1}^v$ are only listed for $v\ge0$ since $\dim E_{n-1}^v = \dim E_{n-1}^{-v}$.}
\label{fig:Tab}
\end{table}
By comparing the numbers for the same fixed value of $n$ in the two tables, one notices a series of identities where 
numbers in the left table can be written as sums of numbers in the right table. For $n=8$ for instance, we observe that 
$35=28+7$, $21=20+1$, $7=7$ and $1=1$, and similarly for $n=9$, we have $70=42+27+1$, $56=48+8$, $28=27+1$, $8=8$ and 
$1=1$. Indeed, the general sum rule
\be
 \dim E_{n-1}^{v} = \hspace{-0.1cm} \sum_{i = 0}^{\big\lfloor\!\frac{n-1-2|v|}4 \!\big\rfloor} \hspace{-0.1cm} \dim V_n^{2|v|+4 i+1}
\label{eq:dimensions}
\ee
is readily established using an 
inductive argument. Naively, this identity suggests that the TL$_n(0)$ module $E_{n-1}^v$ 
could decompose in terms of standard modules as
\be
 E_{n-1}^v \overset{?}{\simeq} \bigoplus_{i=0}^{\big\lfloor\!\frac{n-1-2|v|}4  \!\big\rfloor} V_n^{2|v|+4 i+1}.
\label{eq:directsum?}
\ee
The resolution is given in the following structure theorem and is a key result of this paper.
\begin{Theoreme}
\label{thm:main}
For $v\in\{-\frac{n-1}{2},-\frac{n-3}{2},\ldots,\frac{n-1}{2}\}$, the structure of the module $E_{n-1}^v$ is as follows:
\begin{itemize}
\item[(1)] 
For $n$ odd, the module $E_{n-1}^v$ is fully reducible and decomposes into irreducible modules as
\be
E_{n-1}^v \simeq\,
\mathcal I_n^{2|v|+1}\oplus\mathcal I_n^{2|v|+5}\oplus\mathcal I_n^{2|v|+9} \oplus \ldots \oplus 
 (\mathcal I_n^{n-2} \;{\rm or} \; \mathcal I_n^n).
\label{EVpV}
\ee
\item[(2a)] 
For $n$ even and $v \ge \frac12$, the module $E_{n-1}^v$ is reducible yet indecomposable and has structure pattern
\smallskip

\be
E_{n-1}^v \simeq \left\{ \begin{array}{ll}
\begin{pspicture}[shift=-0.4](-0.8,-0.5)(6,0.5)
\rput(0,-0.5){$\mathcal I_n^{2v+1}$}
\rput(1.0,0.5){$\mathcal I_n^{2v+3}$}
\rput(1.9,-0.5){$\mathcal I_n^{2v+5}$}
\rput(3.2,0.5){$\dots$}
\rput(4.55,-0.5){$\mathcal I_n^{n-2}$}
\rput(5.5,0.5){$\mathcal I_n^{n}$}
\psline[linewidth=.8pt,arrowsize=3pt 2]{->}(0.6,0.25)(0.1,-0.2)
\psline[linewidth=.8pt,arrowsize=3pt 2]{->}(1,0.25)(1.5,-0.2)
\psline[linewidth=.8pt,arrowsize=3pt 2]{->}(2.6,0.25)(2.1,-0.2)
\psline[linewidth=.8pt,arrowsize=3pt 2]{->}(3.7,0.25)(4.2,-0.2)
\psline[linewidth=.8pt,arrowsize=3pt 2]{->}(5.3,0.25)(4.8,-0.2)
\end{pspicture}
\qquad\ \ &(\frac{n-1}{2}-v)\ \mathrm{odd}, 
\\ \\ \\ 
\begin{pspicture}[shift=-0.4](-0.8,-0.5)(6.5,0.5)
\rput(0,-0.5){$\mathcal I_n^{2v+1}$}
\rput(1.0,0.5){$\mathcal I_n^{2v+3}$}
\rput(1.9,-0.5){$\mathcal I_n^{2v+5}$}
\rput(3.2,0.5){$\dots$}
\rput(4.55,-0.5){$\mathcal I_n^{n-4}$}
\rput(5.7,0.5){$\mathcal I_n^{n-2}$}
\rput(6.40,-0.5){$\mathcal I_n^{n}$}
\psline[linewidth=.8pt,arrowsize=3pt 2]{->}(0.6,0.25)(0.1,-0.2)
\psline[linewidth=.8pt,arrowsize=3pt 2]{->}(1,0.25)(1.5,-0.2)
\psline[linewidth=.8pt,arrowsize=3pt 2]{->}(2.6,0.25)(2.1,-0.2)
\psline[linewidth=.8pt,arrowsize=3pt 2]{->}(3.7,0.25)(4.2,-0.2)
\psline[linewidth=.8pt,arrowsize=3pt 2]{->}(5.3,0.25)(4.8,-0.2)
\psline[linewidth=.8pt,arrowsize=3pt 2]{->}(5.7,0.25)(6.2,-0.2)
\end{pspicture}
\qquad &(\frac{n-1}2-v)\ \mathrm{even}. 
\end{array}\right.
\label{EII}
\ee
\bigskip\noindent

\item[(2b)] 
For $n$ even and $v \le -\frac12$, the module $E_{n-1}^v$ is contragredient to $E_{n-1}^{-v}$. That is,
the two modules have the same irreducible composition factors,
 but the structure pattern of $E_{n-1}^v$ is obtained from the one of $E_{n-1}^{-v}$ given in (\ref{EII}) by reversing all the arrows.
\end{itemize}
\end{Theoreme}

\vspace{5truemm}

As the standard modules for $n$ odd are irreducible, it follows that the decomposition of $E_{n-1}^v$ in (\ref{EVpV}) can be written as
\be
 E_{n-1}^v \simeq\, V_n^{2|v|+1} \oplus V_n^{2|v|+5} \oplus V_n^{2|v|+9} \oplus \ldots \oplus (V_n^{n-2} \;{\rm or} \; V_n^n).
\label{eq:directsum!}
\ee
The structure pattern of $E_{n-1}^v$ for $n$ even and 
$v \le -\frac12$ can likewise be expressed in terms of standard modules, as we have
\be
E_{n-1}^v \simeq\, \Big(V_n^{2|v|+1} 
\begin{pspicture}(0,0)(1,0)
\psline[linewidth=.8pt,arrowsize=3pt 2]{<-}(0.2,0.1)(0.8,0.1)
\end{pspicture} 
V_n^{2|v|+5}
\begin{pspicture}(0,0)(1,0)
\psline[linewidth=.8pt,arrowsize=3pt 2]{<-}(0.2,0.1)(0.8,0.1)
\end{pspicture} 
V_n^{2|v|+9}
\begin{pspicture}(0,0)(1,0)
\psline[linewidth=.8pt,arrowsize=3pt 2]{<-}(0.2,0.1)(0.8,0.1)
\end{pspicture} 
\,\ldots
\begin{pspicture}(0,0)(1,0)
\psline[linewidth=.8pt,arrowsize=3pt 2]{<-}(0.2,0.1)(0.8,0.1)
\end{pspicture} 
(V_n^{n-2} \;{\rm or} \; V_n^n)\Big) \qquad (v \le -\tfrac12).
\label{EVV}
\ee
For $v \ge \frac12$, the structure pattern of $E_{n-1}^v$ is obtained from \eqref{EVV} by reversing the arrows and replacing the standard modules 
$V_n^{2|v| + 4k+1}$ by their contragredient counterparts,
\be
 (V_n^0)^\star\,\simeq\,\mathcal I_n^2, \qquad (V_n^d)^\star\,\simeq\,\Big(\mathcal I_n^d 
\begin{pspicture}(0,0)(1,0)
\psline[linewidth=.8pt,arrowsize=3pt 2]{<-}(0.2,0.1)(0.8,0.1)
\end{pspicture} 
\mathcal I_n^{d+2}\Big)\quad (d = 2,4, \dots, n-2),
\qquad (V_n^n)^\star\,\simeq\,\mathcal I_n^n.
\ee 
The naive proposal \eqref{eq:directsum?} therefore holds for $n$ odd, and for $n$ even if 
$v = -\frac{n-1}2, -\frac{n-3}2$ or $\frac{n-1}2$. 

The following two sections are devoted to the proof of the above structure theorem. Section~\ref{sec:Intertwiners} sets the stage 
by introducing a set of intertwiners used in the bulk of the proof which is subsequently presented in Section~\ref{sec:Structure}.

\section{Intertwiners}
\label{sec:Intertwiners}

The proof of Theorem~\ref{thm:main} presented in Section~\ref{sec:Structure} is obtained by relating the module structures of the dimer 
and link state representations. Important roles are played by the three families of intertwiners defined in the following.

\subsection{Spin-spin intertwiner}
\label{sec:J}

As an operator acting on $(\mathbb C^2)^{\otimes(n-1)}$, $J$ is defined as
\be
 J = \sum_{j=1}^{n-2}(-1)^{j-1} \sigma^-_j  \sigma^-_{j+1}
\label{eq:Jeta}
\ee
and is seen to decrease the value of the magnetisation $v$ by two units, 
\be
 J \; : \; E_{n-1}^v \,\to\, E_{n-1}^{v-2}\,.
\ee 
A key property of $J$ is given in the following lemma. 
\begin{Lemma}
The operator $J$ commutes with the spin-chain representation $\tau$, 
\be
 [J,\tau(c)] =0, \qquad  c \in {\rm TL}_n(0),
\label{eq:Jtau}
\ee
and intertwines the dimer representations $\tau_v$ and $\tau_{v-2}$,
\be
 J \, \tau_v = \tau_{v-2} \, J,\qquad v=-\tfrac{n-1}{2},-\tfrac{n-3}{2},\ldots,\tfrac{n-1}{2},
\label{Jtau}
\ee
where $\tau_{v-2}\equiv0$ for $v-2<-\frac{n-1}{2}$.
\label{sec:intertJ}
\end{Lemma} 
\noindent {\sc Proof }
It is straightforward to verify (\ref{eq:Jtau}) and (\ref{Jtau}) when specialised to $c = I$ or $c = e_j$. 
The properties for general $c$ then follow from the fact that $\tau$ is a representation, see Proposition~\ref{prop:tau}.

\subsection{Link-spin intertwiners}  

For every fixed value of $n$ and $v \ge -\frac12$ (with $v$ respectively integer and half-integer for $n$ odd and even), we introduce a 
map 
\be
 h_{v}\; : \;  W_n^{2v+1} \,\to\, E_{n-1}^{v}
\label{eq:hv}
\ee
sending link states with $p$ or $p-1$ half-arcs to spin states with $p$ down-arrows (where $p = \frac{n-1}2-v$). 
Its action is initially defined on the link states in $B_n^{2v+1} \cup B_n^{2v+3}$, as described in the following, 
and then linearly extended to $W_n^{2v+1}$. 

First, we label the half-arcs of the link states in $B_n^{2v+1} \cup B_n^{2v+3}$.
For $w\in B_n^{2v+1}$, we thus assign a label $k \in \{1, \dots, p\}$ to each of its half-arcs, whereas for $w\in B_n^{2v+3}$, 
$k \in \{1, \dots, p-1\}$. 
We then encode the connections of the half-arcs of $w\in B_n^{2v+1} \cup B_n^{2v+3}$ in the set of pairs
\be
 \psi(w) = \{(i_1, j_1), (i_2, j_2), \dots \},
\label{eq:psi}
\ee
where $i_k$ and $j_k$ denote respectively the left and right endpoints of the $k$-th half-arc.
The order of the pairs in $\psi(w)$ is irrelevant for what follows. 

The action of $h_v$ on $w \in B_n^{2v+1}$ is now defined as
\be 
h_{v}(w) = \prod_{(i,j) \, \in \, \psi(w)} \hspace{-0.3cm}
t_{i,j} \: |u \rangle, 
\qquad 
t_{i,j} = \sigma^-_{i-1} + \sigma^-_{j}, \qquad 
|u \rangle = | \! \uparrow \uparrow \dots \uparrow \, \rangle,
\label{eq:hdef}
\ee
where $\sigma_0^- \equiv \sigma_n^-\equiv0$.
For every half-arc $(i,j)$, the map $h_v$ thus assigns an operator $t_{i,j}$ that decreases the magnetisation by one unit. 
Ultimately, this yields a spin state with $p$ down-arrows. 

For a link state $w \in B_n^{2v+3}$, we denote by $a(w)$ the position of the rightmost (wavy) defect, see Section~\ref{sec:lsr}. 
Compared to (\ref{eq:hdef}), the action of $h_v$ on $w\in B_n^{2v+3}$ includes an extra operator $\sigma_{a(w)-1}^-$ and is given by
\be
 h_{v}(w) = \sigma_{a(w)-1}^- \hspace{-0.3cm} \prod_{(i,j) \, \in \, \psi(w)} \hspace{-0.3cm} t_{i,j} \: |u \rangle. 
\ee 
Each of the $(p-1)$ half-arcs contributes one negative unit of magnetisation, 
as does the wavy defect, again yielding a linear combination of spin states with $p$ down-arrows.

To illustrate, let us consider the case $n=6$ and $v=\frac12$. The states 
$\psset{unit=0.5}
\begin{pspicture}[shift=-0.](-0.0,0)(2.4,0.5)
\psline{-}(0,0)(2.4,0)
\psline[linecolor=coloroflink,linewidth=1.5pt]{-}(0.2,0)(0.2,0.5)
\psline[linecolor=coloroflink,linewidth=1.5pt]{-}(2.2,0)(2.2,0.5)
\psarc[linecolor=coloroflink,linewidth=1.5pt]{-}(1.2,0){0.2}{0}{180}
\psbezier[linecolor=coloroflink,linewidth=1.5pt]{-}(0.6,0)(0.6,0.7)(1.8,0.7)(1.8,0)
\end{pspicture}$,\!
$\psset{unit=0.5}
\begin{pspicture}[shift=-0.](-0.0,0)(2.4,0.5)
\psline{-}(0,0)(2.4,0)
\psline[linecolor=coloroflink,linewidth=1.5pt]{-}(0.2,0)(0.2,0.5)
\psline[linecolor=coloroflink,linewidth=1.5pt]{-}(1.4,0)(1.4,0.5)
\psline[linecolor=coloroflink,linewidth=1.5pt]{-}(1.8,0)(1.8,0.5)
\rput(2.2,0){\wobblytwo}
\psarc[linecolor=coloroflink,linewidth=1.5pt]{-}(0.8,0){0.2}{0}{180}
\end{pspicture} \in W_6^2$,
for example, have their connections encoded by
\be 
\psi(\psset{unit=0.5}
\begin{pspicture}[shift=-0.05](-0.0,0)(2.4,0.5)
\psline{-}(0,0)(2.4,0)
\psline[linecolor=coloroflink,linewidth=1.5pt]{-}(0.2,0)(0.2,0.5)
\psline[linecolor=coloroflink,linewidth=1.5pt]{-}(2.2,0)(2.2,0.5)
\psarc[linecolor=coloroflink,linewidth=1.5pt]{-}(1.2,0){0.2}{0}{180}
\psbezier[linecolor=coloroflink,linewidth=1.5pt]{-}(0.6,0)(0.6,0.7)(1.8,0.7)(1.8,0)
\end{pspicture}) = \{(2,5),(3,4)\}, \qquad
\psi(
\begin{pspicture}[shift=-0.](-0.0,0)(2.4,0.5)
\psline{-}(0,0)(2.4,0)
\psline[linecolor=coloroflink,linewidth=1.5pt]{-}(0.2,0)(0.2,0.5)
\psline[linecolor=coloroflink,linewidth=1.5pt]{-}(1.4,0)(1.4,0.5)
\psline[linecolor=coloroflink,linewidth=1.5pt]{-}(1.8,0)(1.8,0.5)
\rput(2.2,0){\wobblytwo}
\psarc[linecolor=coloroflink,linewidth=1.5pt]{-}(0.8,0){0.2}{0}{180}
\end{pspicture}) = \{(2,3)\},
\ee
and are mapped to the following states in $E_5^{1/2}$:
\begin{alignat}{2}
h_{\frac12}(\psset{unit=0.5}
\begin{pspicture}[shift=-0.05](-0.0,0)(2.4,0.5)
\psline{-}(0,0)(2.4,0)
\psline[linecolor=coloroflink,linewidth=1.5pt]{-}(0.2,0)(0.2,0.5)
\psline[linecolor=coloroflink,linewidth=1.5pt]{-}(2.2,0)(2.2,0.5)
\psarc[linecolor=coloroflink,linewidth=1.5pt]{-}(1.2,0){0.2}{0}{180}
\psbezier[linecolor=coloroflink,linewidth=1.5pt]{-}(0.6,0)(0.6,0.7)(1.8,0.7)(1.8,0)
\end{pspicture}) &= (\sigma^-_1+\sigma^-_5)(\sigma^-_2+\sigma^-_4)\, | \!\uparrow \uparrow \uparrow \uparrow \uparrow \,\rangle =  
| \!\downarrow \downarrow \uparrow \uparrow \uparrow  \,\rangle +
| \!\downarrow \uparrow  \uparrow  \downarrow \uparrow  \,\rangle +
| \!\uparrow \downarrow \uparrow \uparrow \downarrow \,\rangle +
| \!\uparrow \uparrow \uparrow \downarrow \downarrow \,\rangle, \\
h_{\frac12}(\psset{unit=0.5}
\begin{pspicture}[shift=-0.](-0.0,0)(2.4,0.5)
\psline{-}(0,0)(2.4,0)
\psline[linecolor=coloroflink,linewidth=1.5pt]{-}(0.2,0)(0.2,0.5)
\psline[linecolor=coloroflink,linewidth=1.5pt]{-}(1.4,0)(1.4,0.5)
\psline[linecolor=coloroflink,linewidth=1.5pt]{-}(1.8,0)(1.8,0.5)
\rput(2.2,0){\wobblytwo}
\psarc[linecolor=coloroflink,linewidth=1.5pt]{-}(0.8,0){0.2}{0}{180}
\end{pspicture}) &= \sigma^-_5 (\sigma^-_1+\sigma^-_3)\, | \!\uparrow \uparrow \uparrow \uparrow \uparrow \,\rangle = 
| \!\downarrow \uparrow \uparrow \uparrow \downarrow \,\rangle +
| \!\uparrow \uparrow \downarrow \uparrow \downarrow \,\rangle.
\end{alignat}

The intertwining property of $h_v$ is described in the following lemma.
\begin{Lemma}
For $v\geq-\frac12$, the map $h_v$ intertwines the representations $\pi_{2v+1}$ and $\tau_{v}$,
\be
 h_v \, \pi_{2v+1}(c) = \tau_{v}(c) \, h_v, \qquad  c \in {\rm TL}_n(0).
\label{eq:intertw}
\ee
\label{sec:intertw}
\end{Lemma} 
{\scshape Proof } To show that $h_v$ is an intertwiner, it suffices to verify that the intertwining property \eqref{eq:intertw} is satisfied for 
$c = I$ and $c=e_i$ acting on link states in $B_n^{2v+1} \cup B_n^{2v+3}$. 
Indeed, the general claim, for $w \in W_n^{2v+1}$ and $c \in {\rm TL}_n(0)$, then follows from the linearity of $h_v$, $\pi_{2v+1}$
and $\tau_v$ and the homomorphism properties of the representations $\pi_{2v+1}$ and $\tau_v$.

For $c = I$, the intertwining property \eqref{eq:intertw} is trivial. For $c = e_i$, the strategy is to show that the maps $\tau_{v}(e_i)$, 
when acting on $h_v(w)$, satisfy local relations consistent with the corresponding action on $W_n^{2v+1}$. 
A complete set of such relations is obtained by considering all possible ways the nodes $i$ and $i+1$ can be
linked to nodes or be occupied by defects (of which the rightmost can be wavy) in $w$.
If the two nodes are not linked together 
or both occupied by defects, one or both of them must be connected to other nodes to the left or right.
On the submodule $V_n^{2v+1} \subset W_n^{2v+1}$, these relations are 
\be 
\begin{array}{l}
\psset{unit=0.7}
\tau_{v}\hspace{-0.06cm}
\left(
\begin{pspicture}[shift=-0.1](-0.05,0)(0.85,0.6)
\psline{-}(0,0)(0.8,0)
\psline{-}(0,0.6)(0.8,0.6)
\psarc[linecolor=blue,linewidth=1.5pt]{-}(0.4,0){0.2}{0}{180}
\psarc[linecolor=blue,linewidth=1.5pt]{-}(0.4,0.6){0.2}{180}{0}
\rput(0.2,-0.2){\tiny$i$}
\end{pspicture}
\right)
  h_{v}(
\begin{pspicture}[shift=-0.05](-0.0,0)(0.8,0.5)
\psline{-}(0,0)(0.8,0)
\psline[linecolor=coloroflink,linewidth=1.5pt]{-}(0.2,0)(0.2,0.5)
\psline[linecolor=coloroflink,linewidth=1.5pt]{-}(0.6,0)(0.6,0.5)
\rput(0.2,-0.2){\tiny$i$}
\end{pspicture}) =0, 
\\[0.4cm]
\psset{unit=0.7}
\tau_{v}\hspace{-0.06cm}
\left(
\begin{pspicture}[shift=-0.1](-0.05,0)(0.85,0.6)
\psline{-}(0,0)(0.8,0)
\psline{-}(0,0.6)(0.8,0.6)
\psarc[linecolor=blue,linewidth=1.5pt]{-}(0.4,0){0.2}{0}{180}
\psarc[linecolor=blue,linewidth=1.5pt]{-}(0.4,0.6){0.2}{180}{0}
\rput(0.2,-0.2){\tiny$i$}
\end{pspicture}
\right)
  h_{v}(
\begin{pspicture}[shift=-0.05](-0.0,0)(0.8,0.5)
\psline{-}(0,0)(0.8,0)
\psarc[linecolor=coloroflink,linewidth=1.5pt]{-}(0.4,0){0.2}{0}{180}
\rput(0.2,-0.2){\tiny$i$}
\end{pspicture}) =0,
\\ [0.4cm]
\psset{unit=0.7}
\tau_{v}\hspace{-0.06cm}
\left(
\begin{pspicture}[shift=-0.1](-0.05,0)(1.25,0.6)
\psline{-}(0,0)(1.2,0)
\psline{-}(0,0.6)(1.2,0.6)
\psarc[linecolor=blue,linewidth=1.5pt]{-}(0.4,0){0.2}{0}{180}
\psarc[linecolor=blue,linewidth=1.5pt]{-}(0.4,0.6){0.2}{180}{0}
\psline[linecolor=blue,linewidth=1.5pt]{-}(1.0,0)(1.0,0.6)
\rput(0.2,-0.2){\tiny$i$}
\rput(1.0,-0.23){\tiny$j$}
\end{pspicture}
\right)
  h_{v}(
\begin{pspicture}[shift=-0.05](-0.0,0)(1.2,0.5)
\psline{-}(0,0)(1.2,0)
\psarc[linecolor=coloroflink,linewidth=1.5pt]{-}(0.8,0){0.2}{0}{180}
\psline[linecolor=coloroflink,linewidth=1.5pt]{-}(0.2,0)(0.2,0.5)
\rput(0.2,-0.2){\tiny$i$}
\rput(1.0,-0.23){\tiny$j$}
\end{pspicture}) =
  h_{v}(
\begin{pspicture}[shift=-0.05](-0.0,0)(1.2,0.5)
\psline{-}(0,0)(1.2,0)
\psarc[linecolor=coloroflink,linewidth=1.5pt]{-}(0.4,0){0.2}{0}{180}
\psline[linecolor=coloroflink,linewidth=1.5pt]{-}(1.0,0)(1.0,0.5)
\rput(0.2,-0.2){\tiny$i$}
\rput(1.0,-0.23){\tiny$j$}
\end{pspicture}),
\\[0.4cm]\psset{unit=0.7}
\tau_{v}\hspace{-0.06cm}
\left(
\begin{pspicture}[shift=-0.1](-0.05,0)(1.25,0.6)
\psline{-}(0,0)(1.2,0)
\psline{-}(0,0.6)(1.2,0.6)
\psarc[linecolor=blue,linewidth=1.5pt]{-}(0.8,0){0.2}{0}{180}
\psarc[linecolor=blue,linewidth=1.5pt]{-}(0.8,0.6){0.2}{180}{0}
\psline[linecolor=blue,linewidth=1.5pt]{-}(0.2,0)(0.2,0.6)
\rput(0.6,-0.2){\tiny$i$}
\rput(0.2,-0.23){\tiny$j$}
\end{pspicture}
\right)
  h_{v}(
\begin{pspicture}[shift=-0.05](-0.0,0)(1.2,0.5)
\psline{-}(0,0)(1.2,0)
\psarc[linecolor=coloroflink,linewidth=1.5pt]{-}(0.4,0){0.2}{0}{180}
\psline[linecolor=coloroflink,linewidth=1.5pt]{-}(1.0,0)(1.0,0.5)
\rput(0.6,-0.2){\tiny$i$}
\rput(0.2,-0.23){\tiny$j$}
\end{pspicture})
 =
  h_{v}(
\begin{pspicture}[shift=-0.05](-0.0,0)(1.2,0.5)
\psline{-}(0,0)(1.2,0)
\psarc[linecolor=coloroflink,linewidth=1.5pt]{-}(0.8,0){0.2}{0}{180}
\psline[linecolor=coloroflink,linewidth=1.5pt]{-}(0.2,0)(0.2,0.5)
\rput(0.6,-0.2){\tiny$i$}
\rput(0.2,-0.23){\tiny$j$}
\end{pspicture}),
\end{array}
\qquad
\begin{array}{c}
\psset{unit=0.7}
\tau_{v}\hspace{-0.06cm}
\left(
\begin{pspicture}[shift=-0.1](-0.05,0)(1.65,0.6)
\psline{-}(0,0)(1.6,0)
\psline{-}(0,0.6)(1.6,0.6)
\psarc[linecolor=blue,linewidth=1.5pt]{-}(0.4,0){0.2}{0}{180}
\psarc[linecolor=blue,linewidth=1.5pt]{-}(0.4,0.6){0.2}{180}{0}
\psline[linecolor=blue,linewidth=1.5pt]{-}(1.0,0)(1.0,0.6)
\psline[linecolor=blue,linewidth=1.5pt]{-}(1.4,0)(1.4,0.6)
\rput(0.2,-0.2){\tiny$i$}
\rput(1.0,-0.23){\tiny$j$}
\rput(1.4,-0.2){\tiny$k$}
\end{pspicture}
\right)
  h_{v}(
\begin{pspicture}[shift=-0.05](-0.0,0)(1.6,0.5)
\psline{-}(0,0)(1.6,0)
\psarc[linecolor=coloroflink,linewidth=1.5pt]{-}(0.8,0){0.2}{0}{180}
\psbezier[linecolor=coloroflink,linewidth=1.5pt]{-}(0.2,0)(0.2,0.6)(1.4,0.6)(1.4,0)
\rput(0.2,-0.2){\tiny$i$}
\rput(1.0,-0.23){\tiny$j$}
\rput(1.4,-0.2){\tiny$k$}
\end{pspicture})
 =
  h_{v}(
\begin{pspicture}[shift=-0.05](-0.0,0)(1.6,0.5)
\psline{-}(0,0)(1.6,0)
\psarc[linecolor=coloroflink,linewidth=1.5pt]{-}(0.4,0){0.2}{0}{180}
\psarc[linecolor=coloroflink,linewidth=1.5pt]{-}(1.2,0){0.2}{0}{180}
\rput(0.2,-0.2){\tiny$i$}
\rput(1.0,-0.23){\tiny$j$}
\rput(1.4,-0.2){\tiny$k$}
\end{pspicture}),
\\[0.4cm]
\psset{unit=0.7}
\tau_{v}\hspace{-0.06cm}
\left(
\begin{pspicture}[shift=-0.1](-0.05,0)(1.65,0.6)
\psline{-}(0,0)(1.6,0)
\psline{-}(0,0.6)(1.6,0.6)
\psarc[linecolor=blue,linewidth=1.5pt]{-}(0.8,0){0.2}{0}{180}
\psarc[linecolor=blue,linewidth=1.5pt]{-}(0.8,0.6){0.2}{180}{0}
\psline[linecolor=blue,linewidth=1.5pt]{-}(0.2,0)(0.2,0.6)
\psline[linecolor=blue,linewidth=1.5pt]{-}(1.4,0)(1.4,0.6)
\rput(0.6,-0.2){\tiny$i$}
\rput(0.2,-0.23){\tiny$j$}
\rput(1.4,-0.2){\tiny$k$}
\end{pspicture}
\right)
  h_{v}(
\begin{pspicture}[shift=-0.05](-0.0,0)(1.6,0.5)
\psline{-}(0,0)(1.6,0)
\psarc[linecolor=coloroflink,linewidth=1.5pt]{-}(0.4,0){0.2}{0}{180}
\psarc[linecolor=coloroflink,linewidth=1.5pt]{-}(1.2,0){0.2}{0}{180}
\rput(0.6,-0.2){\tiny$i$}
\rput(0.2,-0.23){\tiny$j$}
\rput(1.4,-0.2){\tiny$k$}
\end{pspicture})
 =
  h_{v}(
\begin{pspicture}[shift=-0.05](-0.0,0)(1.6,0.5)
\psline{-}(0,0)(1.6,0)
\psarc[linecolor=coloroflink,linewidth=1.5pt]{-}(0.8,0){0.2}{0}{180}
\psbezier[linecolor=coloroflink,linewidth=1.5pt]{-}(0.2,0)(0.2,0.6)(1.4,0.6)(1.4,0)
\rput(0.6,-0.2){\tiny$i$}
\rput(0.2,-0.23){\tiny$j$}
\rput(1.4,-0.2){\tiny$k$}
\end{pspicture}),
\\[0.4cm] \psset{unit=0.7}
\tau_{v}\hspace{-0.06cm}
\left(
\begin{pspicture}[shift=-0.1](-0.05,0)(1.65,0.6)
\psline{-}(0,0)(1.6,0)
\psline{-}(0,0.6)(1.6,0.6)
\psarc[linecolor=blue,linewidth=1.5pt]{-}(1.2,0){0.2}{0}{180}
\psarc[linecolor=blue,linewidth=1.5pt]{-}(1.2,0.6){0.2}{180}{0}
\psline[linecolor=blue,linewidth=1.5pt]{-}(0.2,0)(0.2,0.6)
\psline[linecolor=blue,linewidth=1.5pt]{-}(0.6,0)(0.6,0.6)
\rput(1.0,-0.2){\tiny$i$}
\rput(0.2,-0.23){\tiny$j$}
\rput(0.6,-0.2){\tiny$k$}
\end{pspicture}
\right)
  h_{v}(
\begin{pspicture}[shift=-0.05](-0.0,0)(1.6,0.5)
\psline{-}(0,0)(1.6,0)
\psarc[linecolor=coloroflink,linewidth=1.5pt]{-}(0.8,0){0.2}{0}{180}
\psbezier[linecolor=coloroflink,linewidth=1.5pt]{-}(0.2,0)(0.2,0.6)(1.4,0.6)(1.4,0)
\rput(1.0,-0.2){\tiny$i$}
\rput(0.2,-0.23){\tiny$j$}
\rput(0.6,-0.2){\tiny$k$}
\end{pspicture})
 =
  h_{v}(
\begin{pspicture}[shift=-0.05](-0.0,0)(1.6,0.5)
\psline{-}(0,0)(1.6,0)
\psarc[linecolor=coloroflink,linewidth=1.5pt]{-}(0.4,0){0.2}{0}{180}
\psarc[linecolor=coloroflink,linewidth=1.5pt]{-}(1.2,0){0.2}{0}{180}
\rput(1.0,-0.2){\tiny$i$}
\rput(0.2,-0.23){\tiny$j$}
\rput(0.6,-0.2){\tiny$k$}
\end{pspicture}),
\end{array}
\label{htau}
\vspace{.1cm}  
\ee
where the unlabeled node in each diagram is in position $i+1$. The positions of the nodes labeled by $j$ and $k$ are only constrained 
by the order indicated in the corresponding diagrams. 
Each relation in (\ref{htau}) translates into an algebraic identity that is straightforward to verify.
For example, for the last relation in the first column, one finds that
\be
 (\sigma^-_{i-1}\sigma^+_i + \sigma_i^+\sigma^-_{i+1}) A(w) (\sigma_{j-1}^- + \sigma_i^-)\, 
  |u\rangle = A(w)(\sigma_{i-1}^- + \sigma_{i+1}^-) \, |u\rangle,\qquad A(w) = \prod t_{k,\ell},
\ee
where the product in $A(w)$ is over $(k,\ell) \in \psi(w)\setminus \{(j,i)\}$.
To extend the proof to all of $W_n^{2v+1}$, one readily establishes the following relations involving the wavy defect:
\be 
\begin{array}{l}
\psset{unit=0.7}
\tau_{v}\hspace{-0.06cm}
\left(
\begin{pspicture}[shift=-0.1](-0.05,0)(0.85,0.6)
\psline{-}(0,0)(0.8,0)
\psline{-}(0,0.6)(0.8,0.6)
\psarc[linecolor=blue,linewidth=1.5pt]{-}(0.4,0){0.2}{0}{180}
\psarc[linecolor=blue,linewidth=1.5pt]{-}(0.4,0.6){0.2}{180}{0}
\rput(0.2,-0.2){\tiny$i$}
\end{pspicture}
\right)
  h_{v}(
\begin{pspicture}[shift=-0.05](-0.0,0)(0.8,0.5)
\psline{-}(0,0)(0.8,0)
\psline[linecolor=coloroflink,linewidth=1.5pt]{-}(0.2,0)(0.2,0.5)
\rput(0.6,0){\wobblytwo}
\rput(0.2,-0.2){\tiny$i$}
\end{pspicture}) =
  h_{v}(
\begin{pspicture}[shift=-0.05](-0.0,0)(0.8,0.5)
\psline{-}(0,0)(0.8,0)
\psarc[linecolor=coloroflink,linewidth=1.5pt]{-}(0.4,0){0.2}{0}{180}
\rput(0.2,-0.2){\tiny$i$}
\end{pspicture}), 
\\ [0.4cm]
\psset{unit=0.7}
\tau_{v}\hspace{-0.06cm}
\left(
\begin{pspicture}[shift=-0.1](-0.05,0)(1.25,0.6)
\psline{-}(0,0)(1.2,0)
\psline{-}(0,0.6)(1.2,0.6)
\psarc[linecolor=blue,linewidth=1.5pt]{-}(0.4,0){0.2}{0}{180}
\psarc[linecolor=blue,linewidth=1.5pt]{-}(0.4,0.6){0.2}{180}{0}
\psline[linecolor=blue,linewidth=1.5pt]{-}(1.0,0)(1.0,0.6)
\rput(0.2,-0.2){\tiny$i$}
\rput(1.0,-0.23){\tiny$j$}
\end{pspicture}
\right)
  h_{v}(
\begin{pspicture}[shift=-0.05](-0.0,0)(1.2,0.5)
\psline{-}(0,0)(1.2,0)
\psarc[linecolor=coloroflink,linewidth=1.5pt]{-}(0.8,0){0.2}{0}{180}
\rput(0.2,0){\wobblytwo}
\rput(0.2,-0.2){\tiny$i$}
\rput(1.0,-0.23){\tiny$j$}
\end{pspicture}) =
  h_{v}(
\begin{pspicture}[shift=-0.05](-0.0,0)(1.2,0.5)
\psline{-}(0,0)(1.2,0)
\psarc[linecolor=coloroflink,linewidth=1.5pt]{-}(0.4,0){0.2}{0}{180}
\rput(1.0,0){\wobblytwo}
\rput(0.2,-0.2){\tiny$i$}
\rput(1.0,-0.23){\tiny$j$}
\end{pspicture}),
\\[0.4cm]\psset{unit=0.7}
\tau_{v}\hspace{-0.06cm}
\left(
\begin{pspicture}[shift=-0.1](-0.05,0)(1.25,0.6)
\psline{-}(0,0)(1.2,0)
\psline{-}(0,0.6)(1.2,0.6)
\psarc[linecolor=blue,linewidth=1.5pt]{-}(0.8,0){0.2}{0}{180}
\psarc[linecolor=blue,linewidth=1.5pt]{-}(0.8,0.6){0.2}{180}{0}
\psline[linecolor=blue,linewidth=1.5pt]{-}(0.2,0)(0.2,0.6)
\rput(0.6,-0.2){\tiny$i$}
\rput(0.2,-0.23){\tiny$j$}
\end{pspicture}
\right)
  h_{v}(
\begin{pspicture}[shift=-0.05](-0.0,0)(1.2,0.5)
\psline{-}(0,0)(1.2,0)
\psarc[linecolor=coloroflink,linewidth=1.5pt]{-}(0.4,0){0.2}{0}{180}
\rput(1.0,0){\wobblytwo}
\rput(0.6,-0.2){\tiny$i$}
\rput(0.2,-0.23){\tiny$j$}
\end{pspicture})
 =
  h_{v}(
\begin{pspicture}[shift=-0.05](-0.0,0)(1.2,0.5)
\psline{-}(0,0)(1.2,0)
\psarc[linecolor=coloroflink,linewidth=1.5pt]{-}(0.8,0){0.2}{0}{180}
\rput(0.2,0){\wobblytwo}
\rput(0.6,-0.2){\tiny$i$}
\rput(0.2,-0.23){\tiny$j$}
\end{pspicture}).
\end{array}
\ee
\hfill $\square$
\medskip

For every $v \ge 0$, the map $h_v$ is non-zero. In fact, an even stronger statement holds according to the following lemma.
\begin{Lemma} 
For $v \ge 0$ and $k = 0, \dots, \lfloor\tfrac{n-1-2v}4\rfloor$, the composed map
\be
 h_{v,k}= J^k\, h_{v+2k} \;:\; W_n^{2v+4k+1} \,\to\, E_{n-1}^{v}
 \label{eq:Jhcomp}
\ee
is an intertwiner and it is non-zero on the submodule $V_n^{2v+4k+1} \subset W_n^{2v+4k+1}$.
\label{sec:hvk}
\end{Lemma}
{\scshape Proof }
The intertwining property of $h_{v,k}$ follows from the 
intertwining properties of $J$ and $h_{v+2k}$ given in Lemmas~\ref{sec:intertJ} 
and~\ref{sec:intertw}. Setting 
$d = 2v+4k+1$ 
and $p = \frac{n-d}2$, we 
demonstrate that the image of $h_{v,k}$ on $V_n^d \subset W_n^d$ is non-zero by computing the matrix element
\be
\Big\langle \hspace{-0.1cm}
\psset{unit=0.7}
 \begin{pspicture}[shift=-0.05](-0.0,0)(0.0,0.5)
\rput(0.55,-0.5){\tiny $\underbrace{\ \hspace{0.3cm}\ }_p$}
\rput(1.92,-0.5){\tiny $\underbrace{\ \hspace{1.00cm}\ }_{p+2k}$}
\rput(3.55,-0.5){\tiny $\underbrace{\ \hspace{0.7cm}\ }_{d-2k-1}$}
\end{pspicture}
\uparrow \uparrow \uparrow
\downarrow \downarrow \downarrow 
\downarrow \downarrow \downarrow \downarrow 
\uparrow\uparrow\uparrow\uparrow\uparrow
\!
\Big| h_{v,k}\big(\,
 \begin{pspicture}[shift=-0.8](-0.0,-0.8)(6.4,0.5)
\psline{-}(0,0)(6.4,0)
\psbezier[linecolor=coloroflink,linewidth=1.5pt]{-}(0.2,0)(0.2,1.0)(2.2,1.0)(2.2,0)
\psbezier[linecolor=coloroflink,linewidth=1.5pt]{-}(0.6,0)(0.6,0.6)(1.8,0.6)(1.8,0)
\psarc[linecolor=coloroflink,linewidth=1.5pt]{-}(1.2,0){0.2}{0}{180}
\psline[linecolor=coloroflink,linewidth=1.5pt]{-}(2.6,0)(2.6,0.6)
\psline[linecolor=coloroflink,linewidth=1.5pt]{-}(3.0,0)(3.0,0.6)
\psline[linecolor=coloroflink,linewidth=1.5pt]{-}(3.4,0)(3.4,0.6)
\psline[linecolor=coloroflink,linewidth=1.5pt]{-}(3.8,0)(3.8,0.6)
\psline[linecolor=coloroflink,linewidth=1.5pt]{-}(4.2,0)(4.2,0.6)
\psline[linecolor=coloroflink,linewidth=1.5pt]{-}(4.6,0)(4.6,0.6)
\psline[linecolor=coloroflink,linewidth=1.5pt]{-}(5.0,0)(5.0,0.6)
\psline[linecolor=coloroflink,linewidth=1.5pt]{-}(5.4,0)(5.4,0.6)
\psline[linecolor=coloroflink,linewidth=1.5pt]{-}(5.8,0)(5.8,0.6)
\psline[linecolor=coloroflink,linewidth=1.5pt]{-}(6.2,0)(6.2,0.6)
\rput(0.6,-0.5){\tiny $\underbrace{\ \hspace{0.5cm}\ }_p$}
\rput(4.4,-0.5){\tiny $\underbrace{\qquad \hspace{1.7cm}\qquad}_d$}
\end{pspicture}
\,\big) \Big\rangle
= k!
\label{eq:nonzerohvd}
\ee
which is seen to be non-vanishing.
\hfill $\square$
\medskip

For later convenience, we denote by $\tilde h_{v,k}$ the restriction of the composed intertwiner $h_{v,k}$ 
to the subspace $V_n^{2v+4k+1} \subset W_n^{2v+4k+1}$.

We conclude this subsection by noting that the value $v = -\frac12$ is excluded from Lemma \ref{sec:hvk} even though \eqref{eq:hdef} 
is well-defined. Indeed, for $v = -\frac12$, our proof of Lemma~\ref{sec:hvk} does not apply because the number $d-2k-1=2(v + k)$ 
appearing in \eqref{eq:nonzerohvd} is negative for $k =0$. It is however positive for $k \ge 1$, implying that the image of $h_{-\frac12, k}$ on $V_n^{4k}\subset W_n^{4k}$ 
is non-zero for $k \ge 1$. For $k=0$, one can instead prove that $h_{-\frac12}$ is identically zero on $V_n^0 \subset W_n^0$. 
More generally, we note that for $v \le -\frac 12$, the definition of $h_{v,k}$ is well-defined if $2v+4k+1 \ge 0$, 
and that it is non-zero on $V_n^{2v+4k+1} \subset W_n^{2v+4k+1}$ if $v+k\ge 0$.

\subsection{Link-link intertwiners}

A complete list of intertwiners between TL standard modules was obtained by Graham and Lehrer \cite{GL98}, for all $\beta \in \mathbb R$. 
Specialising to $\beta = 0$, the proof of Lemma~\ref{prop:lastnonzero} below is based on a family of such intertwiners for $n$ even, each 
interlacing a pair of adjacent standard modules,
\be
 g_d \;:\; V_n^{d+2} \,\to\, V_n^d, \qquad 0 \le d \le n-2, \qquad d \; {\rm even}. 
\ee
This section defines these maps explicitly, and to keep the proof of Theorem \ref{thm:main} self-contained, 
their intertwining properties are established independently.

We first note that the state $y_n \in V_n^{n-2}$ defined as
\be
\psset{unit=0.5}
y_n  = \, \begin{pspicture}[shift=-0.05](-0.0,0)(4.0,0.5)
\psline{-}(0,0)(4.0,0)
\psarc[linecolor=coloroflink,linewidth=1.5pt]{-}(0.4,0){0.2}{0}{180}
\psline[linecolor=coloroflink,linewidth=1.5pt]{-}(1.0,0)(1.0,0.5)
\psline[linecolor=coloroflink,linewidth=1.5pt]{-}(1.4,0)(1.4,0.5)
\psline[linecolor=coloroflink,linewidth=1.5pt]{-}(1.8,0)(1.8,0.5)
\psline[linecolor=coloroflink,linewidth=1.5pt]{-}(2.2,0)(2.2,0.5)
\psline[linecolor=coloroflink,linewidth=1.5pt]{-}(2.6,0)(2.6,0.5)
\rput(3.0,0.25){.\hspace{-0.025cm}.\hspace{-0.025cm}.}
\psline[linecolor=coloroflink,linewidth=1.5pt]{-}(3.4,0)(3.4,0.5)
\psline[linecolor=coloroflink,linewidth=1.5pt]{-}(3.8,0)(3.8,0.5)
\end{pspicture} \, - \, 
 \begin{pspicture}[shift=-0.05](-0.0,0)(4.0,0.5)
\psline{-}(0,0)(4.0,0)
\psarc[linecolor=coloroflink,linewidth=1.5pt]{-}(1.2,0){0.2}{0}{180}
\psline[linecolor=coloroflink,linewidth=1.5pt]{-}(0.2,0)(0.2,0.5)
\psline[linecolor=coloroflink,linewidth=1.5pt]{-}(0.6,0)(0.6,0.5)
\psline[linecolor=coloroflink,linewidth=1.5pt]{-}(1.8,0)(1.8,0.5)
\psline[linecolor=coloroflink,linewidth=1.5pt]{-}(2.2,0)(2.2,0.5)
\psline[linecolor=coloroflink,linewidth=1.5pt]{-}(2.6,0)(2.6,0.5)
\rput(3.0,0.25){.\hspace{-0.025cm}.\hspace{-0.025cm}.}
\psline[linecolor=coloroflink,linewidth=1.5pt]{-}(3.4,0)(3.4,0.5)
\psline[linecolor=coloroflink,linewidth=1.5pt]{-}(3.8,0)(3.8,0.5)
\end{pspicture} \, + \,
\begin{pspicture}[shift=-0.05](-0.0,0)(4.0,0.5)
\psline{-}(0,0)(4.0,0)
\psarc[linecolor=coloroflink,linewidth=1.5pt]{-}(2.0,0){0.2}{0}{180}
\psline[linecolor=coloroflink,linewidth=1.5pt]{-}(0.2,0)(0.2,0.5)
\psline[linecolor=coloroflink,linewidth=1.5pt]{-}(0.6,0)(0.6,0.5)
\psline[linecolor=coloroflink,linewidth=1.5pt]{-}(1.0,0)(1.0,0.5)
\psline[linecolor=coloroflink,linewidth=1.5pt]{-}(1.4,0)(1.4,0.5)
\psline[linecolor=coloroflink,linewidth=1.5pt]{-}(2.6,0)(2.6,0.5)
\rput(3.0,0.25){.\hspace{-0.025cm}.\hspace{-0.025cm}.}
\psline[linecolor=coloroflink,linewidth=1.5pt]{-}(3.4,0)(3.4,0.5)
\psline[linecolor=coloroflink,linewidth=1.5pt]{-}(3.8,0)(3.8,0.5)
\end{pspicture} - \dots   
+ (-1)^{\frac{n-2}2} \,
\begin{pspicture}[shift=-0.05](-0.0,0)(4.0,0.5)
\psline{-}(0,0)(4.0,0)
\psarc[linecolor=coloroflink,linewidth=1.5pt]{-}(3.6,0){0.2}{0}{180}
\psline[linecolor=coloroflink,linewidth=1.5pt]{-}(0.2,0)(0.2,0.5)
\psline[linecolor=coloroflink,linewidth=1.5pt]{-}(0.6,0)(0.6,0.5)
\psline[linecolor=coloroflink,linewidth=1.5pt]{-}(1.0,0)(1.0,0.5)
\psline[linecolor=coloroflink,linewidth=1.5pt]{-}(1.4,0)(1.4,0.5)
\psline[linecolor=coloroflink,linewidth=1.5pt]{-}(1.8,0)(1.8,0.5)
\psline[linecolor=coloroflink,linewidth=1.5pt]{-}(2.2,0)(2.2,0.5)
\psline[linecolor=coloroflink,linewidth=1.5pt]{-}(3.0,0)(3.0,0.5)
\rput(2.6,0.25){.\hspace{-0.025cm}.\hspace{-0.025cm}.}
\end{pspicture}
\ee
satisfies
\be
 e_j y_n = 0, \qquad j = 1, \dots, n-1
\label{eq:eym0}
\ee
with respect to the standard action.
In fact, the equality $c\, y_n = 0$ holds for all $c \in {\rm TL}_n(0)$ 
except for the identity element for which $I y_n = y_n$. Equivalently, the map $g_{n-2}: V_n^n \to V_n^{n-2}$ defined by $g_{n-2}(\,
\psset{unit=0.5}
\begin{pspicture}[shift=-0.05](-0.0,0)(2.4,0.5)
\psline{-}(0,0)(2.4,0)
\psline[linecolor=coloroflink,linewidth=1.5pt]{-}(0.2,0)(0.2,0.5)
\psline[linecolor=coloroflink,linewidth=1.5pt]{-}(0.6,0)(0.6,0.5)
\psline[linecolor=coloroflink,linewidth=1.5pt]{-}(1.0,0)(1.0,0.5)
\psline[linecolor=coloroflink,linewidth=1.5pt]{-}(1.4,0)(1.4,0.5)
\rput(1.8,0.25){.\hspace{-0.025cm}.\hspace{-0.025cm}.}
\psline[linecolor=coloroflink,linewidth=1.5pt]{-}(2.2,0)(2.2,0.5)
\end{pspicture}\,
) = y_n$ is an intertwiner between 
(the representations corresponding to) the standard modules
$V_n^n$ and $V_n^{n-2}$,
\be
\rho_{n-2}(c) g_{n-2}= g_{n-2} \rho_{n}(c),\qquad c \in {\rm TL}_n(0).
\ee
Hereafter, we will denote $y_n$ diagrammatically by 
\be
y_n= 
\psset{unit=0.5}
\begin{pspicture}[shift=-0.05](-0.0,0)(3.2,0.5)
\psline{-}(0,0)(3.2,0)
\pspolygon[fillstyle=solid,fillcolor=lightpurple](0.2,0)(3,0)(3,0.5)(0.2,0.5)
\rput(1.6,0.25){$_n$}
\end{pspicture}\, .
\ee

For $0\le d < n-2$, the action of $g_d$ on $w\in V_n^{d+2}$ is defined as follows. 
First, one temporarily erases the $\frac{n-d-2}2$ half-arcs of $w$, replaces the $d+2$ defects by $y_{d+2}$, and reinstates the 
half-arcs in their original positions. This procedure is illustrated by 
\be
g_2(\, 
\psset{unit=0.5}
\begin{pspicture}[shift=-0.05](-0.0,0)(3.2,0.5)
\psline{-}(0,0)(3.2,0)
\psarc[linecolor=coloroflink,linewidth=1.5pt]{-}(1.6,0){0.2}{0}{180}
\psarc[linecolor=coloroflink,linewidth=1.5pt]{-}(2.4,0){0.2}{0}{180}
\psline[linecolor=coloroflink,linewidth=1.5pt]{-}(0.2,0)(0.2,0.5)
\psline[linecolor=coloroflink,linewidth=1.5pt]{-}(0.6,0)(0.6,0.5)
\psline[linecolor=coloroflink,linewidth=1.5pt]{-}(1.0,0)(1.0,0.5)
\psline[linecolor=coloroflink,linewidth=1.5pt]{-}(3.0,0)(3.0,0.5)
\end{pspicture}\,)
 = \,
 \begin{pspicture}[shift=-0.05](-0.0,0)(3.2,1.6)
\psline{-}(0,0)(3.2,0)
\psline[linecolor=gray]{-}(0,0.5)(3.2,0.5)
\psarc[linecolor=coloroflink,linewidth=1.5pt]{-}(1.6,0){0.2}{0}{180}
\psarc[linecolor=coloroflink,linewidth=1.5pt]{-}(2.4,0){0.2}{0}{180}
\psline[linecolor=coloroflink,linewidth=1.5pt]{-}(0.2,0)(0.2,0.5)
\psline[linecolor=coloroflink,linewidth=1.5pt]{-}(0.6,0)(0.6,0.5)
\psline[linecolor=coloroflink,linewidth=1.5pt]{-}(1.0,0)(1.0,0.5)
\psline[linecolor=coloroflink,linewidth=1.5pt]{-}(3.0,0)(3.0,0.5)
\rput(0.8,1.2){
\psline{-}(-0.2,0)(1.8,0)
\pspolygon[fillstyle=solid,fillcolor=lightpurple](0,0)(1.6,0)(1.6,0.5)(0,0.5)
\rput(0.8,0.245){$_{_4}$}
}
\psbezier[linecolor=coloroflink,linewidth=1.5pt]{-}(0.2,0.5)(0.2,0.85)(1.0,0.85)(1.0,1.2)
\psbezier[linecolor=coloroflink,linewidth=1.5pt]{-}(0.6,0.5)(0.6,0.85)(1.4,0.85)(1.4,1.2)
\psbezier[linecolor=coloroflink,linewidth=1.5pt]{-}(1.0,0.5)(1.0,0.85)(1.8,0.85)(1.8,1.2)
\psbezier[linecolor=coloroflink,linewidth=1.5pt]{-}(3.0,0.5)(3.0,0.85)(2.2,0.85)(2.2,1.2)
\end{pspicture}
 \,= \,
\begin{pspicture}[shift=-0.05](-0.0,0)(3.2,0.5)
\psline{-}(0,0)(3.2,0)
\psarc[linecolor=coloroflink,linewidth=1.5pt]{-}(0.4,0){0.2}{0}{180}
\psarc[linecolor=coloroflink,linewidth=1.5pt]{-}(1.6,0){0.2}{0}{180}
\psarc[linecolor=coloroflink,linewidth=1.5pt]{-}(2.4,0){0.2}{0}{180}
\psline[linecolor=coloroflink,linewidth=1.5pt]{-}(1.0,0)(1.0,0.5)
\psline[linecolor=coloroflink,linewidth=1.5pt]{-}(3.0,0)(3.0,0.5)
\end{pspicture}
\, - \,
\begin{pspicture}[shift=-0.05](-0.0,0)(3.2,0.5)
\psline{-}(0,0)(3.2,0)
\psarc[linecolor=coloroflink,linewidth=1.5pt]{-}(1.6,0){0.2}{0}{180}
\psarc[linecolor=coloroflink,linewidth=1.5pt]{-}(2.4,0){0.2}{0}{180}
\psline[linecolor=coloroflink,linewidth=1.5pt]{-}(0.2,0)(0.2,0.5)
\psline[linecolor=coloroflink,linewidth=1.5pt]{-}(0.6,0)(0.6,0.5)
\psbezier[linecolor=coloroflink,linewidth=1.5pt]{-}(1.0,0)(1.0,0.7)(3.0,0.7)(3.0,0)
\end{pspicture}\, .
\label{eq:depiction}
\ee
The map $g_d$ thus outputs the alternating sum of the 
$\frac{d+2}{2}$ link states labeled by $k=1,3,\ldots,d+1$ and obtained by capping
the $k$-th and $(k+1)$-th defects of $w$ with an arch.

The intertwining property of $g_d$ is made manifest in the following lemma.
\begin{Lemma}
For $n$ even, the map $g_d$ intertwines the 
standard representations $\rho_{d}$ and $\rho_{d+2}$,
\be
\rho_{d}(c) g_{d} =  g_{d} \rho_{d+2}(c), \qquad  c \in {\rm TL}_n(0),\qquad d=0,2,\ldots,n-2.
\label{eq:intertg}
\ee
\end{Lemma}
{\scshape Proof } 
Below, we demonstrate that \eqref{eq:intertg} holds on any link state $w \in B_n^{d+2}$, for $c = I$ and $c = e_j$, $j = 1, \dots, n-1$. 
As this extends to $V_n^{d+2}$ by linearity of the actions of $\rho_d$, $\rho_{d+2}$ and $g_d$, and
to all $c \in {\rm TL}_n(0)$ by the linearity and homomorphism properties of $\rho_d$ and $\rho_{d+2}$, the proof is then complete.

For $c = I$, the relation \eqref{eq:intertg} holds trivially. For $c=e_j$ and $w \in B_n^{d+2}$, the proof splits into the three cases (0), (1) 
and (2), according to the total number of defects in positions $j$ and $j+1$ in $w$. In all three cases, we show that 
\be 
 e_j g_d(w) = g_d(e_j w), 
\label{eq:gc=cg}
\ee 
where the action of $e_j$ on each side is the corresponding standard action.

Case (0): 
If the nodes $j$ and $j+1$ of $w$ are 
occupied by half-arcs, the action of $e_j$ on $g_d(w)$ either modifies the imbrication pattern of 
half-arcs or, in the case where $j$ and $j+1$ are connected, 
returns the same link state multiplied by $\beta$. 
In either case, the positions of the nodes connected to $y_{d+2}$ 
remain unchanged. The same final result is obtained if the order of the 
applications of $g_d$ and $e_j$ is reversed: the alterations of the half-arcs are carried out first, after which the defects, still unchanged, 
are replaced by $y_{d+2}$. We illustrate this by two examples for $n=8$, $d=2$ and $j=5$, using the diagrammatic 
representation \eqref{eq:depiction} of $g_d(w)$:
\begin{alignat}{2}
&e_5
g_2(\, 
\psset{unit=0.5}
\begin{pspicture}[shift=-0.05](-0.0,0)(3.2,0.5)
\psline{-}(0,0)(3.2,0)
\psarc[linecolor=coloroflink,linewidth=1.5pt]{-}(1.6,0){0.2}{0}{180}
\psarc[linecolor=coloroflink,linewidth=1.5pt]{-}(2.4,0){0.2}{0}{180}
\psline[linecolor=coloroflink,linewidth=1.5pt]{-}(0.2,0)(0.2,0.5)
\psline[linecolor=coloroflink,linewidth=1.5pt]{-}(0.6,0)(0.6,0.5)
\psline[linecolor=coloroflink,linewidth=1.5pt]{-}(1.0,0)(1.0,0.5)
\psline[linecolor=coloroflink,linewidth=1.5pt]{-}(3.0,0)(3.0,0.5)
\end{pspicture}\,) = \,
\begin{pspicture}[shift=-0.05](-0.0,0)(3.2,2.3)
\psline{-}(0,0)(3.2,0)
\psline[linecolor=coloroflink,linewidth=1.5pt]{-}(0.2,0)(0.2,0.7)
\psline[linecolor=coloroflink,linewidth=1.5pt]{-}(0.6,0)(0.6,0.7)
\psline[linecolor=coloroflink,linewidth=1.5pt]{-}(1.0,0)(1.0,0.7)
\psline[linecolor=coloroflink,linewidth=1.5pt]{-}(1.4,0)(1.4,0.7)
\psline[linecolor=coloroflink,linewidth=1.5pt]{-}(2.6,0)(2.6,0.7)
\psline[linecolor=coloroflink,linewidth=1.5pt]{-}(3.0,0)(3.0,0.7)
\psarc[linecolor=coloroflink,linewidth=1.5pt]{-}(2.0,0){0.2}{0}{180}
\psarc[linecolor=coloroflink,linewidth=1.5pt]{-}(2.0,0.7){-0.2}{0}{180}
\rput(0,0.7){
\psline{-}(0,0)(3.2,0)
\psline[linecolor=gray]{-}(0,0.5)(3.2,0.5)
\psarc[linecolor=coloroflink,linewidth=1.5pt]{-}(1.6,0){0.2}{0}{180}
\psarc[linecolor=coloroflink,linewidth=1.5pt]{-}(2.4,0){0.2}{0}{180}
\psline[linecolor=coloroflink,linewidth=1.5pt]{-}(0.2,0)(0.2,0.5)
\psline[linecolor=coloroflink,linewidth=1.5pt]{-}(0.6,0)(0.6,0.5)
\psline[linecolor=coloroflink,linewidth=1.5pt]{-}(1.0,0)(1.0,0.5)
\psline[linecolor=coloroflink,linewidth=1.5pt]{-}(3.0,0)(3.0,0.5)
\rput(0.8,1.2){
\psline{-}(-0.2,0)(1.8,0)
\pspolygon[fillstyle=solid,fillcolor=lightpurple](0,0)(1.6,0)(1.6,0.5)(0,0.5)
\rput(0.8,0.245){$_{_4}$}
}
\psbezier[linecolor=coloroflink,linewidth=1.5pt]{-}(0.2,0.5)(0.2,0.85)(1.0,0.85)(1.0,1.2)
\psbezier[linecolor=coloroflink,linewidth=1.5pt]{-}(0.6,0.5)(0.6,0.85)(1.4,0.85)(1.4,1.2)
\psbezier[linecolor=coloroflink,linewidth=1.5pt]{-}(1.0,0.5)(1.0,0.85)(1.8,0.85)(1.8,1.2)
\psbezier[linecolor=coloroflink,linewidth=1.5pt]{-}(3.0,0.5)(3.0,0.85)(2.2,0.85)(2.2,1.2)
}
\end{pspicture} 
\, = \,
\begin{pspicture}[shift=-0.05](-0.0,0)(3.2,1.7)
\psline{-}(0,0)(3.2,0)
\psline[linecolor=gray]{-}(0,0.5)(3.2,0.5)
\psarc[linecolor=coloroflink,linewidth=1.5pt]{-}(2.0,0){0.2}{0}{180}
\psbezier[linecolor=coloroflink,linewidth=1.5pt]{-}(1.4,0)(1.4,0.5)(2.6,0.5)(2.6,0)
\psline[linecolor=coloroflink,linewidth=1.5pt]{-}(0.2,0)(0.2,0.5)
\psline[linecolor=coloroflink,linewidth=1.5pt]{-}(0.6,0)(0.6,0.5)
\psline[linecolor=coloroflink,linewidth=1.5pt]{-}(1.0,0)(1.0,0.5)
\psline[linecolor=coloroflink,linewidth=1.5pt]{-}(3.0,0)(3.0,0.5)
\rput(0.8,1.2){
\psline{-}(-0.2,0)(1.8,0)
\pspolygon[fillstyle=solid,fillcolor=lightpurple](0,0)(1.6,0)(1.6,0.5)(0,0.5)
\rput(0.8,0.245){$_{_4}$}
}
\psbezier[linecolor=coloroflink,linewidth=1.5pt]{-}(0.2,0.5)(0.2,0.85)(1.0,0.85)(1.0,1.2)
\psbezier[linecolor=coloroflink,linewidth=1.5pt]{-}(0.6,0.5)(0.6,0.85)(1.4,0.85)(1.4,1.2)
\psbezier[linecolor=coloroflink,linewidth=1.5pt]{-}(1.0,0.5)(1.0,0.85)(1.8,0.85)(1.8,1.2)
\psbezier[linecolor=coloroflink,linewidth=1.5pt]{-}(3.0,0.5)(3.0,0.85)(2.2,0.85)(2.2,1.2)
\end{pspicture} 
\, = 
g_2(
\begin{pspicture}[shift=-0.05](-0.0,0)(3.2,0.5)
\psline{-}(0,0)(3.2,0)
\psarc[linecolor=coloroflink,linewidth=1.5pt]{-}(2.0,0){0.2}{0}{180}
\psbezier[linecolor=coloroflink,linewidth=1.5pt]{-}(1.4,0)(1.4,0.5)(2.6,0.5)(2.6,0)
\psline[linecolor=coloroflink,linewidth=1.5pt]{-}(0.2,0)(0.2,0.5)
\psline[linecolor=coloroflink,linewidth=1.5pt]{-}(0.6,0)(0.6,0.5)
\psline[linecolor=coloroflink,linewidth=1.5pt]{-}(1.0,0)(1.0,0.5)
\psline[linecolor=coloroflink,linewidth=1.5pt]{-}(3.0,0)(3.0,0.5)
\end{pspicture}\,)
=
g_2(e_5\, 
\begin{pspicture}[shift=-0.05](-0.0,0)(3.2,0.5)
\psline{-}(0,0)(3.2,0)
\psarc[linecolor=coloroflink,linewidth=1.5pt]{-}(1.6,0){0.2}{0}{180}
\psarc[linecolor=coloroflink,linewidth=1.5pt]{-}(2.4,0){0.2}{0}{180}
\psline[linecolor=coloroflink,linewidth=1.5pt]{-}(0.2,0)(0.2,0.5)
\psline[linecolor=coloroflink,linewidth=1.5pt]{-}(0.6,0)(0.6,0.5)
\psline[linecolor=coloroflink,linewidth=1.5pt]{-}(1.0,0)(1.0,0.5)
\psline[linecolor=coloroflink,linewidth=1.5pt]{-}(3.0,0)(3.0,0.5)
\end{pspicture}\,),\\
& e_5g_2(\, 
\psset{unit=0.5}
\begin{pspicture}[shift=-0.05](-0.0,0)(3.2,0.5)
\psline{-}(0,0)(3.2,0)
\psarc[linecolor=coloroflink,linewidth=1.5pt]{-}(0.8,0){0.2}{0}{180}
\psarc[linecolor=coloroflink,linewidth=1.5pt]{-}(2.0,0){0.2}{0}{180}
\psline[linecolor=coloroflink,linewidth=1.5pt]{-}(0.2,0)(0.2,0.5)
\psline[linecolor=coloroflink,linewidth=1.5pt]{-}(1.4,0)(1.4,0.5)
\psline[linecolor=coloroflink,linewidth=1.5pt]{-}(2.6,0)(2.6,0.5)
\psline[linecolor=coloroflink,linewidth=1.5pt]{-}(3.0,0)(3.0,0.5)
\end{pspicture}\,) = \,
\begin{pspicture}[shift=-0.05](-0.0,0)(3.2,2.3)
\psline{-}(0,0)(3.2,0)
\psline[linecolor=coloroflink,linewidth=1.5pt]{-}(0.2,0)(0.2,0.7)
\psline[linecolor=coloroflink,linewidth=1.5pt]{-}(0.6,0)(0.6,0.7)
\psline[linecolor=coloroflink,linewidth=1.5pt]{-}(1.0,0)(1.0,0.7)
\psline[linecolor=coloroflink,linewidth=1.5pt]{-}(1.4,0)(1.4,0.7)
\psline[linecolor=coloroflink,linewidth=1.5pt]{-}(2.6,0)(2.6,0.7)
\psline[linecolor=coloroflink,linewidth=1.5pt]{-}(3.0,0)(3.0,0.7)
\psarc[linecolor=coloroflink,linewidth=1.5pt]{-}(2.0,0){0.2}{0}{180}
\psarc[linecolor=coloroflink,linewidth=1.5pt]{-}(2.0,0.7){-0.2}{0}{180}
\rput(0,0.7){
\psline{-}(0,0)(3.2,0)
\psline[linecolor=gray]{-}(0,0.5)(3.2,0.5)
\psarc[linecolor=coloroflink,linewidth=1.5pt]{-}(0.8,0){0.2}{0}{180}
\psarc[linecolor=coloroflink,linewidth=1.5pt]{-}(2.0,0){0.2}{0}{180}
\psline[linecolor=coloroflink,linewidth=1.5pt]{-}(0.2,0)(0.2,0.5)
\psline[linecolor=coloroflink,linewidth=1.5pt]{-}(1.4,0)(1.4,0.5)
\psline[linecolor=coloroflink,linewidth=1.5pt]{-}(2.6,0)(2.6,0.5)
\psline[linecolor=coloroflink,linewidth=1.5pt]{-}(3.0,0)(3.0,0.5)
\rput(0.8,1.2){
\psline{-}(-0.2,0)(1.8,0)
\pspolygon[fillstyle=solid,fillcolor=lightpurple](0,0)(1.6,0)(1.6,0.5)(0,0.5)
\rput(0.8,0.245){$_{_4}$}
}
\psbezier[linecolor=coloroflink,linewidth=1.5pt]{-}(0.2,0.5)(0.2,0.85)(1.0,0.85)(1.0,1.2)
\psline[linecolor=coloroflink,linewidth=1.5pt]{-}(1.4,0.5)(1.4,1.2)
\psbezier[linecolor=coloroflink,linewidth=1.5pt]{-}(2.6,0.5)(2.6,0.85)(1.8,0.85)(1.8,1.2)
\psbezier[linecolor=coloroflink,linewidth=1.5pt]{-}(3.0,0.5)(3.0,0.85)(2.2,0.85)(2.2,1.2)
}
\end{pspicture} 
\, = 
\beta \, \,
\begin{pspicture}[shift=-0.05](-0.0,0)(3.2,1.7)
\psline{-}(0,0)(3.2,0)
\psline[linecolor=gray]{-}(0,0.5)(3.2,0.5)
\psarc[linecolor=coloroflink,linewidth=1.5pt]{-}(0.8,0){0.2}{0}{180}
\psarc[linecolor=coloroflink,linewidth=1.5pt]{-}(2.0,0){0.2}{0}{180}
\psline[linecolor=coloroflink,linewidth=1.5pt]{-}(0.2,0)(0.2,0.5)
\psline[linecolor=coloroflink,linewidth=1.5pt]{-}(1.4,0)(1.4,0.5)
\psline[linecolor=coloroflink,linewidth=1.5pt]{-}(2.6,0)(2.6,0.5)
\psline[linecolor=coloroflink,linewidth=1.5pt]{-}(3.0,0)(3.0,0.5)
\rput(0.8,1.2){
\psline{-}(-0.2,0)(1.8,0)
\pspolygon[fillstyle=solid,fillcolor=lightpurple](0,0)(1.6,0)(1.6,0.5)(0,0.5)
\rput(0.8,0.245){$_{_4}$}
}
\psbezier[linecolor=coloroflink,linewidth=1.5pt]{-}(0.2,0.5)(0.2,0.85)(1.0,0.85)(1.0,1.2)
\psline[linecolor=coloroflink,linewidth=1.5pt]{-}(1.4,0.5)(1.4,1.2)
\psbezier[linecolor=coloroflink,linewidth=1.5pt]{-}(2.6,0.5)(2.6,0.85)(1.8,0.85)(1.8,1.2)
\psbezier[linecolor=coloroflink,linewidth=1.5pt]{-}(3.0,0.5)(3.0,0.85)(2.2,0.85)(2.2,1.2)
\end{pspicture} 
\, = 
\beta \, g_2(\, 
\begin{pspicture}[shift=-0.05](-0.0,0)(3.2,0.5)
\psline{-}(0,0)(3.2,0)
\psarc[linecolor=coloroflink,linewidth=1.5pt]{-}(0.8,0){0.2}{0}{180}
\psarc[linecolor=coloroflink,linewidth=1.5pt]{-}(2.0,0){0.2}{0}{180}
\psline[linecolor=coloroflink,linewidth=1.5pt]{-}(0.2,0)(0.2,0.5)
\psline[linecolor=coloroflink,linewidth=1.5pt]{-}(1.4,0)(1.4,0.5)
\psline[linecolor=coloroflink,linewidth=1.5pt]{-}(2.6,0)(2.6,0.5)
\psline[linecolor=coloroflink,linewidth=1.5pt]{-}(3.0,0)(3.0,0.5)
\end{pspicture}\,) = 
g_2(e_5\, 
\begin{pspicture}[shift=-0.05](-0.0,0)(3.2,0.5)
\psline{-}(0,0)(3.2,0)
\psarc[linecolor=coloroflink,linewidth=1.5pt]{-}(0.8,0){0.2}{0}{180}
\psarc[linecolor=coloroflink,linewidth=1.5pt]{-}(2.0,0){0.2}{0}{180}
\psline[linecolor=coloroflink,linewidth=1.5pt]{-}(0.2,0)(0.2,0.5)
\psline[linecolor=coloroflink,linewidth=1.5pt]{-}(1.4,0)(1.4,0.5)
\psline[linecolor=coloroflink,linewidth=1.5pt]{-}(2.6,0)(2.6,0.5)
\psline[linecolor=coloroflink,linewidth=1.5pt]{-}(3.0,0)(3.0,0.5)
\end{pspicture}\,).
\end{alignat}

Case (1):
If the nodes $j$ and $j+1$ of $w$ are occupied by one defect and one half-arc, the action of $e_j$ on $g_d(w)$ modifies the position of 
one node connected to $y_{d+2}$ and changes one half-arc. Acting on $w$ with $e_j$ first does the same for the corresponding defect 
and half-arc, and applying $g_d$ yields the same result. For example,
\be
e_5g_2(\, 
\psset{unit=0.5}
\begin{pspicture}[shift=-0.05](-0.0,0)(3.2,0.5)
\psline{-}(0,0)(3.2,0)
\psarc[linecolor=coloroflink,linewidth=1.5pt]{-}(1.2,0){0.2}{0}{180}
\psbezier[linecolor=coloroflink,linewidth=1.5pt]{-}(0.6,0)(0.6,0.5)(1.8,0.5)(1.8,0)
\psline[linecolor=coloroflink,linewidth=1.5pt]{-}(0.2,0)(0.2,0.5)
\psline[linecolor=coloroflink,linewidth=1.5pt]{-}(2.2,0)(2.2,0.5)
\psline[linecolor=coloroflink,linewidth=1.5pt]{-}(2.6,0)(2.6,0.5)
\psline[linecolor=coloroflink,linewidth=1.5pt]{-}(3.0,0)(3.0,0.5)
\end{pspicture}\,) = \,
\begin{pspicture}[shift=-0.05](-0.0,0)(3.2,2.3)
\psline{-}(0,0)(3.2,0)
\psline[linecolor=coloroflink,linewidth=1.5pt]{-}(0.2,0)(0.2,0.7)
\psline[linecolor=coloroflink,linewidth=1.5pt]{-}(0.6,0)(0.6,0.7)
\psline[linecolor=coloroflink,linewidth=1.5pt]{-}(1.0,0)(1.0,0.7)
\psline[linecolor=coloroflink,linewidth=1.5pt]{-}(1.4,0)(1.4,0.7)
\psline[linecolor=coloroflink,linewidth=1.5pt]{-}(2.6,0)(2.6,0.7)
\psline[linecolor=coloroflink,linewidth=1.5pt]{-}(3.0,0)(3.0,0.7)
\psarc[linecolor=coloroflink,linewidth=1.5pt]{-}(2.0,0){0.2}{0}{180}
\psarc[linecolor=coloroflink,linewidth=1.5pt]{-}(2.0,0.7){-0.2}{0}{180}
\rput(0,0.7){
\psline{-}(0,0)(3.2,0)
\psline[linecolor=gray]{-}(0,0.5)(3.2,0.5)
\psarc[linecolor=coloroflink,linewidth=1.5pt]{-}(1.2,0){0.2}{0}{180}
\psbezier[linecolor=coloroflink,linewidth=1.5pt]{-}(0.6,0)(0.6,0.5)(1.8,0.5)(1.8,0)
\psline[linecolor=coloroflink,linewidth=1.5pt]{-}(0.2,0)(0.2,0.5)
\psline[linecolor=coloroflink,linewidth=1.5pt]{-}(2.2,0)(2.2,0.5)
\psline[linecolor=coloroflink,linewidth=1.5pt]{-}(2.6,0)(2.6,0.5)
\psline[linecolor=coloroflink,linewidth=1.5pt]{-}(3.0,0)(3.0,0.5)
\rput(0.8,1.2){
\psline{-}(-0.2,0)(1.8,0)
\pspolygon[fillstyle=solid,fillcolor=lightpurple](0,0)(1.6,0)(1.6,0.5)(0,0.5)
\rput(0.8,0.245){$_{_4}$}
}
\psbezier[linecolor=coloroflink,linewidth=1.5pt]{-}(0.2,0.5)(0.2,0.85)(1.0,0.85)(1.0,1.2)
\psbezier[linecolor=coloroflink,linewidth=1.5pt]{-}(2.2,0.5)(2.2,0.85)(1.4,0.85)(1.4,1.2)
\psbezier[linecolor=coloroflink,linewidth=1.5pt]{-}(2.6,0.5)(2.6,0.85)(1.8,0.85)(1.8,1.2)
\psbezier[linecolor=coloroflink,linewidth=1.5pt]{-}(3.0,0.5)(3.0,0.85)(2.2,0.85)(2.2,1.2)
}
\end{pspicture} 
\, = \,
\begin{pspicture}[shift=-0.05](-0.0,0)(3.2,1.7)
\psline{-}(0,0)(3.2,0)
\psline[linecolor=gray]{-}(0,0.5)(3.2,0.5)
\psarc[linecolor=coloroflink,linewidth=1.5pt]{-}(1.2,0){0.2}{0}{180}
\psarc[linecolor=coloroflink,linewidth=1.5pt]{-}(2.0,0){0.2}{0}{180}
\psline[linecolor=coloroflink,linewidth=1.5pt]{-}(0.2,0)(0.2,0.5)
\psline[linecolor=coloroflink,linewidth=1.5pt]{-}(0.6,0)(0.6,0.5)
\psline[linecolor=coloroflink,linewidth=1.5pt]{-}(2.6,0)(2.6,0.5)
\psline[linecolor=coloroflink,linewidth=1.5pt]{-}(3.0,0)(3.0,0.5)
\rput(0.8,1.2){
\psline{-}(-0.2,0)(1.8,0)
\pspolygon[fillstyle=solid,fillcolor=lightpurple](0,0)(1.6,0)(1.6,0.5)(0,0.5)
\rput(0.8,0.245){$_{_4}$}
}
\psbezier[linecolor=coloroflink,linewidth=1.5pt]{-}(0.2,0.5)(0.2,0.85)(1.0,0.85)(1.0,1.2)
\psbezier[linecolor=coloroflink,linewidth=1.5pt]{-}(0.6,0.5)(0.6,0.85)(1.4,0.85)(1.4,1.2)
\psbezier[linecolor=coloroflink,linewidth=1.5pt]{-}(2.6,0.5)(2.6,0.85)(1.8,0.85)(1.8,1.2)
\psbezier[linecolor=coloroflink,linewidth=1.5pt]{-}(3.0,0.5)(3.0,0.85)(2.2,0.85)(2.2,1.2)
\end{pspicture} 
\, = 
\, g_2(\, 
\begin{pspicture}[shift=-0.05](-0.0,0)(3.2,0.5)
\psline{-}(0,0)(3.2,0)
\psarc[linecolor=coloroflink,linewidth=1.5pt]{-}(1.2,0){0.2}{0}{180}
\psarc[linecolor=coloroflink,linewidth=1.5pt]{-}(2.0,0){0.2}{0}{180}
\psline[linecolor=coloroflink,linewidth=1.5pt]{-}(0.2,0)(0.2,0.5)
\psline[linecolor=coloroflink,linewidth=1.5pt]{-}(0.6,0)(0.6,0.5)
\psline[linecolor=coloroflink,linewidth=1.5pt]{-}(2.6,0)(2.6,0.5)
\psline[linecolor=coloroflink,linewidth=1.5pt]{-}(3.0,0)(3.0,0.5)
\end{pspicture}\,) = 
g_2(e_5\, 
\begin{pspicture}[shift=-0.05](-0.0,0)(3.2,0.5)
\psline{-}(0,0)(3.2,0)
\psarc[linecolor=coloroflink,linewidth=1.5pt]{-}(1.2,0){0.2}{0}{180}
\psbezier[linecolor=coloroflink,linewidth=1.5pt]{-}(0.6,0)(0.6,0.5)(1.8,0.5)(1.8,0)
\psline[linecolor=coloroflink,linewidth=1.5pt]{-}(0.2,0)(0.2,0.5)
\psline[linecolor=coloroflink,linewidth=1.5pt]{-}(2.2,0)(2.2,0.5)
\psline[linecolor=coloroflink,linewidth=1.5pt]{-}(2.6,0)(2.6,0.5)
\psline[linecolor=coloroflink,linewidth=1.5pt]{-}(3.0,0)(3.0,0.5)
\end{pspicture}\,).
\ee

Case (2):
If the nodes $j$ and $j+1$ are both occupied by defects, it follows from \eqref{eq:eym0} that $e_j g_d(w) =0$, while 
$g_d(e_j w)=0$ because $e_j$ closes two defects of $w$. For instance,
\be
e_5g_2(\, 
\psset{unit=0.5}
\begin{pspicture}[shift=-0.05](-0.0,0)(3.2,0.5)
\psline{-}(0,0)(3.2,0)
\psarc[linecolor=coloroflink,linewidth=1.5pt]{-}(0.4,0){0.2}{0}{180}
\psarc[linecolor=coloroflink,linewidth=1.5pt]{-}(2.8,0){0.2}{0}{180}
\psline[linecolor=coloroflink,linewidth=1.5pt]{-}(1.0,0)(1.0,0.5)
\psline[linecolor=coloroflink,linewidth=1.5pt]{-}(1.4,0)(1.4,0.5)
\psline[linecolor=coloroflink,linewidth=1.5pt]{-}(1.8,0)(1.8,0.5)
\psline[linecolor=coloroflink,linewidth=1.5pt]{-}(2.2,0)(2.2,0.5)
\end{pspicture}\,) = \,
\begin{pspicture}[shift=-0.05](-0.0,0)(3.2,2.2)
\psline{-}(0,0)(3.2,0)
\psline[linecolor=coloroflink,linewidth=1.5pt]{-}(0.2,0)(0.2,0.7)
\psline[linecolor=coloroflink,linewidth=1.5pt]{-}(0.6,0)(0.6,0.7)
\psline[linecolor=coloroflink,linewidth=1.5pt]{-}(1.0,0)(1.0,0.7)
\psline[linecolor=coloroflink,linewidth=1.5pt]{-}(1.4,0)(1.4,0.7)
\psline[linecolor=coloroflink,linewidth=1.5pt]{-}(2.6,0)(2.6,0.7)
\psline[linecolor=coloroflink,linewidth=1.5pt]{-}(3.0,0)(3.0,0.7)
\psarc[linecolor=coloroflink,linewidth=1.5pt]{-}(2.0,0){0.2}{0}{180}
\psarc[linecolor=coloroflink,linewidth=1.5pt]{-}(2.0,0.7){-0.2}{0}{180}
\rput(0,0.7){
\psline{-}(0,0)(3.2,0)
\psline[linecolor=gray]{-}(0,0.5)(3.2,0.5)
\psarc[linecolor=coloroflink,linewidth=1.5pt]{-}(0.4,0){0.2}{0}{180}
\psarc[linecolor=coloroflink,linewidth=1.5pt]{-}(2.8,0){0.2}{0}{180}
\psline[linecolor=coloroflink,linewidth=1.5pt]{-}(1.0,0)(1.0,0.5)
\psline[linecolor=coloroflink,linewidth=1.5pt]{-}(1.4,0)(1.4,0.5)
\psline[linecolor=coloroflink,linewidth=1.5pt]{-}(1.8,0)(1.8,0.5)
\psline[linecolor=coloroflink,linewidth=1.5pt]{-}(2.2,0)(2.2,0.5)
\rput(0.8,1.2){
\psline{-}(-0.2,0)(1.8,0)
\pspolygon[fillstyle=solid,fillcolor=lightpurple](0,0)(1.6,0)(1.6,0.5)(0,0.5)
\rput(0.8,0.245){$_{_4}$}
}
\psline[linecolor=coloroflink,linewidth=1.5pt]{-}(1.0,0.5)(1.0,1.2)
\psline[linecolor=coloroflink,linewidth=1.5pt]{-}(1.4,0.5)(1.4,1.2)
\psline[linecolor=coloroflink,linewidth=1.5pt]{-}(1.8,0.5)(1.8,1.2)
\psline[linecolor=coloroflink,linewidth=1.5pt]{-}(2.2,0.5)(2.2,1.2)
}
\end{pspicture} 
\, = 0 = 
g_2\big(\,
\begin{pspicture}[shift=-0.05](-0.0,0)(3.2,2.3)
\psline{-}(0,0)(3.2,0)
\psline[linecolor=coloroflink,linewidth=1.5pt]{-}(0.2,0)(0.2,0.7)
\psline[linecolor=coloroflink,linewidth=1.5pt]{-}(0.6,0)(0.6,0.7)
\psline[linecolor=coloroflink,linewidth=1.5pt]{-}(1.0,0)(1.0,0.7)
\psline[linecolor=coloroflink,linewidth=1.5pt]{-}(1.4,0)(1.4,0.7)
\psline[linecolor=coloroflink,linewidth=1.5pt]{-}(2.6,0)(2.6,0.7)
\psline[linecolor=coloroflink,linewidth=1.5pt]{-}(3.0,0)(3.0,0.7)
\psarc[linecolor=coloroflink,linewidth=1.5pt]{-}(2.0,0){0.2}{0}{180}
\psarc[linecolor=coloroflink,linewidth=1.5pt]{-}(2.0,0.7){-0.2}{0}{180}
\rput(0,0.7){
\psline{-}(0,0)(3.2,0)
\psarc[linecolor=coloroflink,linewidth=1.5pt]{-}(0.4,0){0.2}{0}{180}
\psarc[linecolor=coloroflink,linewidth=1.5pt]{-}(2.8,0){0.2}{0}{180}
\psline[linecolor=coloroflink,linewidth=1.5pt]{-}(1.0,0)(1.0,0.5)
\psline[linecolor=coloroflink,linewidth=1.5pt]{-}(1.4,0)(1.4,0.5)
\psline[linecolor=coloroflink,linewidth=1.5pt]{-}(1.8,0)(1.8,0.5)
\psline[linecolor=coloroflink,linewidth=1.5pt]{-}(2.2,0)(2.2,0.5)
}
\end{pspicture}\,\big)  = 
g_2(e_5\, 
\begin{pspicture}[shift=-0.05](-0.0,0)(3.2,0.5)
\psline{-}(0,0)(3.2,0)
\psarc[linecolor=coloroflink,linewidth=1.5pt]{-}(0.4,0){0.2}{0}{180}
\psarc[linecolor=coloroflink,linewidth=1.5pt]{-}(2.8,0){0.2}{0}{180}
\psline[linecolor=coloroflink,linewidth=1.5pt]{-}(1.0,0)(1.0,0.5)
\psline[linecolor=coloroflink,linewidth=1.5pt]{-}(1.4,0)(1.4,0.5)
\psline[linecolor=coloroflink,linewidth=1.5pt]{-}(1.8,0)(1.8,0.5)
\psline[linecolor=coloroflink,linewidth=1.5pt]{-}(2.2,0)(2.2,0.5)
\end{pspicture}\,).
\ee
This concludes the proof. \hfill $\square$ 
\medskip

Because $g_d$ is evidently non-zero and intertwines the standard modules 
\be
 V_n^{d+2}\,\simeq\, \Big(\mathcal I_n^{d+2} 
\begin{pspicture}(0,0)(1,0)
\psline[linewidth=.8pt,arrowsize=3pt 2]{->}(0.2,0.1)(0.8,0.1)
\end{pspicture} 
\mathcal I_n^{d+4}\Big),\qquad
V_n^d\,\simeq\,\Big(\mathcal  I_n^d 
\begin{pspicture}(0,0)(1,0)
\psline[linewidth=.8pt,arrowsize=3pt 2]{->}(0.2,0.1)(0.8,0.1)
\end{pspicture} 
\mathcal I_n^{d+2}\Big)
\ee
where $\mathcal I_n^{d}$, $\mathcal I_n^{d+2}$ and $\mathcal I_n^{d+4}$ are non-isomorphic irreducible modules, we readily obtain the 
following corollary which states that the image of $g_d$ is isomorphic to the submodule $\mathcal I_n^{d+2} \subset V_n^d$.
\begin{Corollaire}
For $0\leq d\leq n-2$ with $d$ and $n$ even, $\mathrm{im}\, g_d\simeq \mathcal I_n^{d+2}$.
\label{sec:gd2}
\end{Corollaire}

For $n$ odd, the construction of an intertwiner like $g_d$ is not possible because the standard modules are inequivalent 
irreducible modules (see (\ref{PVI})) and because the only homomorphism between two such modules is the zero homomorphism.

\section{Proof of the module structure theorem}
\label{sec:Structure}

The intertwiners $J$, $h_{v}$ and $g_d$ introduced in Section~\ref{sec:Intertwiners} are essential to the proof of Theorem~\ref{thm:main} 
presented in this section. Likewise important are the following basic properties of homomorphisms. Let
$f : M \rightarrow N$ be a homomorphism from $M$ to $N$. Then, its kernel $\ker f$ is a submodule of $M$, while its image 
$\textrm{im}\, f$ is a submodule of $N$. 
These properties were already used to obtain Corollary~\ref{sec:gd2}. 
In the following, the kernels and images of $h_{v}$ and $J$ will play crucial roles.

\begin{Proposition}
For $n$ odd, $E_{n-1}^v$ decomposes into the direct sum 
(\ref{EVpV}).
\label{prop:structurenodd}
\end{Proposition}
\noindent{\scshape Proof } 
As argued earlier, it suffices to determine the structure of $E_{n-1}^{v}$ for $v\ge 0$, $E_{n-1}^{-v}$ being contragredient to $E_{n-1}^v$. 
From Lemma~\ref{sec:hvk}, for every $k = 0, \dots, \lfloor \frac{n-1-2v}4 \rfloor$, the map $\tilde h_{v,k}$ is a non-zero homomorphism into 
$E_{n-1}^v$ whose kernel is a submodule that cannot be all of $V_n^{2v+4k+1}$. From the irreducibility of $V_n^{2v+4k+1}$ for $n$ 
odd, $\ker \tilde h_{v,k}$ must therefore be trivial. The image of $\tilde h_{v,k}$ is then an invariant subspace in $E_{n-1}^{v}$ isomorphic 
to $\mathcal I_n^{2v+4k+1}$,
\be
 {\rm im}\,\tilde h_{v,k}\,\simeq\,\mathcal I_n^{2v+4k+1}.
\label{imhI}
\ee
Because irreducible modules $\mathcal I_n^{d}$ with different $d=2v+4k+1$ labels are non-isomorphic, the 
subspaces produced from 
$\tilde h_{v,k}$ at different values of $k$ have no overlap. The direct sum of these submodules is also a submodule of 
$E_{n-1}^v$, and from \eqref{eq:dimensions}, this submodule exhausts the dimension of $E_{n-1}^v$.  \hfill $\square$ 
\medskip 

As the module decompositions for $n$ even are richer than those for $n$ odd, 
cf.~Theorem~\ref{thm:main}, it is no surprise that proving them is also more involved. 
As for $n$ odd, the module $E_{n-1}^{-v}$ is contragredient to $E_{n-1}^{v}$ for $n$ even, so we focus on $E_{n-1}^{v}$ with 
$v\ge \frac12$. To prove the structure of $E_{n-1}^{v}$, we use induction in $v$, 
starting with $v=\frac{n-1}2,\frac{n-3}2$ and decreasing in steps of $2$ until $v = \frac12$ and $v = \frac32$ are reached. 
The induction step establishing the structure of $E_{n-1}^v$ with $\frac12 \le v \le \frac{n-5}2$ is thus based on the 
induction assumption that the structure of $E_{n-1}^{v+2}$ is given by the corresponding zigzag module in \eqref{EII}. 
The step also relies on the following lemma ensuring that the intertwiner $J : E_{n-1}^{v+2} \rightarrow E_{n-1}^v$ is injective. 
In fact, the lemma establishes the injectivity of $J$ for both parities of $n$.

\begin{Lemma}
For $n \geq 3$ and $1 \le v \le \frac{n-1}2$, the operator $J$ is injective on the subspace $E_{n-1}^v$.
\label{LemmaJ}
\end{Lemma} 
\noindent {\sc Proof } 
We first observe that the injectivity of $J$ for $n$ even follows from the injectivity for $n$ odd. Indeed, suppose for $n$ 
even that $|s\ra \in E_{n-1}^v$ is in the kernel of $J$ for some $v \geq \frac32$. 
Focusing on the value of the first spin, one can write 
\be 
 |s\ra = |\!\uparrow,s_1\ra + |\!\downarrow,s_2\ra, \qquad s_1 \in E_{n-2}^{v-1/2}, \quad s_2 \in E_{n-2}^{v+1/2}.
\ee 
The relation $0=\sigma_1^- J|s\ra = -|\!\downarrow,Js_1\ra$ then implies $J|s_1\ra = 0$, and thus $|s_1\ra=0$ by the assumed injectivity 
of $J$ for $n$ odd. From $J|s\ra = -|\!\downarrow,Js_2\ra = 0$, we similarly find $|s_2\ra=0$. It follows that $|s\ra=0$ and hence that $J$ 
is injective. 

Turning to $n$ odd, Proposition~\ref{prop:structurenodd} shows that each composition factor 
$\mathcal I_{n,2v+4k+1}$ of $E_{n-1}^v$ is realised as the image of one of the restricted intertwiners $\tilde h_{v,k}$,
\be
 E_{n-1}^v \simeq \,\textrm{im}\, \tilde h_{v,0} \oplus \,\textrm{im}\, \tilde h_{v,1}  \oplus \dots 
  \oplus  \Big(\,\textrm{im}\, \tilde  h_{v,\frac{n-3-2v}4} \ \ \textrm{or} \ \ \,\textrm{im}\, \tilde  h_{v,\frac{n-1-2v}4} \Big).
\ee
From the definition \eqref{eq:Jhcomp} of the composed intertwiners, applying $J$ from the left 
on both sides yields
\be
 J E_{n-1}^v \simeq \,\textrm{im}\, \tilde h_{v-2,1} \oplus \,\textrm{im}\, \tilde h_{v-2,2}  \oplus \dots 
  \oplus  \Big(\,\textrm{im}\, \tilde  h_{v-2,\frac{n-3-2(v-2)}4} \ \ \textrm{or} \ \ \,\textrm{im}\, \tilde  h_{v-2,\frac{n-1-2(v-2)}4} \Big).
\ee
For $v \ge 2$, Lemma~\ref{sec:hvk} shows that each term in this decomposition, indicated by 
im$\,\tilde h_{v-2,k'}$ with $k' = 1, \dots, \lfloor \frac{n-1-2(v-2)}4\rfloor$, 
is non-zero and therefore irreducible because im$\,\tilde h_{v,k'-1}\simeq\,\mathcal I_n^{2v+4k'-3}$ was itself irreducible. 
The map $J:E_{n-1}^v \rightarrow E_{n-1}^{v-2}$ therefore has a trivial kernel for $v \ge 2$, making it injective.
For $v = 1$, the injectivity follows from the discussion below Lemma~\ref{sec:hvk}.
\hfill $\square$
\medskip 

We are now ready to start the proof of the module decomposition \eqref{EII} for $n$ even. The next lemma shows the structure of $E_{n-1}^v$ 
for the two initial conditions $v=\frac{n-1}2$ and $v=\frac{n-3}2$. It follows, in particular, that the module $E_{n-1}^{(n-3)/2}$ 
is contragredient to the standard module $V_n^{n-2}$. The induction step outlined above Lemma~\ref{LemmaJ} is discussed 
in Proposition~\ref{prop:indecomp}.

\begin{Lemme}
$E_{n-1}^{(n-1)/2} \simeq \mathcal I_n^n$ and $E_{n-1}^{(n-3)/2} \simeq \Big(\mathcal I_n^{n}
\begin{pspicture}(0,0)(1,0)
\psline[linewidth=.8pt,arrowsize=3pt 2]{->}(0.2,0.1)(0.8,0.1)
\end{pspicture} 
\mathcal I_n^{n-2}\Big)$.
\label{sec:n-1}
\end{Lemme}
{\scshape Proof } 
The case $v = \frac{n-1}2$ is trivial since the two modules are one-dimensional and intertwined by $h_{(n-1)/2}$, implying that
\be 
 E_{n-1}^{(n-1)/2}\,\simeq\,V_n^n\,\simeq\,\mathcal I_n^n.
\ee 
For $v = \frac{n-3}2$, the homomorphism $h_{(n-3)/2}$ maps $W_n^{n-2} \simeq \mathcal P_n^{n}\/$ into $E_{n-1}^{(n-3)/2}$. 
Here, $\mathcal P_n^n$ has three composition factors, 
see \eqref{eq:Ps}. Because 
the restricted intertwiner $\tilde h_{(n-3)/2}$ is non-zero, the kernel of the full map $h_{(n-3)/2}$ is either the submodule 
$\mathcal I_n^n$ of $W_{n}^{n-2}$ or trivial. Using
\be
 \dim W_n^{n-2} = \dim E_{n-1}^{(n-3)/2} + 1,
\ee 
dimension counting shows that the kernel cannot be trivial, thus excluding the second option. 
From the intertwining property of $h_{(n-3)/2}$ in Lemma~\ref{sec:intertw} and the structure of $\mathcal P_n^n$, it follows that the image of $h_{(n-3)/2}$ is isomorphic to $W_n^{n-2}/\ker h_{(n-3)/2} 
\simeq \mathcal P_n^{n}/\mathcal I_n^n \simeq \big(\mathcal I_n^{n}
\begin{pspicture}(0,0)(1,0)
\psline[linewidth=.8pt,arrowsize=3pt 2]{->}(0.2,0.1)(0.8,0.1)
\end{pspicture} 
\mathcal I_n^{n-2}\big)$ 
and is a submodule of $E_{n-1}^{(n-3)/2}$ exhausting its dimension. It follows that  
\be 
 E_{n-1}^{(n-3)/2}\simeq \,\Big(\mathcal I_n^{n}
\begin{pspicture}(0,0)(1,0)
\psline[linewidth=.8pt,arrowsize=3pt 2]{->}(0.2,0.1)(0.8,0.1)
\end{pspicture} 
\mathcal I_n^{n-2}\Big),
\ee 
as announced. 
\hfill $\square$
\medskip

Turning to the induction step, we first note that for $n = 2,4$, the previous lemma and the contragredience of $E_{n-1}^v$ and 
$E_{n-1}^{-v}$ give the module structure of $E_{n-1}^v$ for all values of $v$. The following proposition 
establishes the module structure of $E_{n-1}^v$ for $n\ge 6$ even.
\begin{Proposition}
For $n\ge 6$ even and $\frac12 \le v \le \frac{n-5}2$, the module $E_{n-1}^v$ is a reducible yet indecomposable ${\rm TL}_n(0)$ module 
with structure pattern given in (\ref{EII}).
\label{prop:indecomp}
\end{Proposition}
{\scshape Proof } 
The injectivity of $J:E_{n-1}^{v+2} \rightarrow E_{n-1}^v$, established in Lemma~\ref{LemmaJ}, 
and the assumed module structure of $E_{n-1}^{v+2}$ tell us that $E_{n-1}^{v}$ has a submodule isomorphic to $E_{n-1}^{v+2}$,
\be
\begin{pspicture}[shift=-0.5](-0.8,-0.6)(3.7,0.6)
\rput(0,-0.5){$\mathcal I_n^{2v+5}$}
\rput(1.0,0.5){$\mathcal I_n^{2v+7}$}
\rput(1.9,-0.5){$\mathcal I_n^{2v+9}$}
\rput(3.2,0.5){$\dots$}
\psline[linewidth=.8pt,arrowsize=3pt 2]{->}(0.6,0.25)(0.1,-0.2)
\psline[linewidth=.8pt,arrowsize=3pt 2]{->}(1,0.25)(1.5,-0.2)
\psline[linewidth=.8pt,arrowsize=3pt 2]{->}(2.6,0.25)(2.1,-0.2)
\end{pspicture}
\subset E_{n-1}^v.
\label{eq:firstsub}
\ee
The intertwiner $h_{v}$ maps $W_n^{2v+1} \simeq \mathcal P_n^{2v+3}$ into $E_{n-1}^{v}$ and 
offers further insight. Because $\mathcal P_n^{2v+3}$ has four composition factors and $\tilde h_{v}$ is non-zero on 
the submodule $V_n^{2v+1}$, there are three possibilities for the kernel of $h_v$, namely 
trivial, $\mathcal I_n^{2v+3}$ and $\big(\mathcal I_n^{2v+5}
\begin{pspicture}(0,0)(1,0)
\psline[linewidth=.8pt,arrowsize=3pt 2]{->}(0.2,0.1)(0.8,0.1)
\end{pspicture} 
\mathcal I_n^{2v+3}\big)
$, 
all of which are submodules of $W_n^{2v+1}$. Noting that 
\be 
 \dim E_{n-1}^v - \dim E_{n-1}^{v+2} = \dim V_n^{2v+1} = \dim \mathcal I_n^{2v+3} + \dim \mathcal I_n^{2v+1},
\ee 
dimension counting excludes the possibility that $\ker h_v$ is trivial since, in that case, the image of $h_v$ would contain
two copies of $\mathcal I_n^{2v+3}$. 
Moreover, Lemma~\ref{prop:lastnonzero} below states that the map $h_v$ is not identically zero on the composition factor 
$\mathcal I_n^{2v+5}$ of $W_n^{2v+1}$. Because $\mathcal I_n^{2v+5}$ is irreducible, it therefore appears, as a whole, in the image of 
$h_v$. This eliminates the third option, 
$\ker h_v \simeq 
\big(\mathcal I_n^{2v+5}
\begin{pspicture}(0,0)(1,0)
\psline[linewidth=.8pt,arrowsize=3pt 2]{->}(0.2,0.1)(0.8,0.1)
\end{pspicture} 
\mathcal I_n^{2v+3}\big)
$, and 
thus yields
\be
{\rm im}\, h_v \simeq 
\begin{pspicture}[shift=-0.5](-0.8,-0.6)(2.5,0.6)
\rput(0,-0.5){$\mathcal I_n^{2v+1}$}
\rput(1.0,0.5){$\mathcal I_n^{2v+3}$}
\rput(1.9,-0.5){$\mathcal I_n^{2v+5}$}
\psline[linewidth=.8pt,arrowsize=3pt 2]{->}(0.6,0.25)(0.1,-0.2)
\psline[linewidth=.8pt,arrowsize=3pt 2]{->}(1,0.25)(1.5,-0.2)
\end{pspicture}
\subset E_{n-1}^v.
\label{eq:secondsub}
\ee
Again by dimension counting, the composition factors $\mathcal I_n^{2v+1}, \mathcal I_n^{2v+3}, \dots, \mathcal I_n^n$, 
that appear in either \eqref{eq:firstsub}, \eqref{eq:secondsub} or both, can only have multiplicity $1$ in $E_{n-1}^v$. 
In particular, even though $\mathcal I_n^{2v+5}$ 
appears in both \eqref{eq:firstsub} and \eqref{eq:secondsub}, it is only present once as a composition factor of $E_{n-1}^v$, so the zigzag 
chain \eqref{eq:firstsub} extends to the left and includes the two extra factors $\mathcal I_n^{2v+1}$ and $\mathcal I_n^{2v+3}$,
as announced in \eqref{EII}.

The previous argument has shown that $\textrm{im}\, h_{v}$ and $\textrm{im}\,J$ constitute a pair of submodules 
of $E_{n-1}^v$ that combine to form an indecomposable submodule of $E_{n-1}^v$ exhausting its dimension. There are therefore no 
other composition factors. Moreover, we have the full set of arrows in the Loewy diagram, as extra arrows would either contradict the fact that 
both $\textrm{im}\, h_v$ and $\textrm{im}\, J$ are submodules, or
already have appeared in either $\textrm{im}\, h_v$ or $\textrm{im}\, J$. This concludes the proof of the module structure of $E_{n-1}^v$. 
\hfill $\square$ 
\medskip

The only remaining thing to show to complete the proof of Theorem~\ref{thm:main} is the following lemma which was used in the proof of 
Proposition~\ref{prop:indecomp}.

\begin{Lemme} 
For $n$ even and $v\geq\frac{1}{2}$, the action of $h_v$ on the composition 
factor $\mathcal I_n^{2v+5}$ in $W_{n}^{2v+1}$ is not identically zero.
\label{prop:lastnonzero}
\end{Lemme}
{\scshape Proof } 
The factor $\mathcal I_n^{2v+5}$ is a submodule of $V_n^{2v+3} \simeq W_n^{2v+1}/V_n^{2v+1}$ and, from Corollary~\ref{sec:gd2}, 
it finds a basis in the image of the map $g_{2v+3}$. We explicitly calculate the action of $h_v g_{2v+3}$ on the state $w$ whose $2v+5$ 
leftmost nodes are occupied by defects while the remaining nodes are linked pairwise by simple half-arcs only,
\begin{alignat}{1}
h_v&\big(g_{2v+3}(w)\big)=h_v\big(g_{2v+3}(\,
\psset{unit=0.5}
\begin{pspicture}[shift=-1.05](-0.0,-1)(6.0,0.5)
\psline{-}(0,0)(6.0,0)
\psline[linecolor=gray,linestyle=dashed,dash=2pt 1pt]{-}(3.2,-0.3)(3.2,0.8)
\psarc[linecolor=coloroflink,linewidth=1.5pt]{-}(3.6,0){0.2}{0}{180}
\psarc[linecolor=coloroflink,linewidth=1.5pt]{-}(4.4,0){0.2}{0}{180}
\psarc[linecolor=coloroflink,linewidth=1.5pt]{-}(5.6,0){0.2}{0}{180}
\psline[linecolor=coloroflink,linewidth=1.5pt]{-}(0.2,0)(0.2,0.5)
\psline[linecolor=coloroflink,linewidth=1.5pt]{-}(0.6,0)(0.6,0.5)
\psline[linecolor=coloroflink,linewidth=1.5pt]{-}(1.0,0)(1.0,0.5)
\psline[linecolor=coloroflink,linewidth=1.5pt]{-}(1.4,0)(1.4,0.5)
\psline[linecolor=coloroflink,linewidth=1.5pt]{-}(1.8,0)(1.8,0.5)
\psline[linecolor=coloroflink,linewidth=1.5pt]{-}(2.6,0)(2.6,0.5)
\psline[linecolor=coloroflink,linewidth=1.5pt]{-}(3.0,0)(3.0,0.5)
\rput(2.2,0.25){.\hspace{-0.025cm}.\hspace{-0.025cm}.}
\rput(5.0,0.25){.\hspace{-0.025cm}.\hspace{-0.025cm}.}
\rput(1.56,-0.7){$\underbrace{\quad \hspace{0.7cm}\quad}_{2v+5}$}
\end{pspicture}
\,)\big) = 
h_v\big(\,
\begin{pspicture}[shift=-0.05](-0.0,0)(6.0,1.0)
\psline{-}(0,0)(6.0,0)
\psline[linecolor=gray,linestyle=dashed,dash=2pt 1pt]{-}(3.2,-0.3)(3.2,0.8)
\psarc[linecolor=coloroflink,linewidth=1.5pt]{-}(3.6,0){0.2}{0}{180}
\psarc[linecolor=coloroflink,linewidth=1.5pt]{-}(4.4,0){0.2}{0}{180}
\psarc[linecolor=coloroflink,linewidth=1.5pt]{-}(5.6,0){0.2}{0}{180}
\psline[linecolor=coloroflink,linewidth=1.5pt]{-}(0.2,0)(0.2,0.5)
\psline[linecolor=coloroflink,linewidth=1.5pt]{-}(0.6,0)(0.6,0.5)
\psline[linecolor=coloroflink,linewidth=1.5pt]{-}(1.0,0)(1.0,0.5)
\psline[linecolor=coloroflink,linewidth=1.5pt]{-}(1.4,0)(1.4,0.5)
\psline[linecolor=coloroflink,linewidth=1.5pt]{-}(1.8,0)(1.8,0.5)
\psline[linecolor=coloroflink,linewidth=1.5pt]{-}(2.6,0)(2.6,0.5)
\psline[linecolor=coloroflink,linewidth=1.5pt]{-}(3.0,0)(3.0,0.5)
\rput(2.2,0.25){.\hspace{-0.025cm}.\hspace{-0.025cm}.}
\rput(5.0,0.25){.\hspace{-0.025cm}.\hspace{-0.025cm}.}
\rput(0.2,0.5){
\pspolygon[fillstyle=solid,fillcolor=lightpurple](-0.1,0)(2.9,0)(2.9,0.5)(-0.1,0.5)
\rput(1.4,0.235){$_{_{2v+5}}$} 
}
\end{pspicture}
\,\big) \nonumber\\
& = h_v\big(\,
\psset{unit=0.5}
\begin{pspicture}[shift=-0.05](-0.0,0)(6.0,0.5)
\psline{-}(0,0)(6.0,0)
\psline[linecolor=gray,linestyle=dashed,dash=2pt 1pt]{-}(3.2,-0.3)(3.2,0.8)
\psarc[linecolor=coloroflink,linewidth=1.5pt]{-}(0.4,0){0.2}{0}{180}
\psarc[linecolor=coloroflink,linewidth=1.5pt]{-}(3.6,0){0.2}{0}{180}
\psarc[linecolor=coloroflink,linewidth=1.5pt]{-}(4.4,0){0.2}{0}{180}
\psarc[linecolor=coloroflink,linewidth=1.5pt]{-}(5.6,0){0.2}{0}{180}
\psline[linecolor=coloroflink,linewidth=1.5pt]{-}(1.0,0)(1.0,0.5)
\psline[linecolor=coloroflink,linewidth=1.5pt]{-}(1.4,0)(1.4,0.5)
\psline[linecolor=coloroflink,linewidth=1.5pt]{-}(1.8,0)(1.8,0.5)
\psline[linecolor=coloroflink,linewidth=1.5pt]{-}(2.6,0)(2.6,0.5)
\rput(3.0,0){\wobblytwo}
\rput(2.2,0.25){.\hspace{-0.025cm}.\hspace{-0.025cm}.}
\rput(5.0,0.25){.\hspace{-0.025cm}.\hspace{-0.025cm}.}
\end{pspicture}
\,\big) - 
h_v\big(\,
\begin{pspicture}[shift=-0.05](-0.0,0)(6.0,0.5)
\psline{-}(0,0)(6.0,0)
\psline[linecolor=gray,linestyle=dashed,dash=2pt 1pt]{-}(3.2,-0.3)(3.2,0.8)
\psarc[linecolor=coloroflink,linewidth=1.5pt]{-}(1.2,0){0.2}{0}{180}
\psarc[linecolor=coloroflink,linewidth=1.5pt]{-}(3.6,0){0.2}{0}{180}
\psarc[linecolor=coloroflink,linewidth=1.5pt]{-}(4.4,0){0.2}{0}{180}
\psarc[linecolor=coloroflink,linewidth=1.5pt]{-}(5.6,0){0.2}{0}{180}
\psline[linecolor=coloroflink,linewidth=1.5pt]{-}(0.2,0)(0.2,0.5)
\psline[linecolor=coloroflink,linewidth=1.5pt]{-}(0.6,0)(0.6,0.5)
\psline[linecolor=coloroflink,linewidth=1.5pt]{-}(1.8,0)(1.8,0.5)
\psline[linecolor=coloroflink,linewidth=1.5pt]{-}(2.6,0)(2.6,0.5)
\rput(3.0,0){\wobblytwo}
\rput(2.2,0.25){.\hspace{-0.025cm}.\hspace{-0.025cm}.}
\rput(5.0,0.25){.\hspace{-0.025cm}.\hspace{-0.025cm}.}
\end{pspicture}
\,\big) + \dots + (-1)^{v+\frac32}
h_v\big(\,
\begin{pspicture}[shift=-0.05](-0.0,0)(6.0,0.5)
\psline{-}(0,0)(6.0,0)
\psline[linecolor=gray,linestyle=dashed,dash=2pt 1pt]{-}(3.2,-0.3)(3.2,0.8)
\psarc[linecolor=coloroflink,linewidth=1.5pt]{-}(2.8,0){0.2}{0}{180}
\psarc[linecolor=coloroflink,linewidth=1.5pt]{-}(3.6,0){0.2}{0}{180}
\psarc[linecolor=coloroflink,linewidth=1.5pt]{-}(4.4,0){0.2}{0}{180}
\psarc[linecolor=coloroflink,linewidth=1.5pt]{-}(5.6,0){0.2}{0}{180}
\psline[linecolor=coloroflink,linewidth=1.5pt]{-}(0.2,0)(0.2,0.5)
\psline[linecolor=coloroflink,linewidth=1.5pt]{-}(0.6,0)(0.6,0.5)
\psline[linecolor=coloroflink,linewidth=1.5pt]{-}(1.0,0)(1.0,0.5)
\psline[linecolor=coloroflink,linewidth=1.5pt]{-}(1.8,0)(1.8,0.5)
\rput(2.2,0){\wobblytwo}
\rput(1.4,0.25){.\hspace{-0.025cm}.\hspace{-0.025cm}.}
\rput(5.0,0.25){.\hspace{-0.025cm}.\hspace{-0.025cm}.}
\end{pspicture}
\,\big).
\label{eq:hgw}
\end{alignat}
Note that a dashed delimiter has been included to indicate the virtual interface between the $2v+5$ leftmost nodes and the remaining 
ones. Each term in (\ref{eq:hgw}) can be evaluated individually,
\begin{alignat}{2}
&h_v\big(\,
\psset{unit=0.5}
\begin{pspicture}[shift=-1.15](-0.4,-1.1)(6.0,0.5)
\psline{-}(-0.4,0)(6.0,0)
\psline[linecolor=gray,linestyle=dashed,dash=2pt 1pt]{-}(3.2,-0.3)(3.2,0.8)
\psarc[linecolor=coloroflink,linewidth=1.5pt]{-}(1.2,0){0.2}{0}{180}
\psarc[linecolor=coloroflink,linewidth=1.5pt]{-}(3.6,0){0.2}{0}{180}
\psarc[linecolor=coloroflink,linewidth=1.5pt]{-}(4.4,0){0.2}{0}{180}
\psarc[linecolor=coloroflink,linewidth=1.5pt]{-}(5.6,0){0.2}{0}{180}
\psline[linecolor=coloroflink,linewidth=1.5pt]{-}(-0.2,0)(-0.2,0.5)
\psline[linecolor=coloroflink,linewidth=1.5pt]{-}(0.6,0)(0.6,0.5)
\psline[linecolor=coloroflink,linewidth=1.5pt]{-}(1.8,0)(1.8,0.5)
\psline[linecolor=coloroflink,linewidth=1.5pt]{-}(2.6,0)(2.6,0.5)
\rput(3.0,0){\wobblytwo}
\rput(0.2,0.25){.\hspace{-0.025cm}.\hspace{-0.025cm}.}
\rput(2.2,0.25){.\hspace{-0.025cm}.\hspace{-0.025cm}.}
\rput(5.0,0.25){.\hspace{-0.025cm}.\hspace{-0.025cm}.}
\rput(1.0,-0.4){\tiny $\uparrow$}
\rput(1.0,-0.8){\scriptsize$_{2j+1}$}
\rput(3.0,-0.4){\tiny $\uparrow$}
\rput(3.0,-0.8){\scriptsize$_{2v+5}$}
\end{pspicture}
\,\big) = \sigma^-_{2v+4}\big(\sigma^-_{2j}+\sigma^-_{2j+2}\big)\sigma^-_{2v+5}\sigma^-_{2v+7} \dots \sigma^-_{n-2}| u \rangle,
\qquad (j = 0, \dots, v + \tfrac12)\\
&h_v\big(\,
\psset{unit=0.5}
\begin{pspicture}[shift=-1.15](-0.4,-1.1)(6.0,0.5)
\psline{-}(-0.4,0)(6.0,0)
\psline[linecolor=gray,linestyle=dashed,dash=2pt 1pt]{-}(3.2,-0.3)(3.2,0.8)
\psarc[linecolor=coloroflink,linewidth=1.5pt]{-}(2.8,0){0.2}{0}{180}
\psarc[linecolor=coloroflink,linewidth=1.5pt]{-}(3.6,0){0.2}{0}{180}
\psarc[linecolor=coloroflink,linewidth=1.5pt]{-}(4.4,0){0.2}{0}{180}
\psarc[linecolor=coloroflink,linewidth=1.5pt]{-}(5.6,0){0.2}{0}{180}
\psline[linecolor=coloroflink,linewidth=1.5pt]{-}(-0.2,0)(-0.2,0.5)
\psline[linecolor=coloroflink,linewidth=1.5pt]{-}(0.2,0)(0.2,0.5)
\psline[linecolor=coloroflink,linewidth=1.5pt]{-}(0.6,0)(0.6,0.5)
\psline[linecolor=coloroflink,linewidth=1.5pt]{-}(1.0,0)(1.0,0.5)
\psline[linecolor=coloroflink,linewidth=1.5pt]{-}(1.8,0)(1.8,0.5)
\rput(2.2,0){\wobblytwo}
\rput(2.2,-0.4){\tiny $\uparrow$}
\rput(2.2,-0.8){\scriptsize$_{2v+3}$}
\rput(1.4,0.25){.\hspace{-0.025cm}.\hspace{-0.025cm}.}
\rput(5.0,0.25){.\hspace{-0.025cm}.\hspace{-0.025cm}.}
\end{pspicture}
\,\big) = \sigma^-_{2v+2}\sigma^-_{2v+3}\sigma^-_{2v+5}\sigma^-_{2v+7} \dots \sigma^-_{n-2}| u \rangle,
\end{alignat}
where $\sigma^-_0 \equiv 0$. It follows that
\begin{alignat}{2}
h_v\big(g_{2v+3}(w)\big)& = \Big[\sigma^-_{2v+4} \Big( \sum_{j=0}^{v+\frac12}(-1)^{j} (\sigma^-_{2j}+\sigma^-_{2j+2})\Big) 
 + (-1)^{v+\frac32} \sigma^-_{2v+2}\sigma^-_{2v+3} \Big] \sigma^-_{2v+5}\sigma^-_{2v+7} \dots \sigma^-_{n-2} |u\rangle
\nonumber \\
& = (-1)^{v+\frac 32} (\sigma^-_{2v+2}-\sigma^-_{2v+4}) \sigma^-_{2v+3}\sigma^-_{2v+5}\sigma^-_{2v+7} \dots \sigma^-_{n-2} |u\rangle
\end{alignat}
which is non-zero. 
\hfill $\square$
\medskip

\section{Conclusion}
\label{sec:Conclusion}

We have introduced a new spin-chain representation of the Temperley-Lieb algebra TL$_n(0)$. It is related to Lieb's transfer matrix in the dimer model and, for $n$ even, decomposes 
as a direct sum of indecomposable zigzag representations. These results immediately beg the question of whether similar constructions are possible for $\beta\neq0$. The results are also likely to yield insight into the continuum scaling limit and conformal properties of Lieb's transfer matrix approach to 
the dimer model. We hope to discuss these important issues elsewhere.

\subsection*{Acknowledgments}

AMD is supported by the National Sciences and Engineering Research Council of Canada as a postdoctoral fellow. 
JR is supported by the Australian Research Council under the Future Fellowship scheme, project number FT100100774. 
PR is Senior Research Associate of the Belgian Fonds National de la Recherche Scientifique (FNRS), and acknowledges the support
of the Belgian Interuniversity Attraction Poles Program P7/18 through the network DYGEST (Dynamical, Geometry and Statistical 
Physics). The authors thank David Ridout and Yvan Saint-Aubin for valuable comments and discussions, as well as an anonymous referee for useful comments.

%

\end{document}